\documentclass[reprint,twocolumn,prx,superscriptaddress,longbibliography]{revtex4-2}
\usepackage{amssymb}
\usepackage{amsmath}
\usepackage{graphicx,bm}
\usepackage{epstopdf}
\usepackage{subfigure}
\usepackage{natbib}
\usepackage{epsfig}
\usepackage{amsfonts}
\usepackage{mathrsfs}
\usepackage{CJK}
\usepackage[toc,page,title,titletoc,header]{appendix}
\usepackage{array,longtable}

\usepackage[usenames, dvipsnames, x11names]{xcolor}

\usepackage[colorlinks]{hyperref}
\hypersetup{
 colorlinks=true,
 citecolor=red,
 linkcolor=blue,
 urlcolor=cyan}

\newcommand{\cmatH}{\check{\mathcal{H}}}
\newcommand{\cmatU}{\check{\mathcal{U}}}

\newcommand{\cmat}[1]{\check{\mathcal{#1}}}

\newcommand{\nlqnr}{ \mathrm{dir}  }

\usepackage{comment}

\begin{document}
\title{Quantum nonreciprocal interactions via dissipative gauge symmetry}

\author{Yu-Xin Wang}
\affiliation{Pritzker School of Molecular Engineering,  University  of  Chicago,  Chicago,  Illinois  60637,  U.S.A.}

\author{Chen Wang}
\affiliation{Department of Physics, University of Massachusetts-Amherst, Amherst, MA, USA}

\author{Aashish A. Clerk}
\affiliation{Pritzker School of Molecular Engineering,  University  of  Chicago,  Chicago,  Illinois  60637,  U.S.A.}

\date{\today}

\begin{abstract}
One-way nonreciprocal interactions between two quantum systems are typically described by a cascaded quantum master equation, and rely on an effective breaking of time-reversal symmetry as well as the balancing of coherent and dissipative interactions. Here, we present a new approach for obtaining nonreciprocal quantum interactions that is \emph{completely distinct} from cascaded quantum systems, and that does not in general require broken TRS.  Our method relies on a local gauge symmetry present in any Markovian Lindblad master equation.  This new kind of quantum nonreciprocity has many implications, including a new mechanism for performing dissipatively-stabilized gate operations on a target quantum system.  We also introduce a new, extremely general quantum-information based metric for quantifying quantum nonreciprocity.      
\end{abstract}

\maketitle

\section{Introduction}

\label{sec:intro}

The study of interactions and scattering that are intrinsically directional (i.e.~nonreciprocal) is at the forefront of many areas of physics.  Such interactions are of fundamental interest:  for example, they can lead to exotic phase transitions in classical active matter 
systems~\cite{Marchetti2020,Vitelli2021,Vitelli2021review}, and can also be used to generate dimerized many-body entangled states~\cite{Zoller2012,Zoller2015,Zoller2017}.  They also have a myriad of practical applications in both classical and quantum information processing tasks, in settings that range from classical photonic and acoustic systems~\cite{Fan2012,Alu2014,Haberman2020},  to quantum circuits and networks~\cite{Devoret2011,Devoret2014diramp,Devoret2015,Verhagen2016,Aumentado2017,Painter2017,Teufel2017,Kippenberg2017,Harris2019,Nunnenkamp2018}.

\begin{figure}[t]
    \centering
    \includegraphics[width=\columnwidth]{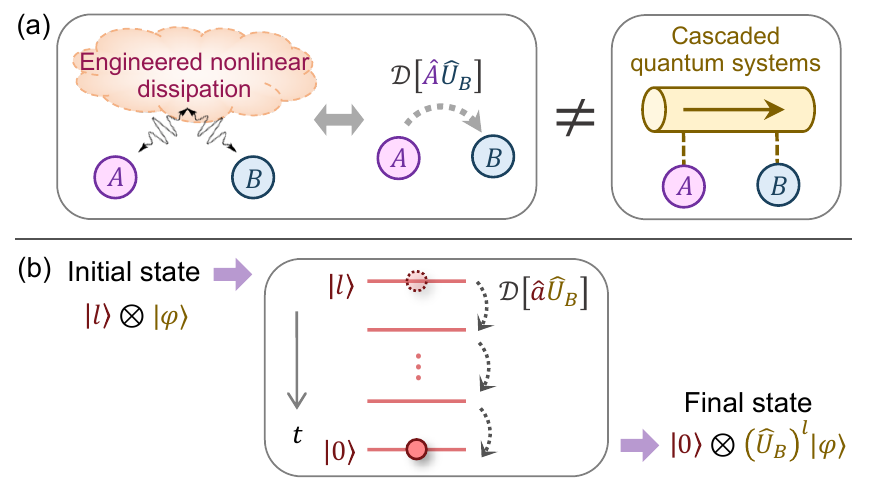}
    \caption{(a) Schematic for quantum nonreciprocal interactions via gauge symmetry.  This approach uses correlated dissipation, but is completely distinct from a cascaded quantum system.  (b) Dissipative stabilization of a tunable unitary gate making use of the nonreciprocal interaction in (a), in a system where subsystem $A$ is a cavity mode $a$ that controls the gate, and subsystem $B$ is the target qubit. 
    One can selectively apply a unitary gate $(\hat U _{B})^{\ell} $ on $B$ by initializing $A$ in the corresponding Fock state 
    $| \ell \rangle$, and letting the system relax to the steady state (see Eq.~\eqref{eq:dissp.ss.qop} and Sec.~\ref{sec:qnr.diss.gate}). }
    \label{fig:schematic}
\end{figure}

While classically, one can describe directional interactions using effective non-Hermitian Hamiltonians, in quantum settings one needs a description that conserves probability and accounts for quantum fluctuations.  The standard quantum description of nonreciprocity is provided by the theory of cascaded quantum systems~\cite{Gardiner1993,Carmichael1993}.  It describes an extremely general class of fully directional interactions between two subsystems $A$ and $B$ that involve a pair of  arbitrary ``local" operators $\hat{A}$ and $\hat{B}$ (i.e.~$\hat{A}$ only acts on subsystem $A$, $\hat{B}$ only acts on subsystem $B$).  The directional dynamics is described by a Lindblad quantum master equation (QME) of the form
\cite{Gardiner1993,Carmichael1993,Clerk2015}:  
\begin{align} 
& \frac{  d \hat \rho }{ dt } 
	 = - i [  
	 \hat H _{AB}
	,  \hat \rho    ] 
	+ \mathcal{D}  
	[ \hat A  - i  \hat B     ] 
	   \hat \rho  
    \equiv  
    \mathcal{L}  _{\mathrm{CS}} 
     \hat \rho
	, 
	\label{eq:cqme.gen} 
\end{align} 
where $\hat{\rho}$ is the system density matrix,  $ \hat H _{AB} =
( \hat A   ^\dag  \hat B  
+ \hat B   ^\dag  \hat A )/2 $, and $  \mathcal{D} [\hat{O}  ]  
\hat{\rho} = 
 (\hat{O}  \hat{\rho} \hat{O} ^\dagger 
- \{\hat{O} ^\dagger \hat{O}
, \hat{\rho}\} /2  ) $ denotes the standard Lindblad dissipator. One can show that the dynamics encodes a fully one-way interaction where subsystem $A$ affects the dynamics and evolution of subsystem $B$, but not vice versa.

Cascaded QMEs were first derived for setups involving an explicitly nonreciprocal element (e.g.~a directional waveguide or circulator).  More recently, it was realized that Eq.~(\ref{eq:cqme.gen}) provides a more general blueprint for engineering nonreciprocal interactions, based on balancing a coherent Hamiltonian interaction (described by $\hat{H}_{AB}$) and a dissipative interaction (described by the dissipator  
$\mathcal{D}  [     
\hat A  - i  \hat B ] 
\hat \rho $) \cite{Clerk2015,Clerk2017}.  This can be realized by engineering suitable drives and couplings to dissipative environments, without using a conventional nonreciprocal element.  This approach has been employed in a variety of experiments including quantum optomechanics (e.g.~\cite{Verhagen2016,Painter2017,Teufel2017,Kippenberg2017,Harris2019}) and superconducting quantum circuits (e.g.~\cite{Devoret2015,Aumentado2017}).  Note that the coefficient $-i$ in the dissipator of Eq.~(\ref{eq:cqme.gen}) ultimately implies that any physical means for realizing this dynamics requires an effective broken time-reversal symmetry (TRS).  This can be achieved by using phases encoded in the drive tones applied to the system, in a manner that generates a synthetic gauge flux.  This time-modulation approach to nonreciprocity is also well studied in completely classical contexts~\cite{Fan2012,Alu2014,Haberman2020}.


One might guess that the general structure of Eq.~(\ref{eq:cqme.gen}) is the only way to obtain fully-directional, Markovian quantum interactions between two systems.  In this work we show that this is not the case.  We introduce a new kind of quantum open system dynamics where correlated dissipation generates nonreciprocal interactions in a manner distinct from the cascaded QME.  As we discuss in detail, we ultimately exploit a basic gauge symmetry present in any Lindblad master equation.  This gives us a mechanism for nonreciprocity that, surprisingly, does not require any notion of broken TRS or a non-trivial synthetic gauge field.  In its simplest form, the new nonreciprocal QME can be written in terms of a generic local operator $ \hat A $ on $A$ and a unitary $ \hat U _{B} $ on $B$, as (see Fig.~\ref{fig:schematic}(a))
\begin{align}
& \frac{  d \hat \rho }{ dt } 
= \Gamma \mathcal{D} [  
  \hat A   \hat U _{B}   ] 
\hat \rho  
\equiv 
\mathcal{L}  _{ \nlqnr } \hat \rho   
	.  
	\label{eq:nl.nr.ME} 
\end{align} 
As we show, this purely dissipative dynamics is strictly unidirectional ($A$ influences $B$ but not vice-versa), and moreover, cannot be written in the cascaded QME form of  Eq.~\eqref{eq:cqme.gen}.  We also show that this structure has a non-trivial generalization to more complex master equations with multiple dissipators.

Our work provides a thorough investigation of this new route to quantum nonreciprocity, including implementation methods.  We also introduce a new, very general metric for quantum nonreciprocity that uses quantum-information theoretic tools, and use this to characterize the nonreciprocity of our mechanism in the presence of imperfections.  Our study also goes beyond just fundamental considerations.    
We discuss a potentially powerful application of our new directional dynamics:  a method for realizing dissipatively-stabilized unitary quantum gate operations.  By ``stabilized", we mean here that the gate operation is encoded in the dissipative steady state of the dynamics.
The basic idea is sketched in Fig.~\ref{fig:schematic}(b).  Starting from a bipartite system having control ($A$) and target ($B$) subsystems, the goal is for the dissipative relaxation of the system (under a dynamics of the form of Eq.~(\ref{eq:nl.nr.ME})) to encode a unitary operation on $B$ whose form is dictated by the initial state of $A$.  Specifically, for a set of initial states for $A$ indexed by $\lambda$, we can achieve: 
\begin{align}
    \hat{\rho}_B(\infty) & = 
\lim 
_{ t \to \infty}
	\text{Tr} _{ A }
	\left [ 
	 e ^{ t \mathcal{L} _{ \nlqnr } } 
	( \hat \rho _{A} 
	(\lambda)
	\otimes 
 \hat \rho _{B} ) 
	\right]  
	    \nonumber \\
	& =   \hat U _{B}(\lambda) 
	\hat \rho _{B} 
	\hat U _{B}(\lambda) 
	^\dag .
	\label{eq:dissp.ss.qop}
\end{align}   
The steady state of $B$ is related to its initial state by a unitary, whose form is dictated by the initial $A$ state.  As we discuss in detail, this mechanism for performing gates has several attractive features:  it does not require any timing control, and its dissipative nature makes it robust against certain kinds of errors.   We stress that the mechanism of Eq.~\eqref{eq:dissp.ss.qop} is completely distinct from previous works exploring alternative dissipative approaches to quantum control
\cite{Verstraete2009,CamposVenuti2014,Jiang2016,Metelmann2020,Sudarshan1977,Pascazio2002,Pascazio2008,Kempf2016}.
  

The remainder of the paper is organized as follows.  In Sec.~\ref{sec:qnr.iso.def}, we set the stage for our discussions of quantum nonreciprocity by introducing a very general quantum information metric that quantifies quantum nonreciprocal (QNR) interactions in an arbitrary system, in a state-independent manner. Sec.~\ref{sec:qnr.sing} introduces our new mechanism for QNR in the simplest setting, and discusses its application to dissipatively-stabilized gate operations. Sec.~\ref{sec:phys.impl} discusses physical implementation strategies that are compatible with state-of-the-art superconducting circuit and quantum optical platforms. Sec.~\ref{sec:qnr.multi} generalizes our new mechanism to more complex cases with multiple dissipators, and demonstrates that these QNR interactions can generate entanglement.  We conclude in Sec.~\ref{sec:discussions}.

\section{Quantifying nonreciprocity of general quantum dynamics}

\label{sec:qnr.iso.def}

\subsection{Basic notions}

Before introducing our new mechanism, we start with a more basic question:  what is a fundamental, system-agnostic way of identifying and quantifying nonreciprocal dynamics?  While in simple linear settings one can just look at the asymmetry of scattering-matrix coefficients at a particular input frequency, we would like a more general metric that can apply even when there is no obvious connection to scattering, and which is not contingent on a particular choice of initial state.  As we now discuss, we can formulate such a metric using well-known quantum information-theoretic quantities.

We start with a generic bipartite system with subsystems $A$ and $B$, whose dynamics are described by the evolution superoperator $\mathcal{E} ^{(AB)} _{t}$.  This superoperator tells us how the system density matrix evolves, i.e.~ 
\begin{align}
\hat{ \rho } _{AB} (t )
&   
 =\mathcal{E} ^{(AB)} _{t}
 \hat{ \rho } _{AB} ( 0 )
	.    
	\label{eq:chan.gen.AB}
\end{align} 
$\mathcal{E} ^{(AB)} _{t}$ is also known as a quantum map or channel; any physical evolution corresponds to such a map, with the requirement that $\mathcal{E} ^{(AB)} _{t}$ be completely positive and trace-preserving (CPTP)~\cite{NielsenChuang2010}.
We stress that this description encompasses a full range of dynamics from non-dissipative unitary evolution, to highly complex non-Markovian dissipative evolution.  

We now want a metric that tells us whether the dynamics describes by $\mathcal{E} ^{(AB)} _{t}$ is nonreciprocal.  
We first introduce the 
{\it isolation function} of subsystem $A$, ${I} ^{(A)}   (t )$, which quantifies how sensitive the evolution of $A$ is over this time interval to the initial state of subsystem $B$.  The isolation function of subsystem $B$, ${I} ^{(B)}   (t )$, will be defined in an analogous manner.  
${I} ^{(A)}   (t )$ can be directly connected to a standard task in quantum information theory.  Suppose we first prepare subsystem $B$ in one of two given states,   
$\left | \phi_{1} 
\right\rangle$ or $\left | \phi_{2} 
\right\rangle$ with equal probability.  We also prepare $A$ in some state $\hat{\rho}_A$.  
We then let the total system evolve for time $t$.  We thus have two possible time-evolution maps for system $A$, contingent on the two subsytem B initial states:
\begin{align}
\mathcal{E} ^{(A)}
_{ \left |\phi _{\mathrm{i} }
\right\rangle } 
( t  )  
\hat \rho _{A} 
	&   \equiv  \text{Tr} _{ B }
	\left [ 
	\mathcal{E} ^{(AB)} _{t} 
	( \hat \rho _{A} 
	\otimes 
 \left |\phi _{\mathrm{i} } 
\right\rangle_{ B }  
\! \left \langle
\phi _{\mathrm{i} } \right |) 
	\right] 
	\label{eq:A.chan.res.def}
	.   
\end{align}

The goal is now to optimally guess which initial state $B$ we started with, using only a {\it single} measurement on the $A$ system at time $t$.  We are interested in the maximum success probability where we optimize over all $A$ initial states as well as the final $A$-subsystem measurement.  This probability $p _{ \mathrm{max }}  
( \{ \left | \phi_{1} 
\right\rangle , 
\left | \phi_{2} 
\right\rangle \} ) $ gives us a measure of how different the $A$ system dynamics is depending on the choice of initial $B$ state.  One finds~\cite{Watrous2018}:
\begin{align}
p _{ \mathrm{max }}  
( \{ \left | \phi_{1} 
\right\rangle , 
\left | \phi_{2} 
\right\rangle \} ) 
	&   =  \frac{1}{2} 
+ \frac{1}{4}
	|| \mathcal{E} ^{(A)}
	_{ 
\left |\phi _{ 1 }
\right\rangle } 
( t  )  
-  \mathcal{E} ^{(A)}
_{ \left |\phi _{ 2 }
\right\rangle } 
( t  )   
	||_{\diamond}
	\label{eq:qchan.dist.prob}
	.     
\end{align} 
Here, $  || \cdot ||_{\diamond}$ denotes the so-called diamond norm and provides a distance measure between two quantum channels~\cite{Kitaev1997}, which is stable with respect to tensor product operations (i.e.~attaching ancillary quantum systems to $A$).  Note that $p_{\rm max}$ must lie in the interval $[0.5,1]$.

Eq.~\eqref{eq:qchan.dist.prob} thus provides a fundamental metric for the sensitivity of the $A$ system dynamics to a change in the intial state of $B$.  This then directly leads to a fundamental notion of how isolated the $A$ system dynamics is from $B$:  further optimize Eq.~\eqref{eq:qchan.dist.prob} over the choice of the $B$ system initial states.  This leads us to define the subsystem $A$ isolation as:
\begin{align} 
{I} ^{(A)} (t ) 
& \equiv  1 - \frac{1}{2} 
\max _{ \left | \phi_{1} 
\right\rangle , 
\left | \phi_{2} 
\right\rangle
\in \mathcal{H} _{B} }
|| \mathcal{E} ^{(A)}
	_{ \left |\phi _{ 1 }
\right\rangle } 
( t  )  
-  \mathcal{E} ^{(A)}
_{ \left |\phi _{ 2 }
\right\rangle } 
( t  )   
	||_{\diamond}
, 
\label{eq:iso.min.def}
\end{align} 
where $\mathcal{H} _{B}$ denotes Hilbert space of $B$. The isolation function ${I} ^{(A)} (t ) $ lies in the interval $[0,1]$, and measures the maximal influence a change of initial subsystem $B$ state could have on the $A$ subsystem dynamics.  The case of complete isolation, ${I} ^{(A)} (t ) =1$, implies that the dynamics of $A$ is completely independent of the initial state of subsystem $B$.  The isolation function for subsystem $B$ is defined in a completely analogous manner.  Note that if $A$ and $B$ are not coupled at all in the dynamics
(i.e. the total channel is a tensor product of independent channels for each subsystem), then both subsystems are fully isolated at all times: ${I}^{(A)} (t )  = {I}^{(B)} (t ) = 1$.  For $t=0$, both systems are also of course always trivially isolated as the total channel is the identity.  

\subsection{Instantaneous nonreciprocity}

These isolation functions now give us a simple way of identifying nonreciprocal dynamics as an evolution map that yields 
${I} ^{(A)}   (t ) \neq {I} ^{(B)}   (t )$, i.e.~a situation where there is an asymmetry in how strongly $A$ influences $B$ versus how strongly $B$ influences $A$.  
One can discuss nonreciprocity for the instantaneous quantum map at a specific time, as well as for the entire evolution. If we focus on a specific time, we can define $\mathcal{E} ^{(AB)} _{t} $ as being reciprocal or nonreciprocal at time $t$ using the isolation functions, i.e.
\begin{align} 
&  {I} ^{(A)} (t )  
= {I} ^{(B)} 
(t )  
\nonumber \\
\Rightarrow & 
 \mathcal{E} ^{(AB)} _{t} 
\text{ is instantaneously reciprocal at } 
t 
;
\\
 & {I} ^{(A)} (t )  
\ne {I} ^{(B)} (t ) 
\nonumber \\
\Rightarrow  
&  \mathcal{E} ^{(AB)} _{t} 
\text{ is instantaneously } 
\text{nonreciprocal at } 
t
. 
\label{eq:qnr.map.def}
\end{align}

\subsection{Global nonreciprocity}

One could also ask about whether the dynamics is nonreciprocal over an entire time interval $[0,t]$.  In this case, we can define reciprocity by insisting the isolations are identical over the entire time interval:
\begin{align} 
\forall t  \in  
( 0,+\infty ),  
&  \, \, 
{I} ^{(A)} (t )  
= {I} ^{(B)} 
(t ) \, 
\nonumber \\
&\Rightarrow 
\, 
\text{dynamics is reciprocal} 
;
\label{eq:def.dyn.recip}
 \\
\exists t \in ( 0,+\infty ) 
\,\, 
& \text{s.t.} \,\, 
{I} ^{(A)} (t )  
\ne {I} ^{(B)} 
(t ) 
\,
\nonumber \\
&\Rightarrow 
\, 
\text{dynamics is nonreciprocal} 
. 
\label{eq:def.nr}
\end{align}

\subsection{Fully nonreciprocal dynamics}

Finally, one is also often interested in identify situations of full nonreciprocity, where one system is unaffected by the other, but is nonetheless still able to influence it.  We first consider quantum map at a specific time, and define instantaneous full nonreciprocity (i.e.~unidirectionality) from $A$ to $B$ for $\mathcal{E} ^{(AB)} _{t} $ at time $t$ as 
\begin{align} 
{I} ^{(A)} (t ) 
= 1, \,
{I} ^{(B)} 
(t )  <1
& \Rightarrow 
\,
\mathcal{E} ^{(AB)} _{t} 
\text{ is instantaneously } 
\nonumber \\
& 
\text{unidirectional } 
 (A \to B) 
.   
\end{align} 
Physically, the conditions on LHS can be understood as ensuring that states of $A$ can affect evolution of $B$, but not vice versa. One can also define maximal unidirectionality from $A$ to $B$ (at time $t$) as any evolution that yields ${I} ^{(A)} (t ) =1 $ and ${I} ^{(B)} (t ) = 0 $. More generally, one can also define fully nonreciprocal dynamics via the condition that dynamics of $A$ is fully isolated at all times, but $B$ is not fully isolated at some time, as 
\begin{align} 
\forall t, \,\, 
{I} ^{(A)} (t ) 
= 1 , 
\text{ and }
\,
\exists t \,\, 
\text{s.t.} \,\, 
{I} ^{(B)} (t )  
<1
.  
\label{eq:qnr.full.def}
\end{align} 
The case of $B$-to-$A$ full nonreciprocity can be similarly defined by interchanging $A$ and $B$ in Eq.~\eqref{eq:qnr.full.def}.  We stress that having fully isolated $A$ dynamics is a necessary but not sufficient condition for full nonreciprocity from $A$ to $B$.  In fact, it is possible to have dynamics generated by nontrivial interactions that is isolated in both directions, i.e.~${I} ^{(A)} (t ) 
= {I} ^{(B)} (t ) =1 $ (for an example, see Sec.~\ref{sec:qnr.sing.other}).

\subsection{Physical intuition and example cases}

For a variety of simple test cases, our formal definitions of reciprocity and nonreciprocity agree with simple intuition.  
For example, it is easy to show that if the starting bipartite system is uncoupled, or is symmetric under permutation of $A$ and $B$ labels, then its dynamics is automatically reciprocal as per the definition in Eq.~\eqref{eq:def.dyn.recip}.
Our definition also does more than simply quantify asymmetry of the bipartite system. As an example, in Appendix~\ref{appsec:iso.recip.disp} we consider a class of highly asymmetric bipartite, non-dissipative systems that are always reciprocal as per our definition in  Eq.~\eqref{eq:def.dyn.recip}.
These systems take the $B$ subsystem to be a qubit, the $A$ system to be {\it arbitrary}, and take the two subsystems to interact via coupling Hamiltonian that commutes with the $B$-only Hamiltonian.  
Another interesting test case is where $A$ and $B$ are both single qubits.  In this case, if the evolution is an arbitrary unitary, then it must be fully reciprocal (see Appendix~\ref{appsec:iso.recip.2qb}).

To gain intuition about the opposite limit of full nonreciprocity, it is useful to examine cases where dynamics of $A$ is fully isolated.  This is of course a necessary condition for fully unidirectional dynamics (see Eq.~\eqref{eq:qnr.full.def}), but is of course not sufficient. One can show if $B$ can be exactly traced out from the total system dynamics, then the $A$ isolation by our definition stays unity throughout the time evolution, i.e. 
\begin{align}
\forall t,	 \,\, 
\text{Tr} _{ B }
	\left [ 
	\mathcal{E} ^{(AB)} _{t} 
	( \hat \rho _{A} 
	\otimes 
\hat \rho _{B} ) 
	\right] 
=  \mathcal{E} ^{(A)} _{t} 
	\hat \rho _{A} 
	\Rightarrow
\forall t,  \,\, 
{I} ^{(A)} (t ) 
\equiv 1  
	.   
	\label{eq:cond.fully.iso}
\end{align}  
Here $\mathcal{E} ^{(A)} _{t}$ is a local superoperator acting on $A$ and is independent of $\hat \rho _{B}$. As a result, dynamics of a generic cascaded quantum systems from $A$ to $B$ (see Eq.~\eqref{eq:cqme.gen}) must be fully isolated in terms of subsystem $A$. Furthermore, because $A$ cannot be exactly traced out from the system dynamics, the dynamics of $B$ can be affected by $A$, so that dynamics generated by Eq.~\eqref{eq:cqme.gen} is fully nonreciprocal by Eq.~\eqref{eq:qnr.full.def}.

\section{Quantum nonreciprocity via generalized gauge symmetry}

\label{sec:qnr.sing}

\subsection{Gauge-invariance nonreciprocity with a single dissipator}

\label{sec:qnr.sing.recipe}

We now introduce our new method for realizing nonreciprocal quantum dynamics via an open system Markovian dynamics that is {\it distinct} from cascaded quantum systems.
We begin with the simplest case of a Lindblad master equation with a single dissipator, leaving generalizations to Sec.~\ref{sec:qnr.multi}. 
We start with a seemingly trivial observation for a single, generic Lindblad dissipator on system $A$.  Such a dynamics is described by 
\begin{align}
&   \mathcal{L} 
_{ A , \mathrm{1}} 
\hat \rho _{A} 
	= 
	\Gamma 
	\mathcal{D}  [   
  \hat A  
  ]  \hat \rho  _{A}
	. 
	\label{eq:qme.sing}
\end{align} 
It is straightforward to see that this Lindbladian is invariant under an arbitrary gauge transformation of the jump operator
$\hat A \to \hat A  
e ^{i \theta 
\left( t  \right) }  $, where $ \theta  \left( t  \right) $ can be an arbitrary time-dependent real function. This invariance can formally corresponds to a local (in time) gauge symmetry of a generic Lindblad dissipator. 

We can use this trivial insensitivity of the dynamics to $\theta(t)$ to now obtain a nonreciprocal interaction between two systems: simply replace the classical time-dependent phase with a quantum operator acting on a different quantum system $B$:  $ \theta  \left( t  \right) 
\to \hat \theta _{B} $. 
As shown in Fig.~\ref{fig:schematic}(a), we now rewrite the phase factor in the jump operator as unitary operator $ \hat U _{B} $ acting on subsystem $B$.  We thus obtain a new QME (see also Eq.~\eqref{eq:nl.nr.ME}):  
\begin{align}
& \mathcal{L}  _{ \nlqnr } 
\hat \rho   
= \Gamma \mathcal{D} [  
  \hat A   \hat U _{B}   ] 
\hat \rho  
	.  
	\label{eq:nl.nr.Lindbladian} 
\end{align} 
One can easily show that gauge invariance property discussed above ensures that the dynamics of $A$ is insensitive to $B$.  More explicitly, consider a general master equation where the interaction between $A$ and $B$ is given by Eq.~\eqref{eq:nl.nr.Lindbladian}: 
$({  d \hat \rho } / { dt } ) 
= ( \mathcal{L} _{A, \mathrm{i}} 
+ \mathcal{L} _{B, \mathrm{i}} )  \hat \rho
+ \Gamma \mathcal{D} [  
  \hat A   \hat U _{B}   ] 
\hat \rho $, with 
$\mathcal{L} _{A(B), \mathrm{i}} $ describing internal dynamics of $A$ ($B$). One can exactly trace out $B$ to obtain a QME for the $A$ reduced density matrix 
$\hat \rho _{A} =  
\text{Tr} _{ B } \hat \rho $ alone, as 
\begin{align}
& \frac{  d \hat \rho _{A} }{ dt } 
= \mathcal{L} _{A, \mathrm{i}} 
\hat \rho _{A} 
+ \Gamma \mathcal{D} [  
\hat A ]  
\hat \rho _{A} 
	.  
	\label{eq:nl.nr.ME.A} 
\end{align}
However, the converse is in general not true:  $B$ will be in general influenced by $A$, i.e.~its evolution is sensitive to the initial state of $A$ as well as 
$ \mathcal{L} _{A, \mathrm{i}}$.  The only exception is the case where
$\hat A $ is proportional to a unitary, see also Sec.~\ref{sec:qnr.sing.other}.

The more formal definitions of nonreciprocity introduced in Sec.~\ref{sec:qnr.iso.def}
also yield an identical picture. Because the equation of motion of the $A$ subsystem is independent of the $B$ state (see Eq.~\eqref{eq:nl.nr.ME.A}), 
it follows that the $A$ isolation must be unity throughout time evolution, 
i.e.~${I} ^{(A)} (t ) 
\equiv 1$ for all $t$. For the $B$ isolation, assuming $\hat A $ is not proportional to a unitary operator, one can generally show that 
${I} ^{(B)} 
(t ) <1$ for some time $t$; see also Sec.~\ref{sec:qnr.sing.a.isoB} for a concrete example with a bosonic lowering operator as  $\hat A $. Thus, according to the new metric based on isolation functions, the QME in Eq.~\eqref{eq:nl.nr.Lindbladian} describes fully nonreciprocal dynamics from $A$ to $B$ as long as we have $\hat A ^\dag \hat A  
\not\propto \hat {\mathbb{I}} $.

We stress that Eq.~\eqref{eq:nl.nr.Lindbladian} describes a generic nonreciprocal open-systems dynamics that is distinct from a cascaded quantum system:  it {\it cannot} be written in the form of a cascaded QME, Eq.~\eqref{eq:cqme.gen}.
Our new approach in Eq.~\eqref{eq:nl.nr.Lindbladian} can be written as a Liouvillian that has no Hamiltonian part, and that has a single dissipator with a jump operator that is a \emph{product} of an $A$ operator and a $B$ operator.  In marked contrast, the cascaded quantum systems QME in  Eq.~\eqref{eq:cqme.gen} has a Hamiltonian in its Liouvillian, and a jump operator that is the {\it sum} of a  subsystem-$A$ operator and a subsystem-$B$ operator.  These cannot be made equivalent.  At a more physical level, the differences in jump operators correspond to different forms of system-bath coupling.  The inequivalence also implies that the nonreciprocal interaction described by Eq.~\eqref{eq:nl.nr.Lindbladian} \textit{cannot} be realized by coupling $A$ and $B$ to a directional waveguide (see Fig.~\ref{fig:schematic}(a)).

\subsection{Example: photon-loss dissipator}

\label{sec:qnr.sing.a.isoB}

To make our ideas more concrete, consider a simple case where the $A$ subsystem in Eq.~\eqref{eq:nl.nr.Lindbladian} is a bosonic mode, and $\hat{A}$ is taken to be the photon lowering operator $\hat{a}$ for this mode.  
Further, take an initial state where $A$ is unentangled with $B$, and is prepared either in the vacuum state $|0 \rangle$, or in the Fock state $| \ell \rangle$ ($\ell > 0$).   
From Eq.~\eqref{eq:iso.min.def}, we can thus obtain an upper limit of the corresponding $B$ isolation in the long-time $t \to \infty$ limit as
\begin{align}
{I} ^{(B)}  ( \infty )
\le 1 - \frac{1}{2}  
\lim _{t \to \infty} 
|| \mathcal{E} ^{(B)}
	_{ \left | 0
\right\rangle } 
( t  )  
-  \mathcal{E} ^{(B)}
_{ \left | \ell 
\right\rangle } 
( t  )   
	||_{\diamond} 
	.
\end{align} 
One can also show (see Sec.~\ref{sec:qnr.diss.gate}) that the subsystem-$B$ evolution maps appearing in this equation have an extremely simple form:
\begin{align}
\lim _{t \to \infty}
\mathcal{E} ^{(B)}
_{ \left | n 
\right\rangle } 
( t  )  
\hat \rho _{B} 
& = \hat U _{B} ^{ n }  
\hat \rho _{B}   
\hat U _{B} ^{ \dag n }   
.  
\end{align}
Intuitively, this describes a dissipative process where each time a photon is lost from the $A$ cavity, subsystem $B$ undergoes a unitary evolution $ \hat U _{B} $. 
We can thus derive an upper bound for the corresponding $B$ isolation in the long-time $t \to \infty$ limit as~\cite{dnorm_isom} 
\begin{align}
  &  
{I} ^{(B)}  ( \infty )
\le   1 -
\sqrt{ 1 - 
\min _{\left |\phi  
\right\rangle \in 
\mathcal{H} _{B} } 
| \langle \phi  
| \hat U _{B} ^{ \ell }
\left |\phi  
\right\rangle   
	| ^{2}
}  	.    
\label{eq:qnr.sing.Biso}
\end{align} 
Letting $ e ^{i \beta  _{m}  }$ denote the eigenvalues $\hat U _{B} $, the RHS of Eq.~\eqref{eq:qnr.sing.Biso} can be further rewritten explicitly as 
\begin{align}
{I} ^{(B)}  ( \infty )
\le 1 -  
\max  
_{ m,n } 
\left | \sin 
\frac{ \ell (\beta  _{m}  
- \beta  _{n}  )}{2} \right | 
. 
\label{eq:qnr.sing.Biso.fin}
\end{align} 
Thus, the $B$ isolation in the long-time limit is less than $1$ for any nontrivial unitary $\hat U _{B} ^{ \ell }$ (i.e.~not proportional to identity map), signalling nontrivial influence from $A$ to $B$. The isolation reaches minimal value of zero if 
$\hat U _{B} ^{ \ell }$ has two eigenvalues with relative $\pi$ phase difference, in which case the long-time  evolution becomes maximally nonreciprocal.

\subsection{Dissipative quantum gates mediated by new form of nonreciprocal interaction}

\label{sec:qnr.diss.gate}

Our new nonreciprocal QME has many interesting features.  Here, we focus on a potentially powerful application: the implementation of dissipatively-stabilized unitary gate operations on subsystem $B$, whose form is controlled by the initial state of subsystem $A$.  The most generic way to realize this is to construct a dynamics of the form of Eq.~\eqref{eq:nl.nr.Lindbladian}, where $\hat{A}$ has a subspace of dark states $\mathcal{D}$:  if $|d \rangle \in \mathcal{D}$, then $\hat{A} | d \rangle = 0$.  Further, let $\mathcal{S}$ denote the set of states in the
intersection between the orthogonal complement of $\mathcal{D}$ and the inverse image of $\mathcal{D}$ under $\hat{A}$.  A given state $|\psi \rangle \in \mathcal{S}$ is both orthogonal to the dark state subspace, and has the property $\hat{A} | \psi \rangle$ is in $\mathcal{D}$ (i.e.~a single action of $\hat{A}$ results in a dark state). We now have a simple way to obtain our dissipative gate:
\begin{itemize}
    \item{At $t=0$ the full system is taken to be in a product state $\hat{\rho}_{AB}(0) = \hat{\rho}_A(0) \otimes \hat{\rho}_B(0)$.}  
    
    \item{If we don't want a gate operation to be performed on $B$, we start subsystem $A$ in an arbitrary dark state in $\mathcal{D}$.  In this case, there is no evolution under Eq.~\eqref{eq:nl.nr.Lindbladian}, and the subsystem $B$ state is unchanged. }
    
    \item{To turn the gate on, we instead prepare subsystem $A$ in an arbitrary state in $\mathcal{S}$.  In this case, there is non-trivial evolution under Eq.~\eqref{eq:nl.nr.Lindbladian}.  To achieve the gate operation, one just waits until the system reaches its steady state.  The dissipative steady state will be
    \begin{align}
        \hat{\rho}_{AB}(\infty) = \hat{\rho}'_A \otimes \left(
            \hat{U}_B \hat{\rho}_B(0) \hat{U}_B^\dagger \right)
            ,
    \end{align}
    where $\hat{\rho}'_A$ is in $\mathcal{D}$, i.e.~it is a dark state.  The final state of $B$ is related to the initial state by the unitary $\hat{U}_B$.  }
\end{itemize}
We stress that this approach realizes a gate operation on system $B$ in the dissipative steady state; no precise timing control is needed.  The only control that is needed is to prepare subsystem $A$ at $t=0$ in a state in the subspace $\mathcal{S}$.  This control also need not be perfect, as any state in this manifold (pure or impure) will lead to the desired gate operation.

An even more versatile kind of controllable dissipative gate is possible if $\hat{A}$ has the general structure of a lowering operator.  By this, we mean that within a given subspace, $\hat{A}$ is a matrix that only has non-zero entries along the super-diagonal.  This is exactly the situation we have if $\hat{A}$ is a bosonic lowering operator $\hat{a}$, hence we consider this case in what follows. As illustrated in Fig.~\ref{fig:schematic}(b), if the control subsystem $A$ is initialized in a Fock state $  | \ell \rangle $, the long-time dynamics of the target system effectively applies 
$\hat U _{B}  $ $\ell$ times on the initial target state, so that we have 
\begin{align} 
& \lim _{ t \to \infty} 
	\text{Tr} _{ A }
	\left \{
	e ^{  \mathcal{L}  _{\nlqnr } t  }
	\left[  | \ell \rangle \langle \ell | 
	\otimes  \hat \rho _{B} 
	\left (  0
	\right) \right] 
	\right \}
	=  ( \hat U _{B} )^{ \ell }  
	\hat \rho _{B}  
	\left (  0
	\right)
	( \hat U _{B}^\dag)^\ell .
\end{align} 
This recipe allows one to apply e.g.~tunable phase gates on a target qubit.

We note that this approach to dissipatively-stabilized gate operations is distinct from previous works exploring dissipative quantum control.  
For example, Ref.~\cite{Verstraete2009} focused on dissipatively stabilizing a {\it unique} steady state that effectively realizes a quantum computational task. In contrast, we are dissipatively stabilizing a unitary operation on subsystem $B$, not a unique state.  Our dissipator in  Eq.~\eqref{eq:nl.nr.Lindbladian} has of course multiple steady states, something that is exploited by our protocol.   
Alternatively, other works utilize dynamics with steady-state degeneracy, and propose to use strong dissipation~\cite{CamposVenuti2014,Jiang2016,Metelmann2020}, measurements~\cite{Sudarshan1977,Pascazio2002,Pascazio2008}, or fast repetitive resets~\cite{Kempf2016} to mimic Hamiltonian evolution of a target system.  Unlike our work, there is no stabilization here:  one needs to shut off the dynamics at a particular time in order to achieve a particular unitary (whereas we achieve the unitary in the long-time steady state).

\subsection{Gauge-invariance nonreciprocity: other generic cases}

\label{sec:qnr.sing.other}

We now discuss the physics of our nonreciprocal dynamics in Eq.~\eqref{eq:nl.nr.Lindbladian} for different generic choices of the subsystem-$A$ operator (beyond the lowering operator case discussed above).  Consider first the case where $\hat{A}$ is itself unitary.  In this case, there is no asymmetry in our dissipator (i.e.~both subsystem operators are unitary), and correspondingly we would expect that there cannot be any directional interaction.  This is indeed what occurs:  in this case, it is easy to confirm that both systems are isolated from one another.  The only way to see signatures of the interaction  would be to consider the evolution of correlations between them.  As discussed in~\cite{maassen1987,viola2019}, this type of dissipative dynamics can be understood as coupling both $A$ and $B$ to the same classical Poisson point process. It is then straightforward to show that time evolution of both of subsystems is independent of the other, and the correlation between classical stochastic processes coupled to $A$ versus $B$ is only discernible if one looks at $AB$ correlators.

Another general case is where the $ \hat  A$ operator in Eq.~\eqref{eq:nl.nr.Lindbladian} is Hermitian. While the dynamics in this case is directional from $A$ to $B$, we can exactly solve for the time evolution in terms of eigenbasis of the jump operator. The system dynamics now allows a simple interpretation, i.e.~$B$ dephases at rates depending on $A$ state, but not vice versa.

\subsection{Connection to measurement-and-feedforward processes}

\label{sec:qnr.sing.mf}

Given that a large class of standard quantum cascaded systems can be intuitively understood as being equivalent to a measurement-and-feedforward (MF) process~\cite{Wiseman1994,Clerk2017}, it is worth discussing the relation between our nonlinear dissipator in Eq.~\eqref{eq:nl.nr.Lindbladian} and MF protocols. As shown in~\cite{Wiseman1994}, dissipators given by Eq.~\eqref{eq:nl.nr.Lindbladian} can be realized via a generalized MF process as $e ^{ \mathcal{L}  _{ \nlqnr }\delta t } \hat \rho  
=   \sum _{\ell=1,2}\hat M_{\ell }  \hat \rho  \hat M_{ \ell } ^{\dag}$. Here, the Kraus operators $ \hat M_{\ell } $ are given by (to order $\delta t$) 
\begin{align}
\hat M_{ 1 } = \sqrt{ \Gamma \delta t} 
\hat A \hat U _{B}, 
\quad 
\hat M_{ 2 } = 1 - \Gamma  
\hat A  ^{\dag} \hat A 
\delta t 
, 
\label{eq:mmff.kraus}
\end{align}
which satisfy the normalization condition
$\sum _{\ell=1,2} \hat M_{ \ell } ^{\dag}\hat M_{\ell } = \hat{\mathbb{I}}$. Intuitively, this stochastic process corresponds to weakly measuring $A$, and subsequently applying unitary transformations on $B$ conditioned on the measurement results.  This interpretation provides a simple, complementary way to understand the directionality of our dynamics in the single dissipator case.  It also tells us that this dynamics can never generate entanglement between $A$ and $B$.  

While for a single dissipator, both the gauge-invariance picture and MF picture let us understand the directionality, the same is not true for the multiple dissipator case analyzed in Sec.~\ref{sec:qnr.multi}.  In this case, the gauge invariance pictures ensures nonreciprocity, but there is no mapping onto a MF process (and in fact, the dynamics can create entanglement).

\subsection{Non-Markovian effects}

\label{sec:nonMarkov.sing.gen}

As we have stressed, the nonreciprocity of the dissipative dynamics in our basic dissipator of Eq.~\eqref{eq:nl.nr.Lindbladian} is directly related to the local gauge invariance of a standard Lindblad dissipator.  This effective gauge symmetry however only emerges in the limit of a Markovian bath, something we discuss in detail in Appendix~\ref{appsec:symm.sing}. Physically, it requires the bath correlation time $\tau _{\mathrm{E}}$ to be much smaller than the timescale associated with variation of the gauge phase $\theta(t)$: 
$ \tau _{\mathrm{E}} \ll 
[\dot{\theta}  \left( t  \right)] ^{-1} $.  If this condition is met, then the bath is effectively only sensitive to the instantaneous value of $\theta(t)$, and there is no difference between a constant in time $\theta(t)$ versus a time varying phase.  Conversely, if changes in the gauge phase $ \theta  \left( t  \right) $ are not negligible during the bath correlation time, the dynamics of $ \theta  \left( t  \right) $ will induce non-Markovian effects in the bath, and the system dynamics will no longer be gauge invariant. 

The above picture can be made rigorous, and one can calculate leading non-Markovian corrections.  In Appendix~\ref{appsec:nonM.sing}, we consider microscopic bath model with a system-environment (SE) interaction Hamiltonian of the form
$ \hat H _{\mathrm{SE}} = 
e ^{i \theta \left( t  \right) } 
\hat \xi ^{\dag }   
\hat A  + \text{H.c.}$.  Here, $\hat \xi $ is the bath operator that couples to the system.  We derive the leading-order correction to Eq.~\eqref{eq:qme.sing} due to a finite bath correlation time $ \tau_{\mathrm{E}} $, a correction which scales as $ \tau_{\mathrm{E}} 
\dot{\theta} 
\left ( t \right) $.
Further, in Sec.~\ref{sec:phy.impl.nonM}, we work with an explicit quantum realization of our dissipative scheme, and use the isolation functions defined in Sec.~\ref{sec:qnr.iso.def} to quantify how a finite bath correlation time causes deviations from full nonreciprocity.



\section{Physical implementation in cavity QED systems}

\label{sec:phys.impl}

\subsection{Basic setup}

\label{sec:phys.setup}


\begin{figure}[t]
    \centering
    \includegraphics[width=\columnwidth]{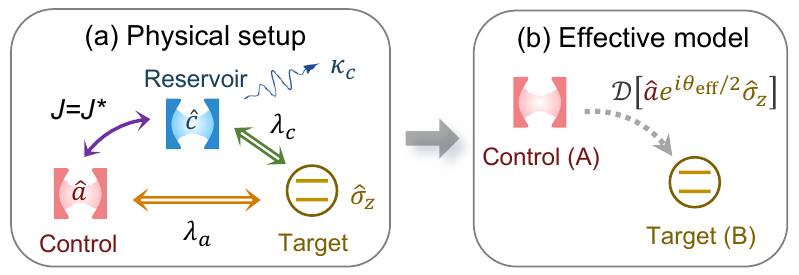}
    \caption{(a) Schematic of a  setup that realizes a nonreciprocal interaction from a cavity mode $a$ ($A$) to a qubit ($B$). Both subsystems are coupled to an auxiliary damped bosonic mode $c$ that plays the role of a reservoir, see Eq.~\eqref{eq:bath.qme.gen}. 
    (b) The effective dissipator describing the nonreciprocal interaction in the Markovian reservoir limit $\kappa_{c} \gg J$.  
    }
    \label{fig:setup_2mode}
\end{figure}

We now discuss methods for implementing the general nonreciprocal dynamics of Eq.~\eqref{eq:nl.nr.Lindbladian} in a quantum optical setup.  One direct approach would be to explicitly break TRS, and use  standard nonreciprocal elements like a circulator or a chiral waveguide.  Formally, such implementations involve starting with a larger system described by a cascaded master equation, eliminating degrees of freedom, and then obtaining the effective dynamics of  Eq.~\eqref{eq:nl.nr.Lindbladian}.    We discuss a generic method for doing this (starting with a chiral waveguide) in App.~\ref{appsec:phys.impl.chiral}.  Note that several recent circuit QED experiments using nonreciprocal elements could be usefully interpreted in this way \cite{Nakamura2018,Wallraff2018,Wallraff2020}
(see App.~\ref{appsec:phys.impl.chiral} for more details).

We focus here on a more intriguing implementation strategy that uses reservoir engineering techniques but {\it does not require any elements that explicitly break time-reversal symmetry}.   We take subsystem $A$ to be a resonator mode with bosonic annihilation operator $  \hat a$, and subsystem $B$ to be a qubit with Pauli $z$ operator $\hat \sigma _{z}$.  We wish to realize our nonreciprocal master equation Eq.~\eqref{eq:nl.nr.Lindbladian}  with the choices $\hat{A} = \hat{a}$, and $\hat{U} _{B} = \exp(-i \theta \hat\sigma_z / 2)$, i.e.
\begin{align}
&  \frac{  d \hat \rho }{ dt }	
=  \mathcal{L}  _{\mathrm{gate}}  
	\hat \rho 
	= \Gamma \mathcal{D}  [   
e ^{ - i \frac{  \theta }{  2  } \hat \sigma_z } 
	 \hat a    ]  \hat \rho  
	. 
	\label{eq:qnr.cavqb.ME}
\end{align} 

To engineer the effective dynamics in Eq.~\eqref{eq:qnr.cavqb.ME}, we couple both the cavity mode $a $ and the qubit to an auxiliary, highly damped bosonic mode $\hat{c}$ (decay rate $ \kappa _{c} $), via tunneling and dispersive interactions respectively (see Fig.~\ref{fig:setup_2mode}(a)). The interaction Hamiltonian 
between the system and the $c$ mode (i.e.~the reservoir) is:
\begin{align}
    \hat{H}_\mathrm{int} =  
    ( J  \hat a  ^\dag \hat c
+ \text{H.c.}  
)  
+ ( \lambda  _{c}  /2 )
\hat \sigma_z \hat c^\dag \hat c
,
\label{eq:phys.imp.Hint}
\end{align}
where $ J $ denotes the complex tunnel coupling rate. 
We will also include a direct Hamiltonian dispersive coupling between the cavity mode $\hat{a}$ and the qubit, 
$ \hat H _\mathrm{S} =
 ( \lambda _{a} /2 )
\hat \sigma_z \hat a^\dag \hat a  $.  We take the $a$, $c$ and qubit to all be resonant, and work in a common rotating frame.  The total dynamics (including the reservoir $c$ mode) is then described by the QME: 
\begin{align}
& \frac{  d \hat \rho _\mathrm{tot} 
}{ dt }
	=  \mathcal{L}  _{\mathrm{SR}}   
	\hat \rho _\mathrm{tot} 
	=
	- i  [ \hat H_\mathrm{S} 
	+ \hat H_\mathrm{int} 
	, \hat \rho _\mathrm{tot}   ] 
	+   \kappa _{c} \mathcal{D} 
    [  \hat c ]  
    \hat \rho _\mathrm{tot} 
	\label{eq:bath.qme.gen}
	.   
\end{align} 

Note that Eq.~\eqref{eq:bath.qme.gen} does not involve any nontrivial gauge phase or breaking of TRS. Even if $J$ is complex, the corresponding hopping phase can always be eliminated by a gauge transformation on $\hat{a}$; hence, the phase of $J$ plays no role.   This can be understood physically from the fact that the setup in Eq.~\eqref{eq:bath.qme.gen} does not host any closed loops enclosing a nontrivial flux. We thus assume henceforth, without loss of generality, that the cavity-reservoir coupling amplitude in Eq.~\eqref{eq:bath.qme.gen} is real and positive, i.e.~$ J = |J| $.

We next consider the limit where reservoir-mode photons decay much faster than their tunneling rate to the cavity, i.e.~$\kappa_{c} \gg J$.  We can then adiabatically eliminate the reservoir
(see App.~\ref{appsec:nonM.bath}), yielding an effective QME for the cavity-qubit density matrix $ \hat \rho $ 
\begin{align}
& \frac{  d \hat \rho }{ dt }  
	= 
	- i \left[ \frac{ \lambda _{a} + \lambda  _{\mathrm{eff}}  }{2}  
	\hat \sigma_z \hat a ^\dag \hat a  , 
	\hat \rho   \right] 
	+  \Gamma  _{\mathrm{eff}}  
	\mathcal{D} \left[  
    e ^{ - i   
    \frac{  \theta _{\mathrm{eff}}  }{2}  
    \hat \sigma_z } 
    \hat a  
	 \right]  \hat \rho 
	\label{eq:1m.eff}.
\end{align} 
For $\kappa_{c} \gg J$, the parameters in this QME are
\begin{align}
    \Gamma  _{\mathrm{eff}}  
        & =  4 J  ^2 
        \kappa _{ c  } 
        /( \kappa _{ c } ^2 
        + \lambda _{ c  } ^2) 
    ,\\
    \lambda_{\mathrm{eff}}  
        & = 
            - 4 J  ^2 \lambda_c 
    /( \kappa _{ c  }   ^2 
    + \lambda _{ c  } ^2)  
    , \\
    \theta  _{\mathrm{eff}}  
        & = 2 \arctan ( 
\lambda _{ c  } 
/   \kappa _{ c } )
.  \label{eq:theta.eff}
\
\end{align}

The various coupling in Eq.~\eqref{eq:1m.eff} can be easily given a physical interpretation.  In the regime $\kappa_{c} \gg J$, the reservoir and the qubit together forms a new effective, Markovian environment for the cavity mode.   The corresponding effective cavity decay rate $\Gamma_{\rm eff}$ matches the Fermi's Golden rule expectation, and is independent of the qubit state.  $ \lambda  _{\mathrm{eff}}$ is an induced dispersive coupling arising from weak hybridization of $a$ and $c$ modes. 
The most interesting parameter is the phase $ \theta  _{\mathrm{eff}} $.   At a heuristic level, whenever a photon hops from the cavity mode to the reservoir mode and subsequently decays, the qubit is rotated by an angle $ \theta  _{\mathrm{eff}} $ about the $z$ axis.  
In the limit  
$\lambda _{ c  } 
\ll \kappa _{ c }   $, this phase shift can be understood as a product between photon dwell time in the reservoir $c$ mode,
$\tau  _{c}  \sim  \kappa _{ c }^{-1 } $, and the bare qubit-reservoir dispersive coupling strength $\lambda _{ c  } $. 

Finally, we imagine tuning the direct dispersive interaction so that it cancels the induced dispersive interaction, i.e. tune $\lambda_a = -  \lambda_{\mathrm {eff} }$.  In this case, we are left only with the dissipator in Eq.~(\ref{eq:1m.eff}), which corresponds exactly to the form in our general directional QME Eq.~\eqref{eq:bath.qme.gen}.  We thus obtain a completely directional dynamics from the cavity to the qubit.  We stress that this physical implementation uses standard forms of qubit-cavity coupling, and does not use any explicitly nonreciprocal elements.

It is worth stepping back to ask what the essential ingredients were here to obtain quantum nonreciprocity without any explicit breaking of TRS.  Like in our general QME in Eq.~(\ref{eq:nl.nr.ME}), it was crucial to have a final dissipator that was a product of $A$ and $B$ operators, corresponding to a nonlinear system-bath interaction (i.e.~the environment couples to the composite operator $\hat{A} \hat{U}_B$).  Further, we needed asymmetry:  the $B$ operator was unitary, the $A$ operator was not.  It is interesting to note that nonlinearity and broken inversion symmetry have been used in a very different manner to engineer nonreciprocal scattering without breaking TRS, both in classical~\cite{Fan2015,Alu2018} and quantum~\cite{Santos2014,Fedorov2018,Alu2022} settings.
However, the scattering in those works are only approximately nonreciprocal, and even then only for a limited range of incident field powers and frequencies. This is very different from our mechanism, which is unidirectional independent of initial state, and which is not equivalent to a simple scattering problem.


\begin{figure}[t]
    \centering
    \includegraphics[width=\columnwidth]{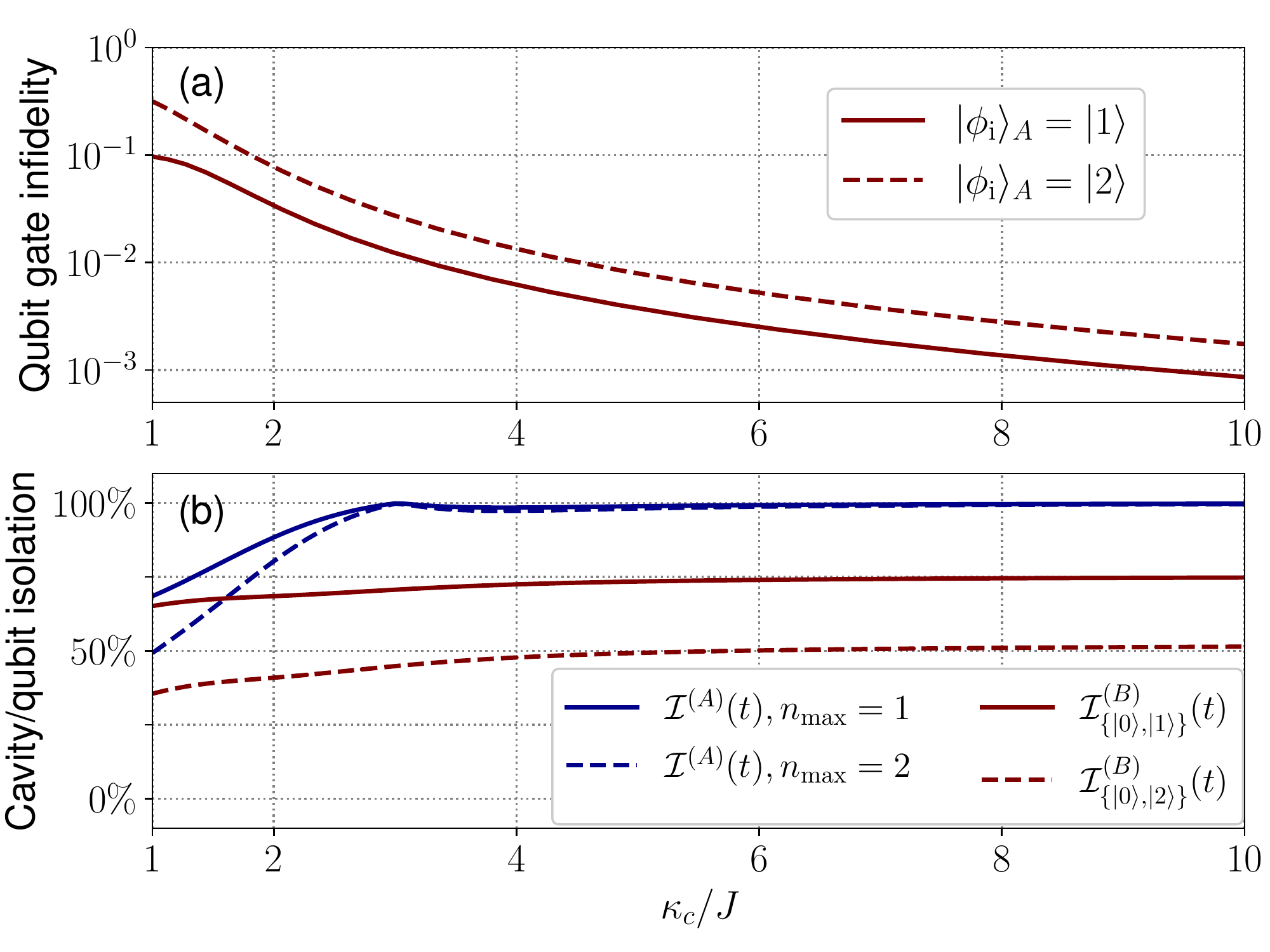}
    \caption{(a) State-averaged infidelity of the cavity-controlled stabilized qubit gate, as a function of the 
    non-Markovianity parameter $\kappa_c / J$.
    We consider different choices of initial cavity state: either the Fock state  $|\phi_{\mathrm{i}} \rangle _{A} 
    = |1 \rangle $ (solid) or $|2 \rangle$ (dashed), and take $\theta_{\rm eff} = \pi/6$.  These different initial cavity states result in different gate operations.  Infidelity tends to zero in the Markovian limit. 
    (b) Isolation functions for both qubit and cavity as a function of the non-Markovianity parameter $\kappa_c / J$.   We take a fixed evolution time $t=\pi/\Gamma_{\mathrm{eff}}$ and 
    $ \theta_{\mathrm{eff}}=\pi/6$.
    Blue curves: cavity isolation ${I} ^{(A)}  (t )  $,  
    c.f.~Eqs.~\eqref{eq:iso.min.def} and \eqref{eq:bath.a.iso}.
    Red curves: conditional qubit isolation function ${I} ^{(B)} 
_{\{ \left |  0
\right\rangle , 
\left | \ell 
\right\rangle \}} (t )$ ($\ell =1,2$) 
which is an upper bound for  the qubit isolation ${I} ^{(B)}  (t ) $ , see Eq.~\eqref{eq:cond.iso.def}.
We see strongly nonreciprocal behaviour
(i.e.~$I^{(A)}(t) \simeq 1$,
$I^{(B)}(t) < 1$) 
even away from the Markovian limit. 
As discussed in the main text, when calculating $I^{(A)}(t)$, we truncate the cavity Fock space to $n_{\rm max}$ photons for simplicity; the solid (dashed) curve is for $n_{\rm max} = 1$ ($n_{\rm max} = 2$). 
    }
    \label{fig:nonM_2mode}
\end{figure}

\subsection{Non-Markovian effects in the qubit-cavity setup}

\label{sec:phy.impl.nonM}

The physical implementation given by Eq.~\eqref{eq:bath.qme.gen} provides a concrete setup where we can quantitatively analyze the effects of non-Markovianity. In this case, the reservoir (i.e.~the highly damped $c$ mode) has a finite correlation time 
$\tau_{c} \sim \kappa_{c} ^{-1}$, 
and the Markovian limit is only reached for large decay rate $\kappa_c$.  As discussed in Sec.~\ref{sec:nonMarkov.sing.gen}, the gauge symmetry that leads to isolation and nonreciprocity in our system only exists in the Markovian limit.  We thus expect to see deviations from ideal behaviour away from this limit.  

In Fig.~\ref{fig:nonM_2mode}(a), we numerically investigate the infidelity~\cite{Nielsen2002} of the steady-state gate operation performed on $B$, as a function of the non-Markovianity parameter 
$\kappa_{c} / J$.  Remaining parameters are also varied to keep the qubit gate angle $\theta$ in  Eq.~\eqref{eq:theta.eff} fixed (here at a value $\theta = \pi/6$).  As expected, the infidelity rapidly drops to zero in the Markovian large $\kappa_c$ limit.  This general trend remains true no matter what the chosen value of $\theta$.

We can also ask how non-Markovian effects impact nonreciprocity in this system.  
The isolation functions introduced in Sec.~\ref{sec:qnr.iso.def} let us quantitatively compare the  nonreciprocity of  Eq.~\eqref{eq:bath.qme.gen} for different values of $\kappa_{c} $. 
  For cavity dynamics due to Eq.~\eqref{eq:bath.qme.gen}, one can show that the cavity isolation function 
can be expressed as
\begin{align} 
    {I}^{(A)}(t) & =  
1 - \frac{1}{2} 
|| \mathcal{E} ^{(A)}
	_{ \left | \uparrow
\right\rangle } 
( t  )  
-  \mathcal{E} ^{(A)}
_{ \left | \downarrow
\right\rangle } 
( t  )   
	||_{\diamond}
. 
\label{eq:bath.a.iso}
\end{align} 
where  
$\left | \uparrow 
\right\rangle , 
\left | \downarrow 
\right\rangle$ are $\hat \sigma_z$  eigenstates.
While the diamond norm can be calculated for quantum maps acting on an infinite-dimensional Hilbert space, to make the problem numerically tractable, we truncate the cavity Hilbert space to have at most one or two photons (as this is already sufficient to illustrate the effect of non-Markovianity).
The numerically calculated cavity isolation for increasing cavity decay rates $\kappa_{c} $ is plotted by the blue curves in Fig.~\ref{fig:nonM_2mode}(b). We see that even for modest values of $\kappa_c$ (i.e.~not strongly in the Markovian regime), the cavity is well isolated.

To characterize nonreciprocity, we also need to consider the qubit isolation ${I} ^{(B)} 
(t ) $.  We can find a simple upper bound for this quantity using the conditional qubit isolation ${I}^{(B)} 
_{\{ \left |  0
\right\rangle , 
\left | \ell 
\right\rangle \}} (t )$,  corresponding to initial control cavity Fock states $\{ \left |  0
\right\rangle , 
\left | \ell 
\right\rangle \}$ ($\ell =1,2$), as 
\begin{align}
    I^{(B)}(t) & \leq
    \left(
    {I} ^{(B)} 
_{\{ \left |  0
\right\rangle , 
\left | \ell 
\right\rangle \}} (t )
  \equiv   1 -
\frac{1}{2} 
	|| \mathcal{E} ^{(B)}
	_{ \left | 0 
\right\rangle } 
( t  )  
-  \mathcal{E} ^{(B)}
_{ \left |  \ell 
\right\rangle } 
( t  )  ||_{\diamond}
\right)
\label{eq:cond.iso.def}
.    
\end{align} 
From definition of qubit isolation (c.f. Eq.~\eqref{eq:iso.min.def}), one sees that Eq.~\eqref{eq:cond.iso.def} must be no less than the actual qubit isolation function  
${I} ^{(B)} 
(t )$.  As shown in Fig.~\ref{fig:nonM_2mode}(b), in the fast reservoir limit  $\kappa_{c} / J \gg 1$, ${I} ^{(B)} 
_{\{ \left |  0
\right\rangle , 
\left | \ell 
\right\rangle \}} (t )$ ($\ell =1,2$) (red curves) 
is considerably smaller than cavity isolation ${I} ^{(A)} 
(t ) $ (blue curves), demonstrating that the reservoir mode mediates an effective unidirectional interaction from control cavity to the qubit.


\section{Generalized gauge-invariance nonrecipropcity: the non-abelian case}

\label{sec:qnr.multi}

We now discuss how our recipe for quantum nonreciprocity based on gauge invarance can be extended from Eq.~\eqref{eq:nl.nr.Lindbladian} to a much broader class of dynamics. This generalized version involves dissipative dynamics with multiple dissipators, and the relevant local gauge symmetry can become non-abelian.  As we show, this generalized version is in general {\it not} equivalent to unconditional evolution under measurement and feedforward, and is capable of generating entanglement.  It is also (like the single dissipator case) distinct from the cascaded quantum systems master equation.   

Similar to Sec.~\ref{sec:qnr.sing.recipe}, we start by considering a single system $A$ undergoing dissipative Lindblad dynamics, but now involving multiple dissipators:
\begin{align}
    \frac{d}{dt} \hat{\rho}_A 
    & = \Gamma 
    \sum _{\ell=1}^{N} 
    \mathcal{D}  
    [ \hat{A} _{ \ell } ] \hat{\rho}_A
    \equiv \mathcal{L} _{A  } \hat{\rho}_A
    . 
\end{align}
As is well known, multi-dissipator Liouvillians like $ \mathcal{L} 
_{A  } $ are invariant under a wide class of transformations that mix the jump operators $\hat{A}_l$.  Let $u_{lm}$ be the matrix elements of an arbitrary $N \times N$ complex unitary matrix $\cmatU$~\cite{comp_mat}.  Then we necessarily have (see e.g.~\cite{oqs2002book}): 
\begin{align}
    \sum _{\ell=1}^{N} 
    \mathcal{D}  
    [ \hat{A} _{ \ell } ]
	&   = 
\sum _{\ell=1}^{N} 
	\mathcal{D}  
\left[ \sum  _{m=1}^{N} 
    u _{\ell m}  
    \hat{A} _{ m }  \right]
    \label{eq:qme.symm.multi}
	.  
\end{align} 

The above invariance of $\mathcal{L}_A$ also trivially continues to hold if we make the  unitary mixing matrix $\cmatU \left( t \right )$ time dependent, even if $\cmatU \left( t \right )$ does not commute with itself at different times.
Similar to Sec.~\ref{sec:qnr.sing.recipe}, this now provides a route to construct a nonreciprocal interaction with a second subsystem $B$:  we make each matrix element $ u _{\ell m}(t) $ of $\cmatU \left( t \right )$  an operator 
$\hat u _{\ell m} $ acting on $B$.  This results in a new master equation acting on the state of the bipartite $A$ plus $B$ system:  \begin{align}
    \frac{d}{dt} \hat{\rho}_{AB}
    & = 
    \Gamma 
	\sum _{\ell=1}^{N} \mathcal{D}   
	\left[ 
    \hat z _{\ell }   \right]
    \hat{\rho} 
    \equiv \mathcal{L} 
_{\mathrm{multi} } 
\hat \rho  , 
	\label{eq:qnr.multi.gen}
\end{align} 
with
\begin{align}
    \hat z _{\ell } & =  
\sum _{m=1}^{N}  
    \hat{A} _{ m }  
	\hat u _{\ell m} 
    .  
    \label{eq:zDissipatorsGeneral}
\end{align}
Note that for given operators $\hat{A} _{ m }  $, Eq.~\eqref{eq:qnr.multi.gen} and~\eqref{eq:zDissipatorsGeneral} still allow an overall normalization factor for all the $\hat u _{\ell m} $ operators. We can fix this ambiguity, without loss of generality, by requiring that in the trivial case where $\hat u _{\ell m} $ are all proportional to the identity operator, they may correspond to elements of a unitary matrix. More specifically, we require the following condition to hold 
\begin{align}
\sum _{\ell, m=1}^{N} 
\text{Tr} [
\left( \hat u _{\ell m} \right)^\dag
\hat u _{\ell m} ] =N 
    . 
    \label{eq:op.uele.scale}
\end{align}

We now come to the central result of this section:  the master equation Eq.~(\ref{eq:qnr.multi.gen}) mediates a fully nonreciprocal interaction from $A$ to $B$ if the $B$ operators satisfy the generalized unitarity constraint
\begin{align} 
 \sum _{\ell=1}^{N} 
\left( \hat u _{\ell m} \right)^\dag
\hat u _{\ell m'}
= \delta _{m m' }
\hat {\mathbb{I}} _{B} 
,
\label{eq:qnr.multi.ortho}
\end{align}
where  $\hat {\mathbb{I}} _{B}$ is the unit operator on subsystem $B$ (see App.~\ref{appsec:qnr.multi} for details).
If Eq.~\eqref{eq:qnr.multi.ortho} is satisfied, it is easy to show that $A$ is completely isolated from $B$:
one can trace out $B$ and derive a closed QME for the dynamics of $A$ that is independent of any additional local dynamics acting on $B$.  
This isolation reflects the underlying gauge symmetry discussed above.  The converse is not true:
$B$ will in general be influenced by $A$.  Note that if Eq.~(\ref{eq:qnr.multi.ortho}) is satisfied, the operators $\hat u _{\ell m} $ can be viewed as matrix elements of a generalized unitary transformation acting on a larger space (see App.~\ref{appsec:qnr.multi}).

\begin{figure}[t]
    \centering
    \includegraphics[width=\columnwidth]{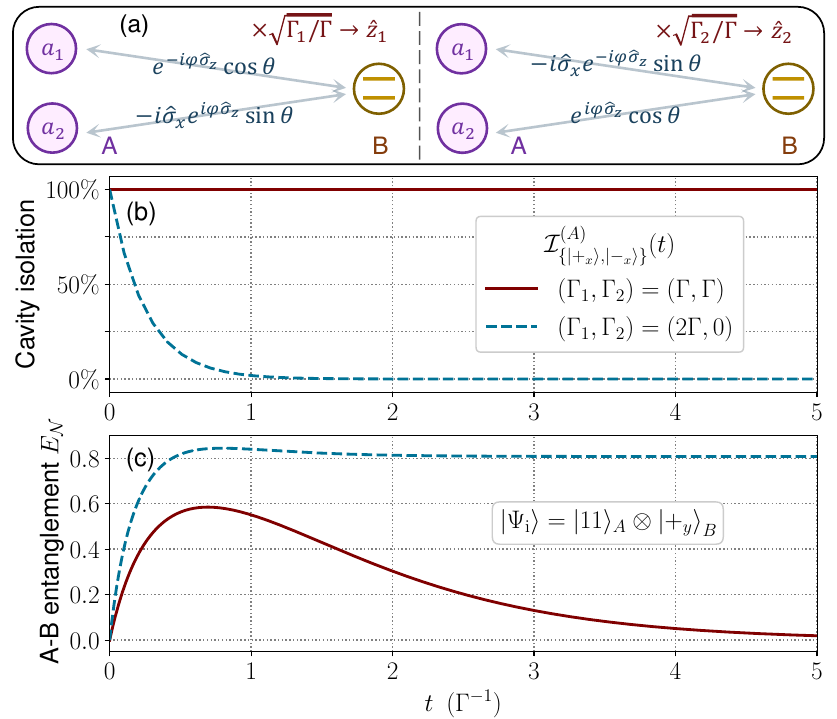}
    \caption{(a) Schematic illustrating $2$ dissipators $\mathcal{D} [\hat z_i] $ ($i=1,2$) that are used to realize a multi-dissipator generalization of nonreciprocal interaction between cavity modes $a_1$,$a_2$ ($A $) and a qubit ($B$), see Eq.~\eqref{eq:qnr.multi.2}. The dynamics generated by their sum 
    $\Gamma \sum  _{i}  \mathcal{D} [\hat z_i]$ becomes fully nonreciprocal if
    $\Gamma _{1} = \Gamma _{2} = \Gamma $. 
    (b) Numerically computed cavity isolation 
    ${I} ^{(A)} 
_{\{ \left | + _{x}
\right\rangle , 
\left | - _{x}
\right\rangle \}} (t ) $ for dynamics generated by the sum of $2$ dissipators with 
$(\Gamma _{1} , \Gamma _{2}) = (  \Gamma ,\Gamma )$ (solid red line), versus a single dissipator,~$(\Gamma _{1} , \Gamma _{2}) = ( 2 \Gamma ,0 )$ (dashed blue curve). The cavity dynamics is fully isolated in the former case, as expected for dissipators generating fully nonreciprocal dynamics.  For illustrative purposes, we restrict to the subspace of at most two total photons in $A$, but our result remains valid for bosonic $A$ modes with infinite levels. (c) Entanglement generation by the dissipations, assuming a product initial state 
$  | 11\rangle _{A} 
\otimes \left | + _{y} 
\right\rangle _{B} $.  Surprisingly, even the fully nonreciprocal interactions generated by two dissipators with   
$(\Gamma _{1} , \Gamma _{2}) = (  \Gamma ,\Gamma )$  can lead to nontrivial entanglement generation (solid red curve). Parameters: $ \theta = \varphi = \pi/4$. }
    \label{fig:multi_diss}
\end{figure}

We stress that Eq.~\eqref{eq:qnr.multi.gen} is not a trivial generalization of the single-dissipator case in Eq.~\eqref{eq:nl.nr.Lindbladian}, since each individual dissipator 
$ \mathcal{D}    
	\left[ 
\hat z _{\ell }   \right]$ need not generate fully nonreciprocal dynamics on its own.  Unidirectionality is thus in general a property of the full Liouvillian, and not of each dissipator on its own.  Further, there is no clever transformation that allows one in general to express the Liouvillian in such a form, i.e.~$\mathcal{L} 
_{\mathrm{multi}  } \hat \rho  
\ne \sum_{\ell } 
\Gamma _{\ell } 
\mathcal{D}  
[ \hat{A}' _{ \ell } 
\hat {U} _{B, \ell } ] 
\hat{\rho} $.

As a concrete example, we consider a bipartite system where the $A$ subsystem is comprised of two bosonic modes $\hat{a}_1, \hat{a}_2$ , and the $B$ subsystem is a single qubit (see Fig.~\ref{fig:multi_diss}(a)).  We take the full system dynamics to be described by the following Lindbladian
\begin{align}
&  \mathcal{L} _{2} \hat \rho  
= \Gamma   
\sum _{\ell=1}^{2} 
\mathcal{D}    
\left[ \hat z _{\ell }
\right]
\hat{\rho}
    \label{eq:qnr.multi.2}
, 
\end{align} 
where the jump operators are given by 
\begin{subequations}
\label{eq:qnr.multi.z}
\begin{align}
& \hat z _{ 1 }
=  
	\sqrt{\Gamma_{1} / \Gamma} e ^{ -i \varphi   
\hat \sigma _{z } } \left( 
    \cos \theta \, \hat{a} _{ 1 }
    - i \sin \theta \,  \hat \sigma _{x }  
  \hat{a} _{ 2 } 
     \right)
 , \\
& \hat z _{ 2 }
    = 
    \sqrt{\Gamma_{2} / \Gamma}
    e ^{ i \varphi 
        \hat \sigma _{z }  } 
        \left( 
    - i \sin \theta  \, \hat \sigma_{x }  
    \hat{a} _{ 1 } 
        + \cos \theta \, \hat{a} _{ 2 } 
    \right)
    . 
\end{align} 
\end{subequations}
From Eq.~\eqref{eq:op.uele.scale}, we thus have 
$\Gamma = (\Gamma_{1}+ \Gamma_{2})/2$. 
Each of these dissipators has a non-trivial action on the composite system that in general will correlate the state of the qubit and cavities
(see Fig.~\ref{fig:multi_diss}(a)).  For example, the jump operator $\hat{z}_1$ has an amplitude for flipping the state of the qubit (correlated with loss from $a_2$) and for not flipping the qubit state (correlated with loss from $a_1$).  The situation is reversed for $\hat{z}_2$.  In general, neither of these jump operators can be written as a product of a qubit operator times a cavity operator.   
For simplicity, we assume $ \theta = \varphi = \pi/4$ in what follows, but our results are valid as long as $ \sin2 \theta \sin2 \varphi \ne 0$. 

The dissipators in the master equation Eq.~(\ref{eq:qnr.multi.2}) have the same form as those in Eq.~(\ref{eq:zDissipatorsGeneral}), if we take $\hat{A}_1 = \hat{a}_1$, $\hat{A}_2 = \hat{a}_2$. 
The $\hat{u}_{\ell m}$ operators are then qubit-only operators.  
A direct computation shows that the 
unitarity condition of Eq.~\eqref{eq:qnr.multi.ortho} is satisfied if 
$\Gamma_{1} = \Gamma_{2} = \Gamma $.  In this case, Eq.~\eqref{eq:qnr.multi.2} {\it necessarily} describes a unidirectional interaction from the subsystem $A$ cavity modes to subsystem $B$, the qubit.  The cavity modes are unaffected by the qubit, and each experience simple loss at rate $\Gamma$.  In contrast, the qubit remains non-trivially influenced by the cavity modes.    

To help see these properties more explicitly, and to see that the directionality is {\it not} a property of each dissipator on its own, we can calculate the isolation function of subsystem $A$ (defined via  Eqs.~\eqref{eq:A.chan.res.def} and \eqref{eq:iso.min.def}). In
Fig.~\ref{fig:multi_diss}(b), we plot the conditional subsystem $A$ isolation ${I} ^{(A)} 
_{\{ \left | + _{x}
\right\rangle , 
\left | - _{x}
\right\rangle \}} (t ) 
$, defined as:
\begin{align}
{I} ^{(A)} 
_{\{ \left | + _{x}
\right\rangle , 
\left | - _{x}
\right\rangle \}} (t ) 
\equiv   1 -
(1/2)
	|| \mathcal{E} ^{(A)}
	_{ \left | + _{x} 
\right\rangle } 
( t  )  
-  \mathcal{E} ^{(A)}
_{ \left |  - _{x} 
\right\rangle } 
( t  )  ||_{\diamond}
. 
\end{align}
This quantity measures how sensitive the subsystem $A$ dynamics is to the initial state of $B$, when $B$ starts in a $\sigma_x$ eigenstate $| \pm_x \rangle$.  It sets an upper bound on the full $A$ subsystem isolation $I^{(A)}(t)$. As shown in  Fig.~\ref{fig:multi_diss}(b), the conditional cavity isolation stays unity at all times when the two dissipators in our master equation Eq.~\eqref{eq:qnr.multi.2} have equal strength ($\Gamma_{1} = \Gamma_{2} = \Gamma $).  This is as expected, as the generalized unitarity constraint is satisfied. As a result, the $I^{(A)}(t)$ should be unity according to Eq.~\eqref{eq:cond.fully.iso}.  In contrast, the $A$ isolation is significantly smaller than $1$ for the case where we only have one of the two required dissipators, 
i.e.~$\Gamma_{1} =2 \Gamma , \Gamma_{2} = 0 $.  In this case, $A$ is not isolated from $B$. This shows concretely that the combined action of both dissipators in Eq.~\eqref{eq:qnr.multi.2} leads to a fully directional dynamics, even though each on its own {\it does not} mediate a one-way interaction.  
Note that in this plot, we have calculated the isolation functions in a restricted cavity Hilbert subspace with at most $2$ total photons.  However, this does not affect the validity of our conclusion, since numerically computed isolation will set an upper bound for the isolation of the bosonic $A$ subsystem with infinite levels. As such, our calculation clearly shows that we cannot get fully nonreciprocal interaction with only one of the dissipators in  Eq.~\eqref{eq:qnr.multi.z}.

Of course, simply showing that subsystem $A$ is isolated does not indicate a nonreciprocal interaction:  we also need to verify that $B$ is influenced by $A$, and that we have not simply cancelled any interaction between the two subsystems.  To show that there is indeed a nonreciprocal interaction, 
in Fig.~\ref{fig:multi_diss}(c) we show that our master equation (with $\Gamma_1 = \Gamma_2 = \Gamma$) can generate entanglement between the two subsystems.  We show in that figure the time-dependent entanglement between $A$ and $B$ starting with an initial product state 
$  | 11\rangle _{A} 
\otimes \left | + _{y} 
\right\rangle _{B} $, with  
$ \left | + _{y}
\right\rangle _{B} $ denoting the qubit $\hat \sigma _{y }   $ eigenstate.  We see that entanglement is generated at intermediate times, even though system $A$ is fully isolated at all times.  This indicates that $A$ must be influencing $B$, and that we have a nontrivial nonreciprocal interaction.  

The fact that our dynamics can generate entanglement also leads to other important conclusions.  It immediately implies that Eq.~\eqref{eq:qnr.multi.2} cannot be re-written as a sum of single nonreciprocal dissipators, each having the form of  Eq.~\eqref{eq:nl.nr.Lindbladian}.  
If such a decomposition were possible, than our dynamics would be equivalent to a local measurement-plus-feedforward protocol, something that cannot generate entanglement (c.f.~discussion below Eq.~\eqref{eq:mmff.kraus}).  We note that this nonreciprocal entanglement generation is unique to the multi-dissipator version of our master equation.

Finally, as in the simpler single dissipator version of our mechanism, deviations from the Markovian limit will also impact the directionality of our interaction; this is discussed in more detail in  Appendix~\ref{appsec:nonM.multi}.


\section{Summary and outlook}

\label{sec:discussions}

In this work, we have introduced and analyzed a new kind of dissipative dynamics that leads to fully nonreciprocal interactions between two quantum systems.  The crucial ingredient was a time-local gauge symmetry inherent in any Markovian, Lindblad master equation.  Surprisingly, the explicit breaking of time-reversal or the use of synthetic gauge fields were not necessary.  As such, our new class of directional quantum master equations do not have the form of a standard cascaded quantum master equation.  

nonreciprocal quantum interactions are being actively studied for both their fundamental and practical implications.  Our results thus greatly expands the toolbox and class of interactions available for such studies.  In terms of application, we have shown how our interactions can be used for a new kind of dissipative quantum gate; the application to more complex kinds of quantum control (using e.g. the multi-dissipator version of our dynamics) could be extremely fruitful.  We note that in a very different context, engineered dissipation has been studied theoretically~\cite{Devoret2014,Jiang2021} and demonstrated experimentally~\cite{Leghtas2020,Devoret2020,Wang2021} as means to realize autonomous error correction.  Here, dissipative processes are designed to mitigate errors by bringing the system back to a desired code space. While our dissipative nonreciprocity and dissipative gates are very different in nature, it would be interesting to see whether these ideas could be combined with autonomous error correction for even more robust forms of quantum information processing.

Our work also has great potential for fundamental studies.  For example, it provides a direct way of designing quantum analogues of classical kinetically constrained models that feature directionality (see e.g.~\cite{Andersen1984}).  Such models could provide a new setting to study glassy dynamics in the quantum regime.  It would also be interesting to study our new kind of dissipative interactions in many body lattice models.  Here, our mechanism could be used to construct a class of directional models that are dissipative analogues of closed systems with dynamical gauge fields.  The latter is a topic of intense interest in a variety of engineered quantum systems (see e.g.~Refs.~\cite{Reznik2015,Grusdt2019,Browaeys2020,Zoller2020}).

\section*{Acknowledgments}

This work was supported by the Army Research Office under Grant 
No.~W911NF-19-1-0380.  AC acknowledges support from the Simons Foundation, through a Simons Investigator award (Grant No.~669487, A.~A.~C.).
C.W. was supported by the U.S. Department of Energy, Office of Science, National Quantum Information Science Research Centers, Co-Design Center for Quantum Advantage under contract DE-SC0012704.


%


\appendix


\section{Examples of Hamiltonians dynamics that is fully reciprocal by Eq.~\eqref{eq:def.dyn.recip}}

In Sec.~\ref{sec:qnr.iso.def} in the main text, we introduce a new metric of isolation to quantify the influence of one subsystem on the dynamics of another subsystem, which leads to a general definition of (non)reciprocity in the quantum regime. It is interesting to ask what the isolation looks like, and how nonreciprocal it is, if one considers fully coherent (i.e.~Hamiltonian) dynamics. In this Appendix, we provide two general class of Hamiltonian dynamics, which can be proven to be reciprocal as per Eq.~\eqref{eq:def.dyn.recip}. It is also intriguing to ask if fully Hamiltonian dynamics of generic systems should always be reciprocal by our definition of (non)reciprocity. While we conjecture that to be case for the definition of (non)reciprocity in Sec.~\ref{sec:qnr.iso.def} to align with (breaking of) Onsager reciprocity relations, we leave a thorough study to future works.

\subsection{Isolation function of reciprocal dynamics generated by Hamiltonians with local qubit ($B$) $\mathbb{Z} _{2} $ symmetry}

\label{appsec:iso.recip.disp}

While our proof can be straightforwardly generalized to Hamiltonian dynamics of generic bipartite systems with $N_A$ ($N_B$)-dimensional subsystem $A$ ($B$), as long as the total $AB$ Hamiltonian has a local symmetry that is $N_B$-dimensional and nondegenerate within $B$ subspace, for the sake of clarity, here we focus on bipartite systems ($AB$) where $B$ is a single qubit. In this subsection, we explicitly derive the isolation functions for dynamics generated purely by Hamiltonians with a local qubit $\mathbb{Z} _{2} $ symmetry, and show that any such dynamics must be fully reciprocal by definition in Eq.~\eqref{eq:def.dyn.recip}.  We stress that such Hamiltonians, albeit having a constrained form, can be generally nonlinear and interacting.  

Consider a generic bipartite system consisting of an $N_{A} $-dimensional system ($A$) and a qubit ($B$). We assume the system evolves under a Hamiltonian $\hat H _{AB}$, where interaction between $A$ and $B$ commutes with the qubit-only Hamiltonian 
$ \hat H _{B,0} 
\equiv \text{Tr} _{A} \hat H _{AB} $.  We can thus define, without loss of generality, the eigenstates of $\hat H _{B,0} $ as qubit $\hat \sigma _{z} $ eigenbasis, and rewrite the system Hamiltonian as follows 
\begin{align}
\hat H _{AB} 
& =  \hat H  _{ A } 
\otimes 
\hat {\mathbb{I}} _{B}
+ \hat \xi _{ A } 
\otimes \hat \sigma _{z} . 
    \label{eq:H.AB.deph}
\end{align}
For notation simplicity, it is convenient to rewrite the total Hamiltonian in terms of projectors onto qubit $\hat \sigma _{z}$ eigenstates, as   
\begin{align}
\hat H _{AB} 
& =  
\hat H ^{(A)}  _{ \uparrow } 
\otimes 
\left |\uparrow
\right\rangle  
\! \left \langle
\uparrow \right | 
+ \hat H ^{(A)}  _{ \downarrow } 
\otimes 
\left |\downarrow
\right\rangle  
\! \left \langle
\downarrow \right | 
    \label{eq:H.AB.deph}
, 
\end{align}
where the conditional $A$ Hamiltonians contingent on qubit states $\left | \sigma
\right\rangle $
($\sigma = \uparrow ,\downarrow $) are given by 
$\hat H ^{(A)}  _{ \uparrow /\downarrow } 
= \hat H  _{ A } \pm 
\hat \xi _{ A } 
$.

Following Eq.~\eqref{eq:A.chan.res.def} in the main text, we can again define evolution superoperator of $A$, depending on initial qubit ($B$) state $ \left |\phi _{\mathrm{i} }
\right\rangle$, as 
\begin{align}
\mathcal{E} ^{(A)}
_{ \left |\phi _{\mathrm{i} }
\right\rangle } 
( t  )  
\hat \rho _{A} 
	&   \equiv 
	\text{Tr} _{B}
	\left [ 
e ^{ - i \hat H _{AB}   t} 
( \hat \rho _{A}  
	\otimes
\left |\phi _{\mathrm{i} } 
\right\rangle_{ B }  
\! \left \langle
\phi _{\mathrm{i} } \right |   ) 
e ^{ i \hat H _{AB}   t} 
	\right] 
	.   
\end{align}   
Substituting Eq.~\eqref{eq:H.AB.deph} into above equation, the conditional $A$ quantum map can be straightforwardly calculated as 
\begin{align}
\mathcal{E} ^{(A)}
_{ \left |\phi _{\mathrm{i} }
\right\rangle } 
( t  )  
\hat \rho _{A} 
	&   = \sum 
_{\sigma = \uparrow ,\downarrow } 
\left | \left \langle
\sigma   |  
 \phi _{\mathrm{i} }
\right\rangle  \right | ^{2}
    e ^{ - i \hat H ^{(A)}  _{ \sigma } t} 
    \hat \rho _{A} 
    e ^{ i \hat H ^{(A)}  _{ \sigma } t} 
    . 
\end{align}   
Since the $A$ evolution in this case can be rewritten as a probabilistic mixture of unitary gates, the $A$ isolation can be shown to have a closed analytical form. Introducing unitary operator acting on $A$ that describes the ``overlap" between the two conditional unitary maps as 
\begin{align}
\hat U 
_{ \downarrow \uparrow } ^{(A)}  
( t  )  
= e ^{ i \hat H ^{(A)}  
    _{ \downarrow } t} 
e ^{ - i \hat H ^{(A)}  
    _{ \uparrow } t} 
    ,
    \label{seq:qnd.UA.diff}
\end{align}  
and defining its eigenvalues as 
$e ^{ i \phi _{\ell} } $ ($\ell = 1,2, \ldots, N_{A}$), one can show that 
\begin{align}
{I} ^{(A)}  (t ) 
= & 1 -  
\max  
_{ 1\le \ell<m
\le N_A } 
\left | \sin 
\frac{ (\phi  _{\ell} 
- \phi  _{m} )}{2} \right | 
. 
\label{seq:qnd.IA.sol}
\end{align}

Conversely, because the qubit experiences simple phase shift and/or dephasing during the time evolution, we can also compute the qubit ($B$) isolation exactly. More specifically, given initial qubit state $\hat \rho _{B} $ and expanding it using $\hat \sigma _{z}$ basis, the qubit populations are conserved throughout the evolution. The qubit coherence due to Eq.~\eqref{eq:H.AB.deph} can be computed as 
\begin{align}
\frac{\left \langle
\uparrow \right | 
\mathcal{E} ^{(B)}
_{ \left |\phi _{\mathrm{i} }
\right\rangle }  
( t  )  
\hat \rho _{B}  \! 
\left |\downarrow 
\right\rangle  
}{
\left \langle
\uparrow \right |  
\hat \rho _{B}  \! 
\left |\downarrow 
\right\rangle  } 
    & = \text{Tr}
\left (
    e ^{ - i \hat H ^{(A)}  
    _{ \uparrow }  t} 
    \left |\phi _{\mathrm{i} } 
\right\rangle_{ A }  
\! \left \langle
\phi _{\mathrm{i} } \right | 
    e ^{ i \hat H ^{(A)}  
    _{ \downarrow }  t} 
\right ) 
\\
    & = \text{Tr}
\left (
    \hat U 
_{ \downarrow \uparrow } ^{(A)}  
( t  )  
    \left |\phi _{\mathrm{i} } 
\right\rangle_{ A }  
\! \left \langle
\phi _{\mathrm{i} } \right |  
\right ) , 
\end{align}
where 
$ \hat U _{ \downarrow \uparrow } 
^{(A)}  $ is again the overlap unitary in Eq.~\eqref{seq:qnd.UA.diff}. Because the qubit map now takes a  pure-dephasing form, the diamond norm between two such qubit maps can be explicitly derived as 
\begin{align}
& 
|| \mathcal{E} ^{(B)}
	_{ \left | \phi_{1} 
\right\rangle } 
( t  )  
-  \mathcal{E} ^{(B)}
_{ \left | \phi_{2} 
\right\rangle } 
( t  )   
	||_{\diamond}
\nonumber \\
= &  
\left | 
\text{Tr}
\left [ 
\hat U _{ \downarrow \uparrow }
^{(A)}  ( t  )  
    (\left |\phi _{1 } 
\right\rangle  
\! \left \langle
\phi _{1} \right |
- \left |\phi _{2 } 
\right\rangle  
\! \left \langle
\phi _{2} \right |)
\right ] 
\right|
\nonumber \\
= &  
\left | 
\left \langle
\phi _{1} \right |
\hat U _{ \downarrow \uparrow }
^{(A)}( t  )  
\left |\phi _{1 } 
\right\rangle  
\! 
- \left \langle
\phi _{2} \right |
\hat U _{ \downarrow \uparrow }
^{(A)}( t  )  
\left |\phi _{2 } 
\right\rangle 
\right|
. 
\label{seq:Hqnd.IB.states}
\end{align} 
Expanding the states $\left |\phi _{  1 } 
\right\rangle ,  
\left |\phi _{ 2 } 
\right\rangle $ in Eq.~\eqref{seq:Hqnd.IB.states} using eigenbasis of 
$\hat U _{ \downarrow \uparrow } 
^{(A)}  $, and minimizing over all $A$ initial states, one can compute the $B$ isolation function ${I} ^{(B)} 
(t ) $ as follows 
\begin{align} 
&  {I} ^{(B)} 
(t ) 
\equiv  1 - \frac{1}{2} 
\max _{ \left | \phi_{1} 
\right\rangle , 
\left | \phi_{2} 
\right\rangle
\in \mathcal{H} _{A} }
|| \mathcal{E} ^{(B)}
	_{ \left |\phi _{ 1 }
\right\rangle } 
( t  )  
-  \mathcal{E} ^{(B)}
_{ \left |\phi _{ 2 }
\right\rangle } 
( t  )   
	||_{\diamond}
\nonumber \\
= & 1 - \frac{1}{2}  
\max _{ \left |\phi _{  1 } 
\right\rangle ,  
\left |\phi _{ 2 } 
\right\rangle \in 
\mathcal{H}_{A} 
}
\left | 
\left \langle
\phi _{1} \right |
\hat U _{ \downarrow \uparrow }
^{(A)}( t  )  
\left |\phi _{1 } 
\right\rangle  
\! 
- \left \langle
\phi _{2} \right |
\hat U _{ \downarrow \uparrow }
^{(A)}( t  )  
\left |\phi _{2 } 
\right\rangle 
\right|
\nonumber \\
= & 1 -  
\max  
_{ 1\le \ell<m
\le N_A } 
\left | \sin 
\frac{ (\phi  _{\ell} 
- \phi  _{m} )}{2} \right | 
= {I} ^{(A)}  (t ) 
\end{align} 
Comparing above expression to Eq.~\eqref{seq:qnd.IA.sol}, we thus have \begin{align}
{I} ^{(A)} 
(t ) 
= &  {I} ^{(B)} (t ) 
. 
\end{align}

\subsection{Proof of reciprocity for arbitrary Hamiltonian dynamics of two-qubit systems}

\label{appsec:iso.recip.2qb}

In this subsection, we restrict to bipartite systems where both $A$ and $B$ are single qubits, and we seek to prove that arbitrary Hamiltonian dynamics of this two-qubit system is reciprocal by Eq.~\eqref{eq:def.dyn.recip}. To start, we note that the isolation function of $A$ ($B$) is invariant under applications of local unitaries at the input and/or output ports of the quantum channel. More specifically, we consider two generic quantum maps 
$\mathcal{E} ^{(AB)} $ and $\mathcal{F} ^{(AB)} $ that are related by the following equation
\begin{align}
\mathcal{E} ^{(AB)} 
\hat{ \rho } _{AB} 
&   
 =
(\hat W _{ A  } 
\otimes 
\hat W _{ B  } )
\left ( 
\mathcal{F} ^{(AB)} 
\hat{ \rho }' _{AB} 
\right )
(\hat W _{ A   } ^\dag
\otimes 
\hat W _{ B   } ^\dag)
	,    
	\label{seq:chan.AB.localunit}
	\\
\hat{ \rho }' _{AB} 
& =  \hat V _{ A  } 
\otimes 
\hat V _{ B  } 
\hat{ \rho } _{AB} 
\hat V _{ A   } ^\dag
\otimes 
\hat V _{ B   } ^\dag
. 
\end{align} 
Making use of the fact that the diamond norm (c.f.~Eq.~\eqref{eq:qchan.dist.prob}) is invariant under unitary transformations on the quantum channels~\cite{Watrous2018}, one can straightforwardly show that the corresponding isolation functions 
${I} ^{(A/B)} 
(\mathcal{E} ^{(AB)} ) $ and 
${I} ^{(A/B)} 
(\mathcal{F} ^{(AB)} ) $ are also equal for two maps, i.e.
\begin{align}
{I} ^{(A/B)} 
(\mathcal{E} ^{(AB)}  )
 = 
{I} ^{(A/B)} 
(\mathcal{F} ^{(AB)}  )
. 
\end{align}
The isolation functions in above equation are defined similarly to Eq.~\eqref{eq:iso.min.def} for the two channels, and we omit any time variables (if applicable) for notation simplicity.

We next observe that, as pointed out in the main text, dynamics generated by any Hamiltonian symmetric under permutation of $A$ and $B$ is automatically reciprocal by our definition. For $2$-qubit systems, and assuming no local Hamiltonians, we conclude that dynamics generated by so-called Heisenberg $XYZ$ interactions should be reciprocal. Rewriting the corresponding unitary evolution superoperator as 
\begin{align}
&  \mathcal{U} ^{(AB)} _{XYZ}
\hat{ \rho } _{AB} 
= \hat U _{ XYZ } 
\hat{ \rho } _{AB} 
\hat U _{ XYZ } ^\dag
,
\\
& \hat U _{ XYZ } 
= 
\exp \left[ -i 
\left(
\sum _{\alpha = x,y,z}
J _{\alpha}
\hat \sigma _{A, \alpha}
\hat \sigma _{B, \alpha}
\right) \right]
, 
\end{align}
we thus have 
\begin{align}
{I} ^{(A)} 
( \mathcal{U} ^{(AB)} _{XYZ} )
 = 
{I} ^{(B)} 
( \mathcal{U} ^{(AB)} _{XYZ} )
. 
\label{seq:iso.symm.XYZ}
\end{align}
Finally, we make use of the standard decomposition of any $2$-qubit rotations into local unitaries and a Heisenberg unitary~\cite{Glaser2001}, which can be viewed as a special case of the KAK decomposition of Lie groups. For a generic $2$-qubit unitary operator $\hat U _{ AB } $, the decomposition states that there exist two local unitary unitaries 
$\hat U _{ A ,\ell  } 
\otimes 
\hat U _{ B ,\ell  }  $ ($\ell = 1,2$), such that the following equality holds 
\begin{align}
&  \hat U _{ AB }
= (\hat U _{ A , 1  } 
\otimes 
\hat U _{ B , 1  })
\hat U _{ \mathrm{H} } 
(\hat U _{ A , 2  } 
\otimes 
\hat U _{ B , 2 })
,
\\
& \hat U _{ \mathrm{H} } 
= 
\exp \left[ -i 
\left(
\sum _{\alpha = x,y,z}
h _{\alpha}
\hat \sigma _{A, \alpha}
\hat \sigma _{B, \alpha}
\right) \right]
.  
\end{align}
Comparing above equation with Eq.~\eqref{seq:chan.AB.localunit}, we see that the unitary map generated by $ \hat U _{ AB }$ has the same isolation functions as that generated by $ \hat U _{ \mathrm{H} }$. From Eq.~\eqref{seq:iso.symm.XYZ}, we can further prove that for any $2$-qubit unitary evolution 
$\mathcal{U} ^{(AB)} _{2}
\hat{ \rho } _{AB} 
= \hat U _{ AB } 
\hat{ \rho } _{AB} 
\hat U _{ AB } ^\dag$, we have
\begin{align}
{I} ^{(A)} 
( \mathcal{U} ^{(AB)} _{2} )
 = 
{I} ^{(B)} 
( \mathcal{U} ^{(AB)} _{2} )
,  
\label{seq:iso.symm.2qb}
\end{align}
so that any $2$-qubit Hamiltonian dynamics must be reciprocal.


\section{The role of Markovianity in emergence of dynamical gauge symmetry in a single-dissipator Lindblad master equation}

\label{appsec:symm.sing}

As discussed in the main text, the unidirectional nature of dynamics generated by the nonlinear dissipator 
$  \mathcal{D}  [   
  \hat A  \hat U _{B}  ] 
\hat \rho $ (see Eq.~\eqref{eq:nl.nr.Lindbladian} in the main text) crucially depends on a fundamental gauge symmetry, which is inherent to Lindblad-form QMEs that describe Markovian environments. In this section, we illustrate this connection using a microscopic model for the quantum dissipation.

Let us consider a generic, single Lindblad dissipator of system $A$, given by 
\begin{align}
&   \mathcal{L} 
_{A , \mathrm{1} } \hat \rho 
	= \Gamma 
	\mathcal{D}  [   
  \hat A     ]  \hat \rho  
	. 
	\label{seq:qme.sing}
\end{align} 
Without loss of generality, we could model the dissipation as due to a bosonic microscopic environment consisting of harmonic oscillator modes $ b _{ \ell} $, so that the system-bath Hamiltonian can be written as 
\begin{align}
& \hat H_\mathrm{tot} 
= 
\hat H_\mathrm{E} 
+ \hat H_\mathrm{SE} 
, \quad 
\hat H_\mathrm{E} 
= \sum _{ \ell} \omega _{ \ell}
\hat b _{ \ell}^\dag 
    \hat b _{ \ell}
, \\
&  \hat H_\mathrm{SE} 
	=  \hat A	
    \hat \xi ^\dag  
	+ \text{H.c.}
	, \quad 
\hat \xi = \sum_{ \ell} 
	g ^{*}_{ \ell}  
    \hat b _{ \ell} 
	\label{seq:Hse.gen}
	. 
\end{align} 
The density of states (DOS) function 
$ \mathcal{J} _{0} \left[   \omega  \right]  $ of this bosonic environment can be explicitly computed in terms of the interaction picture bath operator 
$ \hat \xi \left ( t \right)
= e ^{ i \hat H_\mathrm{E}  t } 
\hat \xi   
e ^{- i \hat H_\mathrm{E}  t } $, as 
\begin{align}
\mathcal{J}  _{0} 
\left[   \omega  \right] 
    &  \equiv  
    \frac{1}{ 2 \pi }
    \int ^{\infty} _{ - \infty}
    \langle [
\hat \xi \left ( t \right)
, \hat \xi^\dag \left ( 0 \right)
] \rangle
e ^{ i  \omega  t   }
    dt 
    \\
    &  = \sum _{ \ell} 
    \left| g_{ \ell} \right| ^{2}
    \delta \left (  \omega 
    - \omega _{ \ell} 
	\right)    . 
	\label{seq:dos.gen}
\end{align} 
In the Markovian limit, the environmental DOS 
$ \mathcal{J}  _{0} 
\left[   \omega  \right]  $ reduces to a constant, i.e.~we have    
\begin{align}
\mathcal{J}   _{0} 
\left[   \omega  \right] 
	\equiv  \Gamma 
	\Leftrightarrow 
	\langle [
\hat \xi \left ( t _{1} \right)
, \hat \xi^\dag \left ( t _{2} \right)
] \rangle
    = \Gamma  
    \delta \left ( 
    t _{1} - t _{2} \right) . 
\end{align} 
In this limit, we can integrate out the bath dynamics to obtain a standard Lindblad equation, as given by Eq.~\eqref{seq:qme.sing}.

We now perform a standard gauge transformation that shifts the jump operator phase by a time-dependent real value 
$ \theta 
\left ( t \right)$, i.e. 
$  \hat A  \to   \hat A ' =  
\hat A  e ^{i\theta 
\left ( t \right)} $, so that the new interaction picture system-bath interaction becomes
\begin{align}
&  \hat H'_\mathrm{SE} \left ( t \right) 
	= \sum_{ \ell} 
	\left ( g_{ \ell} 
    \hat A
	\hat b _{ \ell}^\dag 
	e ^{ i  \omega _{ \ell} t 
	+ i\theta \left ( t \right) }
	+ \text{H.c.}
	\right) 
	. 
\end{align} 
Mathematically, the gauge symmetry of the Lindblad dissipator can be understood from the fact that the microscopic bath DOS 
$ \mathcal{J}' \left[   \omega  \right] $ stays invariant under the gauge transformation. More specifically, we can show rigorously     
\begin{align}
\mathcal{J}' \left[   \omega  \right] 
    & = \frac{1}{ 2 \pi }
    \int ^{\infty} _{ - \infty}
    \sum _{ \ell} 
    \left| g_{ \ell} \right| ^{2}
    e ^{ i  \omega  t 
    -  i  \omega _{ \ell} t  }
e ^{ 
- i\theta \left ( t \right) 
+  i\theta \left ( 0 \right) }
    dt 
    \\
& = \Gamma  
\int ^{\infty} _{ - \infty}
    \delta \left ( t \right) 
    e ^{ - i\theta 
    \left ( t \right) 
    +  i\theta 
    \left ( 0 \right)  }
    dt 
	= \Gamma . 
	\label{seq:dos.gauge.exact}
\end{align} 
This derivation formally shows that the microscopic origin of the (dynamical) gauge symmetry of the Lindbladian in Eq.~\eqref{seq:qme.sing} is due to a completely flat bath DOS function, i.e. a Markovian bath.

As mentioned, above derivation requires the bath is perfectly Markovian, so that the bath correlation function 
$ \langle [
\hat \xi  \left ( t \right)
, \hat \xi ^\dag \left ( 0 \right)
] \rangle $ is proportional to a delta function. We now discuss an intuitive way to understand the role of Markovianity in the emergence of dynamical gauge symmetry. Noting the gauge phase 
$ e  ^{ i\theta 
\left ( t \right) } $ enters the interaction picture Hamiltonian in exactly the same manner as dynamical phases of the environmental modes 
$ e ^{ i  \omega _{ \ell} t } $, we could formally absorb the gauge phase into the definition of environmental chemical potential, by going to the following rotating frame 
\begin{align}
& \hat H'_\mathrm{E} 
= \hat H_\mathrm{E} 
\to  \hat H ''_\mathrm{E} 
= \hat H_\mathrm{E} 
- \hat H_\mathrm{rot}
, \quad
\hat H_\mathrm{rot}
= 
- \dot{\theta } \left ( t \right) 
\sum _{ \ell}  
\hat b _{ \ell}^\dag 
\hat b _{ \ell} 
, \\
& \hat b' _{ \ell}  
= \hat b _{ \ell}  
\to   \hat b'' _{ \ell} =  
e ^{ i \hat H_\mathrm{rot} t } 
\hat b _{ \ell}  
e ^{- i \hat H_\mathrm{rot} t } 
= \hat b _{ \ell}  	
e ^{ i\theta \left ( t \right) }
.  
\end{align} 
We remark that the system-bath interaction in the rotating frame takes the same form as the original interaction Hamiltonian in Eq.~\eqref{seq:Hse.gen}. More specifically, we can write the Hamiltonian in the lab frame as 
\begin{align}
& \hat H ''_\mathrm{E} 
= \sum _{ \ell}  
\left [
\omega _{ \ell} 
+ \dot{\theta } \left ( t \right)
\right ]
\hat b _{ \ell}^\dag 
\hat b _{ \ell}
, \\
&  \hat H''_\mathrm{SE} 
	= \sum_{ \ell} 
	\left ( g_{ \ell} 
    \hat A 
	\hat b _{ \ell}^\dag  
	+ \text{H.c.}
	\right) . 
\end{align} 
Noting that the bath Hamiltonian now varies in time, we can formally {\emph{define}} a time-dependent bath DOS function $\mathcal{J} \left[ \omega ; t \right]  $, as
\begin{align}
\mathcal{J} \left[   \omega ; t \right]  
    &  \equiv
    \sum _{ \ell} 
    \left| g_{ \ell} \right| ^{2} 
    \delta \left (  \omega 
    - \omega _{ \ell} 
    - \dot{\theta } \left ( t \right) 
	\right) 
	,  
\end{align} 
which can be related to the original bath DOS function $\mathcal{J} _{0} 
[ \omega  ] $ in Eq.~\eqref{seq:dos.gen} as 
\begin{align}
\mathcal{J} \left[   \omega ; t \right] 
    = \mathcal{J} _{0} 
    [ \omega - \dot{\theta } \left ( t \right) ] 
    . 
\end{align} 
In the Markovian limit we thus have 
$ \mathcal{J} \left[   \omega ; t \right] \equiv \mathcal{J} _{0} 
[ \omega  ] = \Gamma $, so that the bath DOS is invariant under the gauge transformation. For a realistic environment, the bath bandwidth needs to be finite and hence cannot be perfectly Markovian. In this case, above analysis would be valid as long as the band correlation time 
$ \tau _\mathrm{E}  $ is much smaller than timescale associated with dynamics of the gauge phase, 
i.e.~$ \tau _\mathrm{E}  \dot{\theta } 
\left ( t \right) \ll 1 $.

\section{Leading-order non-Markovian corrections to standard master equations due to broken gauge symmetry}

\subsection{Single-dissipator case}

\label{appsec:nonM.sing}

The discussion in preceding section relates perfect Markovianity of the bath to an emergent gauge symmetry in resulting dissipative dynamics. It is interesting to ask to what extent could this gauge symmetry be broken, if the bath deviates from the Markovian limit. We again start with the interaction picture system-bath Hamiltonian under gauge transformation, which can be generally written as 
\begin{align}
\hat H'_\mathrm{SE} \left ( t \right) 
	&  = \hat A	
    \hat  \xi  ^\dag  
    e ^{  
    i\theta \left ( t \right) }
	+ \text{H.c.}
	\\
	&  = \sum_{ \ell} 
	\left ( g_{ \ell} 
    \hat A
	\hat b _{ \ell}^\dag 
	e ^{ i  \omega _{ \ell} t 
	+ i\theta \left ( t \right) }
	+ \text{H.c.}
	\right) 
	. 
\end{align} 
Assuming the bosonic bath is in the vacuum state, we can derive a Markovian evolution equation for the reduced density matrix of the system as 
\begin{align}
&   \frac{  d }{ dt }	
\hat \rho  \left ( t \right) 
	= - i [ \Sigma \left ( t \right) 
	\hat A ^{\dag} \hat A  
	, \hat \rho  \left ( t \right) ] 
	+ \Gamma _{\mathrm{BR}} \left ( t \right) 
	\mathcal{D}  [   
    \hat A  ]  
    \hat \rho  \left ( t \right)
	\label{seq:breq.gen}
	.  
\end{align} 
Note that above equation is an example of the so-called Bloch-Redfield equation, which generalizes standard Lindblad master equation by incorporating effects due to a finite bath correlation time. The Bloch-Redfield equation still assumes the bath is Gaussian, but allows the bath to be non-Markovian. The first term on the RHS of Eq.~\eqref{seq:breq.gen} is analogous to the Lamb shift, and describes a correction to the coherent system Hamiltonian due to coupling to the environment. The second term takes the same form as the standard Lindblad dissipator, but the dissipator strength can now be negative because we have included effects from a non-Markovian bath. 

To compute the coefficients 
$ \Sigma \left ( t \right) $ and 
$ \Gamma _{\mathrm{BR}} 
\left ( t \right) $ in the Markovian equation, we first introduce the $2$-point correlation function of the new (gauge transformed) bath as 
\begin{align}
\mathcal{E}  _{\theta}
	\left ( t _{1} , t _{2}  \right) 
	&  \equiv 
	\langle \hat  \xi  \left ( t _{1} \right)
	\hat  \xi ^\dag \left ( t _{2} \right) \rangle 
	e ^{  
    - i \theta \left ( t _{1} \right)
	+ i \theta \left ( t _{2} \right)}  
	\\
	& = \sum _{ \ell} 
    \left| g_{ \ell} \right| ^{2}
    e ^{ - i \omega _{ \ell} 
    \left ( 
    t _{1} - t _{2} \right) 
    - i \theta \left ( t _{1} \right)
	+ i \theta \left ( t _{2} \right)}  
	\label{seq:autocor.gauge.gen}
	.  
\end{align} 
It is important to note that while the original bath is stationary, the new bath after gauge transformation is generally nonstationary, unless the gauge phase 
$ \theta \left ( t   \right) $ is only a linear function of time. More specifically, the correlation function 
$ \mathcal{E}  _{\theta} 
\left ( t _{1} , t _{2}  
\right) $ above 
would depend on both the time difference 
$ t _{1} - t _{2} $ and the ``center of times" 
$ (t _{1} + t _{2})/2 $, unless we have 
$ \theta \left ( t \right) 
= \theta \left ( 0 \right) + 
t \dot{\theta} \left ( 0 \right) $.
The Lamb shift coefficient 
$ \Sigma \left ( t \right) $ can now be related to the bath correlation function as  
\begin{align}
 \Sigma \left ( t \right)  
   &   =  \text{Im} \int_0^{t} dt _{1} 
\langle 
\hat  \xi  \left ( t  \right)
\hat  \xi ^\dag \left ( t _{1} \right) 
\rangle
    e ^{  
    - i \theta \left ( t  \right)
	+ i \theta \left ( t _{1} \right)}  
    \\
    &   =  \text{Im} \int_0^{t} dt _{1} 
\mathcal{E}  _{\theta} 
	\left ( t , t _{1}  \right) 
	\label{seq:breq.Sigma.gen}
	,  
\end{align} 
and the dissipator strength 
$ \Gamma _{\mathrm{BR}} 
\left ( t \right) $ is given by 
\begin{align}
&   \Gamma _{\mathrm{BR}} \left ( t \right)  
    = 2 \text{Re} \int_0^{t} dt _{1} 
	\mathcal{E}  _{\theta} 
	\left ( t, t _{1} \right)
	\label{seq:breq.Gamma.gen}
	.   
\end{align}

If the original bath has a finite but small correlation time 
$\tau_{\mathrm{E}}$ (the definition of short will become clear in discussions that follow), we can compute the leading order non-Markovian correction in the master equation, which depends on 
$ \dot{\theta} 
\left ( t \right) $. We first rewrite the correlation function of the (gauge transformed) bath in terms of the 
autocorrelation function of the old bath  
$ \mathcal{E}  
\left ( t _{1} , t _{2}  \right) $ (setting 
$ \theta \left ( t \right) 
\equiv 0 $ in Eq.~\eqref{seq:autocor.gauge.gen}) as 
\begin{align}
\mathcal{E}  _{\theta}  
	\left ( t _{1} , t _{2}  \right) 
	&  
	= \mathcal{E}  
\left ( t _{1} , t _{2}  \right)
    e ^{ 
    - i \theta \left ( t _{1} \right)
	+ i \theta \left ( t _{2} \right)}  
	.  
\end{align} 
Without loss of generality, we assume the bath has a finite correlation time 
$ \tau_{\mathrm{E}} $, so that the stationary bath  autocorrelation function can be written as 
\begin{align}
\mathcal{E}  
	\left ( t _{1} , t _{2}  \right) 
	&  = \left. 
	\mathcal{E}  _{\theta}  
	\left ( t _{1} , t _{2}  \right) 
	\right|_{ \theta 
	\left ( t \right) 
\equiv 0 }
	= 
	g_{ \mathrm{eff} }  ^{2} 
    e ^{ - \left | t _{1} 
    - t _{2}  \right| 
    / \tau_{\mathrm{E}} }  
	,  
\end{align} 
where 
$ g_{ \mathrm{eff} } = 
\sqrt{\mathcal{E}  
\left ( 0 , 0  \right) }$ is a real coupling coefficient that characterizes the net bath coupling strength. 
If the gauge phase changes much slower than the bath correlation time, i.e.~we have 
\begin{align}
\tau_{\mathrm{E}}^{2} 
	\ddot{\theta} 
	\left ( t \right)
	\ll 
\tau_{\mathrm{E}} 
\dot{\theta} \left ( t \right)
    \ll 1 
    , \quad \forall t
	,   
\end{align} 
and if we only cares about system dynamics at timescales that are much larger than the bath correlation time 
($t / \tau_{\mathrm{E}} \gg 1 $), then we can approximate the integral entering Eqs.~\eqref{seq:breq.Sigma.gen} and \eqref{seq:breq.Gamma.gen} as 
\begin{align}
&\int_0^{t} dt _{1} 
\mathcal{E}  _{\theta} 
	\left ( t , t _{1}  \right) 
	\nonumber \\
	= & g_{ \mathrm{eff} }  ^{2} 
	\int_0^{t} 
	 e ^{ - \left ( t - t _{1}  \right)  
    / \tau_{\mathrm{E}} } 
    e ^{ 
    - i \theta \left ( t \right)
	+ i \theta \left ( t _{1} \right)}  
	dt _{1} 
	\\
\simeq 	& g_{ \mathrm{eff} }  ^{2} 
	\int_{-\infty} ^{t} 
	 e ^{ - \left ( t 
	 - t _{1}  \right)   
    / \tau_{\mathrm{E}} } 
    e ^{ 
    - i \dot{\theta} \left ( t \right) 
    \left ( t - t _{1} \right)
    + \frac{i}{2}
    \ddot{\theta} \left ( t \right) 
    \left ( t - t _{1} \right)^{2} } 
	dt _{1} 
	\label{seq:breq.kernel.approx}
	\\
	\simeq & 
	g_{ \mathrm{eff} } ^{2} 
	\tau_{\mathrm{E}}  
	[ 1- i \tau_{\mathrm{E}}
	\dot{\theta} 
	\left ( t \right) 
	+ i \tau_{\mathrm{E}}^{2} 
	\ddot{\theta} 
	\left ( t \right) 
	- \tau_{\mathrm{E}}^{2} 
	\dot{\theta} 
	\left ( t \right)^{2} ] 
	. 
\end{align} 
Thus, we obtain first two leading-order contributions to the Lamb shift and  time-dependent decay rate in terms of the small parameter $ \tau_{\mathrm{E}} 
\dot{\theta} \left ( t \right)$ as
\begin{align}
&  \Sigma \left ( t \right)  
    \simeq 
    - g_{ \mathrm{eff} } ^{2} 
	\tau_{\mathrm{E}} ^{2} 
	[ \dot{\theta} 
	\left ( t \right)
	- \tau_{\mathrm{E}} 
	\ddot{\theta} 
	\left ( t \right) ]
    , \\
& \Gamma _{\mathrm{BR}} \left ( t \right)  
    \simeq 2 
     g_{ \mathrm{eff} } ^{2} 
	\tau_{\mathrm{E}}  
	\left( 1 -  [
	\tau_{\mathrm{E}} 
	\dot{\theta} 
	\left ( t \right)
	 ]^{2} \right ) 
    ,  
\end{align} 
and we can rewrite Eq.~\eqref{seq:breq.gen} as  
\begin{align}
\frac{  d }{ dt }	
\hat \rho  \left ( t \right) 
	\simeq & 
	i g_{ \mathrm{eff} } ^{2} 
	\tau_{\mathrm{E}} ^{2} 
	[ \dot{\theta} 
	\left ( t \right)
	- \tau_{\mathrm{E}} 
	\ddot{\theta} 
	\left ( t \right) ]
	[  
	\hat A ^{\dag} \hat A  
	, \hat \rho  
	\left ( t \right) ]
	\nonumber \\
	& +  2  
	g_{ \mathrm{eff} } ^{2} 
	\tau_{\mathrm{E}}  
	\left( 1 -  [
	\tau_{\mathrm{E}} 
	\dot{\theta} 
	\left ( t \right)
	 ]^{2} \right ) 
	\mathcal{D}  [   
    \hat A  ]  
    \hat \rho  \left ( t \right)
  \label{seq:breq.nM.approx}
	.   
\end{align} 
It is also convenient to rewrite this in terms of original bath DOS $ \mathcal{J}  _{0}  $ given in Eq.~\eqref{seq:dos.gen}, which can now be calculated as 
\begin{align}
\mathcal{J}  _{0} 
\left[   \omega  \right] 
    &  \equiv  
    \frac{1}{ 2 \pi }
    \int ^{\infty} _{ - \infty}
    \langle [
\hat B \left ( t \right)
, \hat B^\dag \left ( 0 \right)
] \rangle
e ^{ i  \omega  t   }
    dt 
    \\
    & =\frac{1}{ 2 \pi }
    \int ^{\infty} _{ - \infty}
    \mathcal{E}  
	\left ( t  , 0 \right) 
e ^{ i  \omega  t   }
    = \frac{1}{ \pi }
    g_{ \mathrm{eff} } ^{2} 
	\tau_{\mathrm{E}}  
	, 
\end{align} 
so that we have 
\begin{align}
\partial_t 
\hat \rho  \left ( t \right) 
	\simeq & i \pi 
	\mathcal{J}  _{0} 
\left[  0 \right]  
	\tau_{\mathrm{E}} 
	[ \dot{\theta} 
	\left ( t \right)
	- \tau_{\mathrm{E}} 
	\ddot{\theta} 
	\left ( t \right) ] 
	[  
	\hat z ^{\dag} \hat z  
	, \hat \rho 
	\left ( t \right) ]
	\nonumber \\
	& + 2 \pi 
    \mathcal{J}  _{0} 
\left[  0 \right] 
    \left( 1 -  [
	\tau_{\mathrm{E}} 
	\dot{\theta} 
	\left ( t \right)
	 ]^{2} \right ) 
	\mathcal{D}  [   
  \hat z     ]  \hat \rho  \left ( t \right)
  \label{seq:breq.nM.app.1}
	.   
\end{align}

\subsection{Generalization to multiple dissipators}

\label{appsec:nonM.multi}

In the main text, we stated that the invariance of a generic Lindbladian under time-dependent unitary transformations $ \cmatU \left ( t \right) $~\cite{comp_mat} on the jump operators is closely related to the Markovian nature of the bath. Here, we illustrate this connection by considering a general class of microscopic environments that can realize such dissipators in the Markovian limit, and further deriving leading-order non-Markovian corrections. 

We start by rewriting a general Lindbladian as (see Eq.~\eqref{eq:qme.symm.multi})
\begin{align}
\mathcal{L} 
_{A  } \hat \rho _{A  }  
	&   =  \Gamma 
\sum _{\ell=1}^{N} 
	\mathcal{D}  
\left[ \sum  _{m=1}^{N} 
    u _{\ell m}  (t ) 
    \hat{A} _{ m }  \right]
    \hat{\rho} _{A  } 
    \label{seq:qme.symm.multi}
	,  
\end{align} 
where $ u _{\ell m}  (t )  $ are again matrix elements of the $N$-dimensional complex unitary matrix $ \cmatU (t ) $. 
For concreteness, we consider a microscopic environment realizing such dissipative dynamics, by coupling the system to a harmonic oscillator bath with independent bath operators 
$ \hat \xi  _{\ell } $. In the interaction picture with respect to bath-only Hamiltonian $\hat H_\mathrm{S} $, the system-bath Hamiltonian is given by   
\begin{align}
\hat H_\mathrm{SE} \left ( t \right) 
	& = \sum _{\ell , m=1}^{N}   
    u _{\ell m} \left ( t \right)  
    \hat{A} _{ m }  
    \hat \xi  _{\ell }  ^\dag  \left ( t \right) 
	+ \text{H.c.}
	,  
\end{align} 
where the bath operators satisfy white-noise statistics   
\begin{align}
\langle 
    \hat \xi   _{\ell }^{(0) \dag } \left ( t _{1}  \right)
\hat \xi   _{\ell' }^{(0) } \left ( t _{2}  \right)
\rangle  
    &   = 0 
, \\
\langle  
\hat \xi   _{\ell }^{(0)  } 
\left ( t _{1}  \right)
 \hat \xi   _{\ell' }^{(0) \dag } 
 \left ( t _{2}  \right)
 \rangle
    &   = \Gamma \delta _{\ell \ell' } 
    \delta \left ( t _{1} 
    - t _{2}  \right) 
    . 
    \label{seq:nsd.white.multi}
\end{align} 
In the Markovian limit, the resulting system ($A$ modes) dynamics will be invariant under any unitary matrix $ \cmatU \left ( t \right) $. To understand the role of Markovianity in this emergent symmetry, we assume that the bath modes have finite but very small correlation times, and we now derive the correction to $A$ system dynamics due to non-Markovian effects. Towards this goal, it is useful to rewrite the system-bath coupling using a new set of bath operators $\hat C _{\ell } $, as 
\begin{align}
\hat H _\mathrm{SE} \left ( t \right) 
	& = \sum _{\ell =1 } ^{N} 
    \hat{A} _{ \ell }  
    \hat C _{\ell }  ^\dag 
    \left ( t \right)  
	+ \text{H.c.}
	,  \quad
	\hat C _{\ell } \left ( t \right) 
	= \sum_{ m =1 } ^{N} 
    u _{m \ell } ^{*}  \left ( t \right) 
    \hat \xi  _{ m } \left ( t \right) 
    . 
\end{align}
In contrast to the original bath operators $\hat \xi   _{ m } \left ( t \right) $, the new bath operators $\hat C _{ m } \left ( t \right) $ due to the presence of the time-dependent coefficients $u _{m \ell } \left ( t \right) $. As we show, such nonstationary noise in combination with a finite bath correlation time will give rise to nontrivial non-Markovian effects in $A$ dynamics.

Making use of the standard Born-Markov approximation, we can integrate out the bath modes to obtain an effective master equation for system $A$ 
\begin{align}
\frac{d \hat \rho  \left ( t \right) }{dt}
	=  &   
	- i \sum _{\ell , m=1}^{N}   
	\Sigma  _{\ell m} \left ( t \right) 
	\left[ 
 \hat{A}_{\ell} ^\dag 
 \hat{A}_{m} 
	,  \hat{\rho} \left ( t \right)\right] 
	 \nonumber \\
	 &   + \sum _{\ell , m=1}^{N}  
 \Gamma  _{\ell m} \left ( t \right) 
    \left(\hat{A}_{m} 
    \hat{\rho} \left ( t \right)  \hat{A}_{\ell} ^\dag - \frac{1}{2} 
    \{\hat{A}_{\ell}^\dag  \hat{A}_{m} , \hat{\rho} \left ( t \right) \} \right)
	.  
\end{align} 
The effective Hamiltonian, also known as the Lamb shift term, and the dissipator coefficient matrices are given by 
\begin{align}
\Sigma   \left ( t \right) 
	&   =  -
	\frac{i}{2} 
	\left [ 
	\mathcal{S}   \left ( t \right) 
	- \mathcal{S}  ^{\dag}
	\left ( t \right) 
	\right]
	, \\ 
\Gamma   \left ( t \right) 
&  = \mathcal{S}  \left ( t \right) 
	+ \mathcal{S} ^{\dag} 
	\left ( t \right)  
	, \\ 
\mathcal{S} _{\ell m} 
\left ( t  \right)
&  =  	\int_0^{t} dt _{1} 
	\langle 
\hat C _{\ell } \left ( t  \right)
\hat C _{m }^\dag \left ( t _{1} \right) 
\rangle  
	.  
\end{align} 
While our discussion is applicable to generic forms of bath correlators in the small correlation time limit, for concreteness we assume that they take the diagonal form as follows 
\begin{align}
\langle 
\hat \xi   _{\ell }^{(0) \dag } \left ( t _{1}  \right)
\hat \xi   _{\ell' }^{(0) } \left ( t _{2}  \right)
\rangle  
    = &   0 
, \\
\langle  
\hat \xi   _{\ell }^{(0)  } 
\left ( t _{1}  \right)
 \hat \xi   _{\ell' }^{(0) \dag } 
 \left ( t _{2}  \right)
 \rangle
    = & \Gamma 
    \delta _{\ell \ell' } 
    \frac{1}{ 2 \tau_{\mathrm{E}, \ell }}
    e ^{ - \left | t _{1} 
    - t _{2}  \right| 
    / \tau_{\mathrm{E}, \ell } }  
    . 
    \label{seq:nsd.corr.multi}
\end{align} 
We can thus compute the master equation coefficients via the bath kernel function $\mathcal{S} _{\ell m} 
\left ( t  \right) $. For convenience, we introduce the time-dependent generator of $ \cmatU \left ( t \right) $ as 
\begin{align}
\cmat{G} 
\left ( t  \right)
&  = - i 
\frac{d\cmatU \left ( t \right) }{dt}
\cmatU ^{\dag} \left ( t \right) 
	.  
\end{align} 
Assuming small bath correlation time $\tau_{\mathrm{E}, \ell } \ll 1 $
\begin{align}
\mathcal{S} _{\ell m} 
\left ( t  \right)
=  &  
\sum _{j, j'=1}^{N}  
u _{j \ell } ^{*}  \left ( t \right) 
\int_0^{t} dt _{1} 
	\langle 
\hat \xi   _{ j } 
\left ( t  \right)
\hat \xi   _{j' }^\dag 
\left ( t _{1} \right) 
\rangle  
u _{j' m }  \left ( t_{1}  \right) 
\\
\simeq &   \frac{1}{2}
\sum _{j, j'=1}^{N}  
u _{j \ell } ^{*}  \left ( t \right) 
u _{j' m }  \left ( t   \right) 
\left [ 1+ i \tau_{\mathrm{E}, j}
	g _{j j' } 
	\left ( t \right) 
	\right. 
	\nonumber \\
&\quad   \left.
	- i \tau_{\mathrm{E},j}^{2} 
	\dot{g} _{j j' } 
	\left ( t \right) 
	- \tau_{\mathrm{E},j }^{2} 
	g _{j a } 
	\left ( t \right) g _{a j' } 
	\left ( t \right) \right ] 
	.  
\end{align} 
The corresponding master equation can be rewritten in terms of a new jump operator basis 
\begin{align}
\hat{A} '_{\ell}  \left ( t   \right) 
= \sum _{m=1}^{N}  
u _{\ell m }  \left ( t   \right) 
\hat{A}_{m} 
	,   
\end{align} 
so that we can write the effective master equation as  
\begin{align}
&   \mathcal{L}  \hat \rho  \left ( t \right) 
	\simeq  
	- i \sum _{\ell , m=1}^{N}   
	\Sigma ' _{\ell m} \left ( t \right) 
	\left[ 
 \hat{A} '_{\ell}  {}^\dag \left ( t   \right) 
 \hat{A} '_{m} \left ( t   \right) 
	,  \hat{\rho} \left ( t \right)\right] 
	\nonumber \\
	+ & \sum _{\ell , m=1}^{N}  
 \Gamma ' _{\ell m} \left ( t \right) 
    \left(
    \hat{A} '_{m} \left ( t   \right) 
    \hat{\rho} \left ( t \right)  \hat{A} '_{\ell}  {}^\dag \left ( t   \right)  
    - \frac{1}{2} 
    \{ \hat{A} '_{\ell}  {}^\dag 
    \left ( t   \right)   
    \hat{A} '_{m} \left ( t   \right) , \hat{\rho} \left ( t \right) \} \right)
	.  
\end{align} 
The Lamb shift term and dissipator coefficients in the new basis are now given by 
\begin{align}
\Sigma  '  _{\ell m} \left ( t \right) 
    = &   
	\frac{1}{4}
	\left [ 
	\left ( 
	\tau_{\mathrm{E}, \ell} 
	+ \tau_{\mathrm{E}, m}
	\right) 
	g _{\ell m } 
	\left ( t \right) 
	-  \left ( 
	\tau_{\mathrm{E}, \ell} ^{2} 
	+ \tau_{\mathrm{E}, m}^{2} 
	\right) 
	\dot{g} _{\ell m } 
	\left ( t \right) 
	\right.
\nonumber\\
	&\left .
	+ i \left ( 
	\tau_{\mathrm{E}, \ell} ^{2} 
	- \tau_{\mathrm{E}, m}^{2} 
	\right) 
	g _{\ell a } 
	\left ( t \right) g _{a m } 
	\left ( t \right) 
	\right]
	, \\ 
\Gamma ' _{\ell m}   \left ( t \right) 
= &  1+ \frac{1}{2} 
	\left [  
	i \left ( 
	\tau_{\mathrm{E}, \ell} 
	- \tau_{\mathrm{E}, m}
	\right) 
	g _{\ell m } 
	\left ( t \right) 
	- i \left ( 
	\tau_{\mathrm{E}, \ell} ^{2} 
	- \tau_{\mathrm{E}, m}^{2} 
	\right) 
	\dot{g} _{\ell m } 
	\left ( t \right) 
	\right.
\nonumber\\
	&\left .
	- \left ( 
	\tau_{\mathrm{E}, \ell} ^{2} 
	+ \tau_{\mathrm{E}, m}^{2} 
	\right) 
	g _{\ell a } 
	\left ( t \right) g _{a m } 
	\left ( t \right) 
	\right]
	.  
\end{align}

Similar to the single-dissipator case, the action of the unitary matrix elements 
$u _{\ell m} \left ( t \right)  $ in the Lindbladian $ \mathcal{L} 
_{ A  } \hat \rho  
=  \sum  _{\ell=1}^{N}   
	\mathcal{D}  
[ \sum _{m=1}^{N}   
    u _{\ell m} \left ( t \right)  
    \hat{A} _{ m } ]
    \hat{\rho} 
$ (see Eq.~\eqref{seq:qme.symm.multi}) can be intuitively understood as shifting the Hamiltonian of a microscopic bath model consisting of harmonic oscillators. More specifically, the dissipators can be realized via the following system-bath Hamiltonian 
\begin{align}
& \hat H_\mathrm{tot} 
= 
\hat H_\mathrm{E} 
+ \hat H_\mathrm{SE} 
, \quad 
\hat H_\mathrm{E} 
=  \sum  _{ \alpha} 
\omega _{ \alpha} 
\sum _{\ell=1}^{N}
\hat b _{ \ell, \alpha }^\dag 
    \hat b _{ \ell, \alpha }
, \\
&  \hat H_\mathrm{SE} \left ( t   \right) 
	=  \sum _{\ell, m=1 }^{N}   
u _{\ell m }  \left ( t   \right) 
\hat{A} _{ m }  
    \hat \xi   _{\ell }  ^\dag   
	+ \text{H.c.}
	, \quad 
\hat \xi   _{\ell } 
= \sum_{ \alpha } 
	g ^{*}_{ \alpha}  
    \hat b _{ \ell, \alpha}  
	. 
\end{align} 
Note that we choose the bath mode frequencies $ \omega _{ \alpha} $ and coupling strengths $g ^{*}_{ \alpha}  $ to be identical for corresponding modes 
$\hat b _{ \ell, \alpha}  $ coupled to different system operators 
$\hat{A} _{ \ell }  $. This bath model is not necessarily physically motivated, but it allows a simple interpretation of the time-dependent coupling coefficients $u _{\ell m }  \left ( t   \right) $. In fact, we can now transform to a properly defined rotating frame
\begin{align}
& \hat \rho _\mathrm{SE} 
\to  
\hat \rho'_\mathrm{SE} 
= \hat U ^{\dag} \left ( t \right) 
\hat \rho _\mathrm{SE} 
\hat U \left ( t \right) 
, \quad
\hat U \left ( t \right) 
= \mathcal{T} e ^{-i 
\int ^{t}_{0} 
\delta \hat H_\mathrm{E} 
\left ( t _{1} \right) 
dt_{1}}
, 
\end{align} 
so that the time dependence in system-bath couplings is converted into an additional term in the bath-only Hamiltonian, as 
\begin{align} 
& \hat H _\mathrm{SE}  
\to  \hat H' _\mathrm{SE} =  
\hat U ^{\dag} \left ( t \right) 
\hat H_\mathrm{SE} 
\hat U  \left ( t \right) 
	=  \sum _{\ell =1 }^{N}   
\hat{A} _{ \ell }  
    \hat \xi   _{\ell }  ^\dag   
	+ \text{H.c.}
\\
& \hat H_\mathrm{E} 
\to  \hat H'_\mathrm{E} 
\left ( t  \right) 
= \hat H_\mathrm{E} 
+ \delta \hat H_\mathrm{E} 
\left ( t  \right) 
.  
\end{align} 
The correction term 
$\delta \hat H_\mathrm{E} 
\left ( t  \right) $ takes the form of a beamsplitter Hamiltonian, and can be written as 
\begin{align}
\delta \hat H_\mathrm{E} 
\left ( t  \right) 
= \sum  _{ \alpha}  
\sum _{\ell,m =1}^{N} 
h _{\ell m } \left ( t \right) 
\hat b _{ \ell, \alpha }^\dag 
    \hat b _{ m , \alpha }
    ,
\end{align} 
where the beamsplitter matrix 
$h _{\ell m } \left ( t \right) 
= \left[ \cmat{H} \left ( t  \right) 
\right ]_{\ell m} $ is related to the unitary matrix 
$\cmatU  \left ( t \right) $ via the following equation  
\begin{align}
\cmat{H} 
\left ( t  \right)
&  =  i 
\frac{d\cmatU ^{\dag} \left ( t \right) }{dt}
\cmatU  \left ( t \right) 
	.  
\end{align}

\section{Reservoir engineering implementation of nonreciprocal interactions in Eq.~\eqref{eq:nl.nr.Lindbladian} via a unidirectional waveguide}

\label{appsec:phys.impl.chiral}

In Sec.~\ref{sec:phys.setup} in the main text, we state that the nonreciprocal single-dissipator master equation in Eq.~\eqref{eq:nl.nr.Lindbladian} can be straightforwardly realized via reservoir engineering, if one also has access to elements explicitly  breaking TRS, e.g.~a unidirectional waveguide. In this Appendix, we provide a detailed discussion about the physical setup in this case and its connection to related experiments~\cite{Nakamura2018,Wallraff2018}.

Recall the desired dissipator in Eq.~\eqref{eq:nl.nr.Lindbladian}, as given by 
$\mathcal{L}  _{ \nlqnr } 
\hat \rho   
= \Gamma \mathcal{D} [  
  \hat A   \hat U _{B}   ] 
\hat \rho  $. For concreteness, in this Appendix we assume the $A$ subsystem is a cavity mode with operator 
$\hat A = \hat a $ being a bosonic lowering operator, but we note that the scheme discussed below can be generalized to other systems as well. The corresponding Lindbladian is thus 
\begin{align}
& \mathcal{L}  _{ \nlqnr } 
\hat \rho   
= \Gamma \mathcal{D} [  
  \hat a   \hat U _{B}   ] 
\hat \rho  
	.  
	\label{seq:nl.nr.ME} 
\end{align} 
For purpose that will become clear later, it is convenient to rewrite the unitary in terms of a Hermitian generating operator $\hat E _{B}$, i.e.
\begin{align}
\hat U _{B}  
= \exp (- i 
\hat E _{B}  ) 
, 
\end{align} 
where eigenvalues of $\hat E _{B}$ are real and lie in the range $( - \pi, \pi ]$. Noting that definition of unitary $\hat U _{B}  $ has a global gauge phase degree of freedom, we can replace it with 
$ e ^{ - i\theta_{0}}\hat U _{B}  $ for arbitrary $\theta_{0}$ without affecting system dynamics; we will choose this phase such that $- 1$ is not in the spectrum of $\hat U _{B}  $, or equivalently, $ \hat E _{B} $ does not contain the eigenvalue $ \pi $.

We now introduce a unidirectional coupling (e.g., mediated by a one-way waveguide) from $A$ to an reservoir mode $c$, as well as a coherent interaction  
$\hat H _{BC}   $ between $B$ and the reservoir of the form 
\begin{align}
\hat H _{BC}  
= \frac{  \lambda  }{2} 
\hat M _{B} 
\hat c^\dag \hat c 
, \label{seq:phys.cas.chi}
\end{align} 
where $\hat M _{B} $ is a dimensionless Hermitian operator on $B$ that we will determine later. The  total system-reservoir (SR) dynamics can thus be described by the following master equation 
\begin{align} 
\frac{d \hat \rho  _{\mathrm{SR}}
}{dt}
	= &  
	- i \left[ 
	\hat H _{AC}  
    + \hat H _{BC}   , 
    \hat \rho   _{\mathrm{SR}} \right] 
	\nonumber \\
	&  +   \mathcal{D} 
	\left[ \sqrt{ \Gamma_ {a} } \hat a  
	- i 
	\sqrt{ \Gamma_{c} } 
	\hat c  \right]  
	\hat \rho  _{\mathrm{SR}} 
	, \label{seq:qme.phys.cas}
	\\
\hat H _{AC}  
    = &  \frac{  \sqrt{ 
    \Gamma_ {a}  \Gamma_{c} }    }{2}  
	\left( \hat a ^\dag 
	\hat c 
	+  \hat c^\dag 
	\hat a   \right) 
	. 
\end{align} 
Note that if we ignore $B$ and its coupling to the ancilla, the remaining setup reduces to the standard cascaded quantum systems, see Eq.~\eqref{eq:cqme.gen}. To realize the dissipator in Eq.~\eqref{seq:nl.nr.ME}, we need to assume the limit where the ancilla serves as a Markovian reservoir for the $AB$ system. More specifically, this requires 
$ \Gamma_{c}  \gg 
\Gamma_ {a} $, in which case we can use standard adiabatic elimination technique~\cite{Gardiner2004book} to obtain a new effective master equation of just $AB$, as 
\begin{align} 
&  \frac{d \hat \rho }{dt}
	=  \Gamma_ {a}   
	\mathcal{D} 
	\left[ \hat a  
	\hat U _{B, \mathrm{eff} }   
	\right]  
	\hat \rho  
	, \label{seq:qme.phys.cas.eff}
	\\
&  \hat U _{B, \mathrm{eff} }    
    = \frac{  
    \Gamma_{c} 
    \hat{\mathbb{I}} _{B} 
    - i \lambda  \hat M _{B} }
    {\Gamma_{c} 
    \hat{\mathbb{I}} _{B} 
    + i \lambda  \hat M _{B}}  
	. 
\end{align} 
It is worth stressing that the validity of Eq.~\eqref{seq:qme.phys.cas.eff} does not depend on having a small coupling between reservoir and $B$ subsystem, i.e.~$\lambda $ can be greater than $\Gamma_{c} $. Hence, we can use this recipe to realize general unitary operators $ \hat U _{B} $.

Comparing Eq.~\eqref{seq:qme.phys.cas.eff} with Eq.~\eqref{seq:nl.nr.ME}, we can choose the $A$ system coupling rate 
$ \Gamma_ {a} $ and subsystem $B$ coupling operator 
$\hat M _{B}$ in the starting setup, such that the two master equations agree, i.e. we require 
\begin{align} 
\Gamma _ {a}  &  = \Gamma
, \\
\hat M _{B}  & = 
    \frac{  \Gamma_{c}  }
    { \lambda   }
    \tan \frac{\hat E _{B} }{2}
    . 
\end{align} 
We thus obtain a general recipe, as given by Eq.~\eqref{seq:qme.phys.cas}, which makes use of directional coupling and reservoir engineering techniques to implement a generic nonreciprocal quantum master equation given by Eq.~\eqref{seq:nl.nr.ME}. For the specific case where $B$ is a single qubit, and assuming the target unitary is a $Z$-rotation
$e ^{ - i \frac{  \theta }{  2  }
\hat \sigma_z } $, we further have 
\begin{align} 
\hat M _{B}  & = 
\frac{  \Gamma_{c}  }
    { \lambda   }
    \left (\tan 
    \frac{ \theta }{4}
    \right) 
    \hat \sigma_z 
    . 
\end{align}

We note that the physical setup discussion in this Appendix is experimentally accessible using e.g.~state-of-the-art superconducting qubit platforms. In fact, in a different context, specific cases of dynamics in Eq.~\eqref{seq:qme.phys.cas} have been implemented for quantum-non-demolition (QND) measurement of itinerant microwave photons~\cite{Nakamura2018,Wallraff2018}. In those works, the QND detector structure consists of a cavity dispersively coupled to a qubit. To perform the QND detection, an itinerant microwave field is reflected off the cavity mode. By measuring the qubit phase shift, one can in turn extract average photon number of the input pulse. Comparing the QND setup to our master equation in Eq.~\eqref{seq:qme.phys.cas}, the cavity and the qubit would be mapped to the reservoir mode $c$ and subsystem $B$, respectively. As a result, external coupling rate of the cavity would correspond to $ \Gamma_{c} $, and the cavity-qubit dispersive interaction to Eq.~\eqref{seq:phys.cas.chi}. As for subsystem $A$, we can replace it with a qubit to describe the single-photon source in Ref.~\cite{Wallraff2018}, and the recipe in Eq.~\eqref{seq:qme.phys.cas} can be straightforwardly modified to describe the experimental system. In Ref.~\cite{Nakamura2018}, the itinerant microwave has a Gaussian pulse shape, which formally can be mimicked via output of a cavity mode ($A$) with a time-dependent coupling rate 
$ \Gamma_ {a} $. In both experiments, the unidirectional coupling is achieved by explicitly using a circulator.

Despite aforementioned similarities between the physical setups used in~\cite{Nakamura2018,Wallraff2018} and our model, we note that those previous works did not really use the master equations in Eq.~\eqref{seq:qme.phys.cas} or~\eqref{seq:qme.phys.cas.eff} to analyze dynamics of their systems. Our derivation thus reveals a striking new feature of dynamics in the setup: in the Markovian limit ($ \Gamma_{c}  \gg 
\Gamma_ {a} $), if one starts from a product state of $A$ cavity Fock state and a generic qubit ($B$) state, the qubit steady state in the long-time limit would  undergo a \textit{coherent} $Z$ rotation from its initial state, with the phase shift controlled by number of photons in the initial $A$ state. The experimental setup in~\cite{Wallraff2018} thus can be directly used to dissipatively stabilize unitary gates (see also Sec.~\ref{sec:qnr.diss.gate} in the main text). 

Finally, we note that the effective master equation in Eq.~\eqref{seq:qme.phys.cas.eff} describes an idealized case; in realistic settings, the fidelity of the unitary gate emerging from such dissipative stabilization process will suffer from a range of imperfections, including non-Markovian effects due to a finite reservoir linewidth, nonzero thermal photon population in the reservoir and cavity modes, imperfect $A$ cavity initial state preparations, etc. Fortunately, in practical systems, it is possible to carefully engineer the setup to suppress those factors and achieve gate fidelity comparable to typical pulse-based gates. Further, given an experimental system, one can incorporate those imperfections into the theory model in Eq.~\eqref{seq:qme.phys.cas} and quantitatively examine how they affect the dissipatively stabilized gate fidelity (e.g.~see Sec.~\ref{sec:phy.impl.nonM} for discussion on non-Markovian effects), which can be useful for designing new setups to improve gate fidelity.

\section{The role of Markovianity in achieving unidirectionality for system in Eq.~\eqref{eq:bath.qme.gen}}

\label{appsec:nonM.bath}

In Sec.~\ref{sec:phys.setup} in the main text, we have shown that one can realize the nonlinear dissipator $  \mathcal{L}  _{\mathrm{gate}} \hat \rho 
	= \Gamma \mathcal{D}  [   
e ^{ - i \frac{  \theta }{  2  } \hat \sigma_z } 
\hat a    ]  
\hat \rho  $ using an intermediate reservoir mode that effectively mediates a one-way interaction from the cavity mode to the qubit (see Eq.~\eqref{eq:bath.qme.gen}), and we state that this protocol crucially relies on the fact that the reservoir operates in the Markovian limit. In this Appendix, we derive this result in more detail.

We start with a general setup consisting of a driven-cavity-qubit system coupled to a reservoir mode. In the rotating frame with respect to the drive frequency, the total system dynamics can be described by the master equation 
$( d \hat \rho /dt) 
= \mathcal{L}  _{\mathrm{0}} 
\hat \rho   $, with the Lindbladian given by 
\begin{align}
\label{seq:sing.bath.setup}
&   \mathcal{L}  _{\mathrm{0}} 
\hat \rho   = 
	- i \left[ \hat H_\mathrm{S} + 
	\hat H_\mathrm{E} + \hat H_\mathrm{int} 
	, \hat \rho   \right] 
	+   \kappa_{a}\mathcal{D} 
	\left[  \hat a  \right]  
	\hat \rho	
	+   \kappa_{c}\mathcal{D} 
	\left[  \hat c  \right]  
	\hat \rho 
	, \\
&\hat H _\mathrm{S} =
	-\Delta _{a}  
	\hat a ^\dag \hat a  
    + \frac{ \lambda  _{a}  }{2}  
	\hat \sigma_z \hat a ^\dag \hat a
	+ f  _\mathrm{dr} ^*   
	\left(  t  \right)  
	 \hat a  
	+  f  _\mathrm{dr}  
	\left(  t  \right)   
	\hat a ^\dag 
	, \\
& \hat H _\mathrm{E} =
	- \Delta_{c} 
	\hat c^\dag \hat c 
    , \quad
    \hat H_\mathrm{int} = 
	\left(  J \hat a ^\dag \hat c 
	+  J ^{*} \hat c^\dag \hat a   \right)  
	+
	\frac{ \lambda _{c} }{2}  \hat \sigma_z \hat c^\dag \hat c
	. 
\end{align} 
We can approximate the reservoir mode as serving as a Markovian environment for the cavity mode if the parameters satisfy the condition $ \kappa_{c} \gg J$. In this section, we discuss the effects due to possible non-Markovianity of the reservoir, by explicitly looking at the system Langevin equations of motion in the parameter regime 
$J \gtrsim \kappa_{c} $. Note that it is also possible to follow procedures similar to standard adiabatic elimination on the master equation in Eq.~\eqref{seq:sing.bath.setup}, and studying non-Markovian corrections.

To understand system dynamics, it is useful to write the corresponding quantum Langevin equations of the system as 
\begin{align}
& i \partial_t  \hat a  =
	  \left(  -\Delta _{a}  
	  + \frac{ \lambda  _{a}  }{2} \hat \sigma_z
	  -i \frac{ \kappa _{ a  } }{2}  
	  \right) \hat a  
	+ J   \hat c 
	- i \sqrt{ \kappa _{ a  }    }  
	\hat a _{  \mathrm{in}} 	 
	+  f  _\mathrm{dr}   \left(  t  \right)   
	,\\
& i \partial_t  \hat c =
	 \left(  -\Delta _{c} 
	 + \frac{ \lambda _{c}}{2} \hat \sigma_z  
	 -i \frac{ \kappa _{ 2  }    }{2}  \right) \hat c 
	 +  J ^*  \hat a
	- i \sqrt{ \kappa _{ c  }    }  
	\hat c _{ \mathrm{in}} 
	,\\
& i \partial_t  \hat \sigma_- = 
	\left(   \lambda _{a}  
	\hat a ^\dag \hat a 
	+ \lambda _{c}  
	\hat c ^\dag \hat c
	\right)
	\hat \sigma_- 
	. 
\end{align} 
While our discussion can be straightforwardly generalized to a generic initial state, for convenience we assume that the cavity mode starts in a coherent state 
$ \left | \alpha  _{0} \right \rangle $, i.e.~with amplitude $ \alpha _{0}  $, and the reservoir is in vacuum at $ t = t_{0}  $. Thus, the system dynamics can be fully determined by solving the following set of linear equations for a set of complex amplitudes 
$ \bar{a} _{  \sigma} $ and 
$ \bar{c} _{  \sigma} $ 
($ \sigma =\uparrow, \downarrow $)
\begin{subequations}
\begin{align}
& i \partial_t  
\bar{a}   _{ \sigma }  =
	  \left(  -\Delta _{a}  
	  + \frac{ \lambda  _{a}  }{2} \sigma_z 
	  -i \frac{ \kappa _{a } }{2}  \right) 
	  \bar{a}   _{ \sigma }
	+ J   \bar{c}   _{ \sigma }
	 +  f  _\mathrm{dr}   \left(  t  \right)   
	,\\
& i \partial_t   
\bar{c}   _{ \sigma }  =
	 \left(  -\Delta _{c} 
	 + \frac{ \lambda _{c} }{2} \sigma_z 
	  -i \frac{ \kappa _{ c   }    }{2}  \right)   
	 \bar{c}   _{ \sigma }  
	 +  J ^*  \bar{a}   _{ \sigma }  
	, 
\end{align}
\end{subequations}
where $  \sigma_z = \pm 1 $ corresponding to $ \sigma = \uparrow, \downarrow $. The qubit coherence function can now be computed by solving the equation
\begin{align}
&  \frac{ 1 }
{ \left \langle \hat \sigma _ - (t) \right \rangle  }
\frac{ d \langle \sigma_-  (t) \rangle }
{ dt  }
= 
	-  i \lambda _{a} 
	 \bar{a} _{  \uparrow}
	\bar{a} ^*_{  \downarrow} 
	-  i \lambda _{c} 
	 \bar{c} _{ \uparrow}
	\bar{c} ^*_{  \downarrow} 
	.   
	\label{seq:2m.eom.qbcoh}
\end{align}

We can formally integrate out the reservoir mode by transforming to the Fourier space and eliminating $\bar{c} _{  \sigma }  \left[   \omega  \right] $ from the equations, using the relation 
\begin{align}
& \bar{c} _{ \sigma} \left[   \omega  \right]
=\frac{  J ^*   }
	{ \omega + \Delta _{ c }
	- \frac{ \lambda _{ c } }{2}  \sigma_z 
	 +i \frac{ \kappa _{ c  }    }{2}} 
\bar{a} _{ \sigma} \left[   \omega  \right] 
	.  
	\label{seq:2m.FT.a2}
\end{align}
For notation convenience, we can rewrite the reservoir amplitude in terms of the bare reservoir mode susceptibility function 
$  \chi ^{(0)} _{c , \sigma } 
\left[   \omega  \right] 
$, as 
\begin{align}
\bar{c} _{ \sigma} \left[   \omega  \right]
& = J ^* \chi ^{(0)} _{c , \sigma } 
\left[   \omega  \right]
\bar{a} _{ \sigma} \left[   \omega  \right] 
, \\
\chi ^{(0)} _{c , \sigma } 
\left[   \omega  \right]
&  \equiv \frac{  1 }
{ \omega + \Delta _{ c }
- \frac{ \lambda _{ c } }{2}  \sigma_z 
	+i \frac{ \kappa _{ c  }    }{2}} 
	, 
\end{align}
so that we obtain 
\begin{align}
\bar{a} _{ \sigma} \left[   \omega  \right]
&  =\frac{  1 }
	   { \omega + \Delta _{a} 
	   - \frac{ \lambda  _{a}  }{2} \sigma_z  
	   + i \frac{ \kappa _{ a } }{2} 
	- \frac{ \left| J  \right|  ^2  }
	{ \omega + \Delta _{ c }  
	- \frac{ \lambda _{ c } }{2}  \sigma_z 
	 +i \frac{ \kappa _{ c  }    }{2}}   }
f  _\mathrm{dr}     \left[   \omega  \right] 
\nonumber \\
&  =\frac{  1 }
	   { \omega + \Delta _{a} 
	   - \frac{ \lambda  _{a}  }{2} \sigma_z  
	   + i \frac{ \kappa _{ a } }{2} 
	- \left| J  \right|  ^2  
	\chi ^{(0)} _{c , \sigma } 
\left[   \omega  \right]
   }
f  _\mathrm{dr}     \left[   \omega  \right] 
.  
\end{align}
For simplicity we assume the drive is resonant with cavity mode $ a  $, and the local cavity mode loss is negligible, so that we have 
$ \Delta _{ a }  = \kappa _{ a } = 0 $. In this case, we can further simplify the Fourier space cavity amplitudes as 
\begin{align}
&  \bar{a} _{ \sigma} \left[   \omega  \right]
=\frac{  1 }
    { \omega  
    - \frac{ \lambda  _{a}  }{2}
    \sigma_z  
	- \left| J  \right|  ^2  
	\chi ^{(0)} _{c , \sigma } 
\left[   \omega  \right]   }
f  _\mathrm{dr}     \left[   \omega  \right] 
	.  
\end{align}
Above equation allows one to express the cavity mode linear response susceptibilities 
$ \chi _{a , \sigma } 
\left[   \omega  \right]$ as 
\begin{align}
&  \bar{a} _{ \sigma}  \left[   \omega  \right]
= \chi _{a , \sigma } 
\left[   \omega  \right]
f  _\mathrm{dr}     
\left[   \omega  \right] 
	, \\
&  \chi _{a , \sigma } 
\left[   \omega  \right]
= \left(  \omega  
    - \frac{ \lambda  _{a}  }{2}
    \sigma_z  
	- \left| J  \right|  ^2  
	\chi ^{(0)} _{c , \sigma } 
\left[   \omega  \right]  
\right) ^{-1}
	 . 
	 \label{seq:bath.a.chi}
\end{align}

Making use of the cavity susceptibilities in Eq.~\eqref{seq:bath.a.chi}, we can define the self-energy of cavity mode as 
$ \mathcal{E}  _{a , \sigma _z  } 
\left[   \omega  \right] \equiv 
 \omega   -
\left(  \chi _{a , \sigma } 
\left[   \omega  \right]   
\right) ^{-1} $, which can be written in terms of qubit-independent and -dependent contributions, i.e.~we have 
\begin{align}
\bar{a} _{ \sigma}  \left[   \omega  \right]
=&  \frac{  1 }
    { \omega      
	- \mathcal{E}  _{a , \sigma _z }
	\left[   \omega  \right] }
f  _\mathrm{dr}     \left[   \omega  \right] 
	,   \\
\mathcal{E}  _{a , \sigma _z  } 
\left[   \omega  \right] 
	= &  \frac{   \omega + \Delta _{ c  }
	 +i \frac{ \kappa _{ c  }    }{2} 
	 }
	{ \left(  \omega + \Delta _{ c  } 
	 +i \frac{ \kappa _{ c  }    }{2}  \right)  ^2
	 - \left(   
	 \frac{ \lambda  _{ c } }{2} \right)  ^2
	} \left| J  \right|  ^2 
    \nonumber \\
    &  + \left(  
    \frac{ \lambda  _{a}  }{2}
    + \frac{   
	\left| J  \right|  ^2  }
	{ \left(  \omega + \Delta _{ c  } 
	 +i \frac{ \kappa _{ c   }    }{2}  \right)  ^2
	 - \left(   \frac{ \lambda  _{ c }  }{2} \right)  ^2
	} \frac{ \lambda  _{ c }  }{2}  \right) 
    \sigma_z 
    . 
\end{align}
If we are only interested in dynamics much slower than reservoir correlation timescale $ \kappa _{ c } ^{-1} $, we can choose a specific $ \lambda  _{ a }  $ to minimize the qubit-state dependent terms
\begin{align}
&  \lambda _{ a } 
    = - \text{Re} \left[  
    \frac{   
	\left| J  \right|  ^2  }
	{ \left(  \Delta _{ c  }
	 +i \frac{ \kappa _{ c } }{2}  \right)  ^2
	 - \left(   
	 \frac{ \lambda  _{ c } }{2} \right)  ^2
	}  \right] 
	 \lambda _{ c } 
    . 
\end{align}
Without loss of generality, we also assume the coupling amplitude between qubit and cavity 
$  J $ is real hereafter. If the cavity mode is resonant with the reservoir mode, i.e.~if 
$\Delta _{ c  } =\Delta _{ a } = 0 $, the qubit-dependent term in the self-energy allows perfect cancellation in the stationary limit at $  \omega =0$. However, even in the resonant case, if we care about transient dynamics, the cavity self-energy can still have nontrivial dependence on qubit states at finite frequency $ \omega  $. We can now further recast the self energy in terms of reservoir linewidth 
$  \kappa _{ c  }  $ and effective parameters 
$  \theta  _{\mathrm{eff}} $ and 
$ \Gamma  _{\mathrm{eff}} $ that describe the stationary limit dissipation, i.e.~letting
\begin{align}
&  \lambda _{ c }  
= \kappa _{ c  }  
\tan \frac{ \theta  _{\mathrm{eff}} }{  2 }   
, \quad 
\Delta _{ c } =0
, \\
&  \Gamma  _{\mathrm{eff}}  
	=  J  ^2 
	\frac{ 4 \kappa _{ c  } }
	{  \kappa _{ c  }   ^2 
	+ \lambda _{ c  } ^2  }  
	\Rightarrow 
J  ^2 =  
\frac{ 1 }{  4 }   
\Gamma  _{\mathrm{eff}}  
\kappa _{ c  }  
\sec ^2 \frac{ \theta  _{\mathrm{eff}} }{  2 }   
    ,  
\\
&  
\lambda  _{a}  
=   J  ^2 
\frac{ 4 \lambda _{ c }  }
{  \kappa _{ c  }   ^2 
    + \lambda_{ c }  ^2 } 
	= \Gamma  _{\mathrm{eff}}  
\tan \frac{ \theta  _{\mathrm{eff}} }{  2  } 
, 
\end{align}
so that the self-energy function can be rewritten as 
\begin{align} 
&  \mathcal{E}  _{a , \sigma _z  } 
\left[   \omega  \right] 
\nonumber \\
	= & \frac{ \Gamma  _{\mathrm{eff}}  }{2} 
	\frac{  \left(  \omega 
	 +i \frac{ \kappa _{ c } }{2}  
	 \right) \frac{ \kappa _{ c } }{2}
	\sec ^2 
    \frac{ \theta  _{\mathrm{eff}} }{  2 } 
    + \omega \left(  \omega 
    +i \kappa _{ c }  \right)  
    \tan \frac{ \theta  _{\mathrm{eff}} }{  2 }
    \sigma_z 
	 }
	{ \left(  \omega  
	 +i \frac{ \kappa _{ c  }    }{2}  \right)  ^2
	 - \left( \frac{ \kappa _{ c  } }{2} 
	 \tan \frac{ \theta  _{\mathrm{eff}} }{  2 }    
	 \right)  ^2
	}  
    . 
\end{align}
For illustrative purposes, we can use a representative value for the effective phase shift 
$  \theta  _{\mathrm{eff}} = \pi /2 $, so that above expression further simplifies as 
\begin{align} 
&  \mathcal{E}  _{a , \sigma _z  } 
\left[   \omega  \right] 
	=  \frac{ \Gamma  _{\mathrm{eff}}  }{2} 
	\frac{  \left(  \omega 
	 +i \frac{ \kappa _{ c } }{2}  \right)
	 \kappa _{ c }  
	+ \omega \left(  \omega 
    +i \kappa _{ c }  \right) 
     \sigma_z  
	 }
	{ \omega  \left(  \omega  
	 +i \kappa _{ c  }  \right) 
    - \frac{ \kappa _{c} ^{2} }{ 2 }  
	}    
    . 
\end{align}
The Markovian limit amounts to requiring that the self-energy scale near resonance, i.e.~$ \Gamma  _{\mathrm{eff}} $, is much smaller than the frequency range over which 
$ \mathcal{E}  _{a , \sigma _z  } 
\left[   \omega  \right]  $ significantly changes, which in turn is set by reservoir mode linewidth $  \kappa _{ c  } $. In this limit where 
$ \Gamma  _{\mathrm{eff}} 
\ll \kappa _{ c  }$, we can expand the self energy function in the vicinity of zero frequency to obtain \begin{align} 
&  \mathcal{E}  _{a , \sigma _z  } 
\left[   \omega  \right] 
	\simeq  -i \frac{ \Gamma  _{\mathrm{eff}}  }{2} 
	\left( 1 +2 
	\frac{ \omega  }
	{ \kappa _{ c }  }  
	 \sigma_z   
    \right)  
    . 
\end{align}
The second term in the parenthesis in above equation represents leading-order non-Markovian corrections, which describes qubit backaction to the cavity and causes deviation from the full-nonreciprocity limit.

Because of the dispersive coupling, the qubit dynamics can be more complicated in the frequency space: taking Fourier transform of Eq.~\eqref{seq:2m.eom.qbcoh}
\begin{align}
&  \omega  \ln \langle \sigma_-  
\left[   \omega  \right]  \rangle
\nonumber \\
= & 
\int ^{ + \infty }_{ - \infty } d {\omega _1 }  
 \left(  \lambda _{a} 
	 \bar{a} _{  \uparrow} 
	  \left[   \omega _1 \right]    
	\bar{a} ^*_{  \downarrow} 
	 \left[ \omega -  \omega _1 \right]    
	 + \lambda _{c}  
	 \bar{c} _{  \uparrow} 
	  \left[   \omega _{1}  \right]    
	\bar{c} ^*_{   \downarrow} 
	 \left[ \omega -  \omega _{1}  \right]  \right) 
	.   
\end{align} 
We can again substitute Eq.~\eqref{seq:2m.FT.a2} into this equation to obtain 
\begin{align}
&  \omega  \ln \langle \sigma_-  
\left[   \omega  \right]  \rangle
= \int ^{ + \infty }_{ - \infty } 
d {\omega _1 }  
\Lambda _{a} \left[ \omega _1 
; \omega \right]
	 \bar{a} _{ \uparrow} 
	  \left[   \omega _1 \right]    
	\bar{a} ^*_{ \downarrow } 
	 \left[ \omega -  \omega _1 \right]  
\\
& \Lambda _{a} \left[ \omega _1 
; \omega \right]
=  \lambda _{a}   
	 + \lambda _{c}    
	 \left| J  \right|  ^2  
	\chi ^{(0)} _{c , \uparrow } 
\left[   \omega  \right] 
\chi ^{(0)} _{c , \downarrow } 
\left[  \omega -  \omega _1  \right] 
	.   
\end{align} 
In the Markovian limit, the cavity amplitude is approximately independent of the qubit state, so that we have 
$ \bar{a} _{  \uparrow}
\left(  t  \right)  
\simeq 	\bar{a}  _{  \downarrow} 
\left(  t  \right) $. In this limit, we remark that the first term in the square bracket above, which is proportional to 
$  \lambda _{a} $, leads to a pure phase shift in qubit dynamics, whereas the second term can induce both phase shift and dephasing to the qubit.


\section{Properties of multiple-dissipator generalization of gauge-invariance-induced nonreciprocal Lindbladians in Eq.~\eqref{eq:qnr.multi.gen} }

In the main text, we have introduced multi-dissipator generalizations of our nonreciprocal Lindbladians, given by (c.f.~Eq.~\eqref{eq:qnr.multi.gen} in the main text) 
\begin{align}
\mathcal{L} 
_{\mathrm{multi}  } \hat \rho  
	&   =  \Gamma 
	\sum _{\ell=1}^{N}   
	\mathcal{D}  
\left[ \sum _{m=1}^{N}  
    \hat u _{\ell m} 
    \hat{A} _{ m }  \right]
    \hat{\rho} 
	.  
	\label{seq:qnr.multi.gen}
\end{align} 
We also state that the $B$ operators $\hat u _{\ell m} $ can be thought of as (operator-valued) matrix elements of a generalized unitary, which acts on the composite linear space 
$\mathbb{C} ^{ N } \otimes 
\mathcal{H} _{B} $ between the \textit{dissipator} space 
$\mathbb{C} ^{ N }$ of the jump operators, and the $B$ system Hilbert space. In this Appendix, we provide more details about both the general structure, as well as some typical examples illustrating the connections and differences between the multi- and single-dissipator Lindbladians.

\subsection{Connection to a generalized unitary operator}
\label{appsec:qnr.multi}

To motivate the generalized unitary, we again start with a multi-dissipator acting only on system $A$, as given by 
$\mathcal{L} 
_{A  } \hat \rho _{A  }  
=  \Gamma \sum _{\ell=1}^{N} 
\mathcal{D}  
[ \hat{A} _{ \ell } ] 
\hat{\rho} _{A  } 
$. As noted in the main text, this Lindbladian is invariant under a generic unitary transformation on the dissipators $\hat{A} _{ \ell }$ 
\begin{align}
\mathcal{L} 
_{A  } \hat \rho _{A  }  
	&   =  \Gamma 
\sum _{\ell=1}^{N} 
	\mathcal{D}  
\left[ \sum  _{m=1}^{N} 
    u _{\ell m}  
    \hat{A} _{ m }  \right]
    \hat{\rho} _{A  } 
	,  
\end{align} 
where $ u _{\ell m}   $ are matrix elements of a $N$-dimensional complex unitary matrix $u _{\ell m} = \left( \cmatU   
\right )_{\ell m}  $. We can also write the unitary matrix explicitly in terms of its Hermitian generator $ \cmatH $, as 
$\cmatU  = 
\exp \left( -i  \cmatH  \right )  $. Note that a generic Hermitian matrix 
$\cmatH $ can be decomposed in terms of $N ^{2} $ basis matrices 
$\cmat{E} _{\ell m} $ as 
$\cmatH =  \sum  _{\ell,m=1}^{N}  
h _{\ell m} 
\cmat{E} _{\ell m} $, where $\cmat{E} _{\ell m} $ satisfy the orthogonality condition $\cmat{E} _{\ell' \ell} ^{\dag}
\cmat{E} _{m' m}
= \delta _{\ell' m'}
\cmat{E} _{\ell m} $.

We now turn to the nonreciprocal Lindbladian given by Eq.~\eqref{seq:qnr.multi.gen}. In analogy to the above case, we can generate $B$ operators 
$ \hat u _{\ell m} $ via a generalized unitary operator acting on the composite linear space $\mathbb{C} ^{ N_{A}} \otimes 
\mathcal{H} _{B} $, as  
\begin{align}
\hat u _{\ell m} 
	&   
	=  
	\left( e 
	^{ -i 
	\sum  _{j,j'=1}^{N }  
	\cmat{E}_{jj' } 
    \hat h _{jj' }   
	 } \right )
	_{\ell m}  
	,  
	\label{seq:qnr.multi.ulm}
\end{align} 
where $\hat h _{jj'  } $ are operating acting on $B$ satisfying $\hat h _{jj'  } 
= \hat h _{j' j } ^{\dag} $. It is straightforward to show that the orthogonality conditions $ \sum_{ \ell }
\hat u _{\ell m} ^\dag
\hat u _{\ell m'}
= \delta _{m m' }
\hat {\mathbb{I}} _{B} $ hold if and only if the
$ \hat u _{\ell m} $ operators can be rewritten as Eq.~\eqref{seq:qnr.multi.ulm}. Making use of the orthogonality condition, we can exactly trace out system $B$ to obtain a closed master equation acting on system $A$ as 
$  (d \hat \rho _{A} / dt) 
= \sum_{\ell } 
 \mathcal{D}  
[ \hat{A} _{ \ell  } 
] \hat{\rho}  _{A} $, so that the multimode master equation in Eq.~\eqref{seq:qnr.multi.gen} is again unidirectional.

\subsection{Cases (in)equivalent to incoherent sum of single fully nonreciprocal dissipators}

\label{appsec:qnr.multi.tosing}

As discussed in the main text, Eq.~\eqref{seq:qnr.multi.gen} describes a much more general class of fully nonreciprocal dynamics compared to the single-dissipator case. Here we discuss in detail conditions for when the former multi-dissipator dynamics can(not) be rewritten as an incoherent sum of unidirectional dissipators, i.e.~if we have 
\begin{align} 
\mathcal{L} 
_{\mathrm{multi}  } \hat \rho  
	&   =  \Gamma \sum_{\ell } 
\mathcal{D}  
\left [ \hat {U} _{B, \ell } 
\hat{A}' _{ \ell } \right ] 
\hat{\rho} 
	.  
	\label{seq:qnr.multi.decomp}
\end{align}  
While it is difficult to comprehensively characterize all the scenarios where it is possible to write this equivalence relation, we discuss two sufficient conditions here. The first case can be most easily understood in terms of the generalized Hermitian generator in Eq.~\eqref{seq:qnr.multi.ulm}. More specifically, if we can diagonalize it using a new local $A$ basis 
$ \cmat{F}_{\ell \ell' } 
= \sum _{ j,j' =1 }^{N }  
v _{\ell j'} 
\cmat{E}_{j' j} 
v _{\ell' j} ^{*}  $ as  
$\sum  _{j,j'=1}^{N }  
	\cmat{E}_{jj' }
    \hat h _{jj' }   
= \sum  _{\ell=1}^{N }  
\cmat{F}_{\ell \ell } 
\hat {\Phi} _{B,\ell}  
$, where $v _{\ell m} 
= \left( \cmat{V}   
\right )_{\ell m}  $ are elements of a unitary matrix, then one can use a few lines of algebra to show that the multi-dissipator Lindbladian is equivalent to a sum of single dissipators as 
\begin{align} 
\mathcal{L} 
_{\mathrm{multi}  } \hat \rho  
	&   =  \Gamma 
\sum _{\ell=1}^{N }  
	\mathcal{D}  
\left[ e ^{-i \hat {\Phi} _{B, \ell } } 
\sum _{m =1}^{N }  
v ^{*}_{\ell m }   
\hat{A} _{ m }  
    \right]
    \hat{\rho} 
	.  
	\label{seq:qnr.multi.diag}
\end{align} 
Note that this is fully equivalent to the starting Lindbladian, but now every jump operator implements a unidirectional interaction from $A$ to $B$.

The second case is if we cannot diagonalize the generator (and hence the generalized unitary) via a local basis transformation on $A$, but the $B$ system only has a single qubit, and all $ \hat u _{\ell m} $ commute with each other. For such Lindbladians, we can again reformulate it as a sum of single nonreciprocal dissipators via a local basis change on $A$ (i.e.~Eq.~\eqref{seq:qnr.multi.diag}). To show this, we first note that the total system dynamics is easily solvable by diagonalizing $ \hat u _{\ell m} $ in a joint $B$ basis (we assume to be $\hat \sigma _{z}$ basis for simplicity), so that time evolution under $\mathcal{L} 
_{\mathrm{multi}  }$ will conserve excitations in that basis. More concretely, the generalized unitary can now be decomposed using projectors $\hat P _{\uparrow (\downarrow)}$ onto $B$ eigenstates as 
\begin{align}
e  	^{ -i 
	\sum  _{j,j'=1}^{N }  
	\cmat{E}_{jj' } 
    \hat h _{jj' }   
	 } 
	&   
	=  \cmat{U} _{ \uparrow } 
	\hat P _{ \uparrow }
	+ \cmat{U} _{ \uparrow }
	\hat P _{ \downarrow }  
	.  
	\label{seq:qnr.22.proj}
\end{align} 
Again, we can transform the local $A$ basis to 
$\cmat{U}'_{\uparrow (\downarrow)} 
= \cmat{W} 
\cmat{U} _{\uparrow (\downarrow)}
\cmat{W} ^{\dag}
$ via a unitary matrix $\cmat{W} $; for the jump operators in this new dissipator frame to be unidirectional, we require that 
$\cmat{U}'_{\uparrow } 
= \exp (- i \cmat{D} )
\cmat{U}'_{\downarrow } 
$, where 
$ \left( \cmat{D}  
\right )_{\ell m}
= \delta _{\ell m} d _{\ell }  $ denotes a real diagonal matrix. This relation can be realized by transforming to the eigenbasis of $\cmat{U} _{\uparrow  }
\cmat{U} _{\downarrow } ^{\dag}
$, so that we have 
\begin{align} 
\mathcal{L} 
_{\mathrm{multi}  } \hat \rho  
	&   =  \Gamma 
	\sum _{\ell=1}^{N}   
\mathcal{D}  
\left [ 
e ^{-i d _{\ell } 
\frac{\hat {\sigma} _{z }}{2} 
} 
\hat{A}' _{ \ell } \right ] 
\hat{\rho} 
	.  
	\label{seq:qnr.22.decomp}
\end{align}  
As a concrete example, we consider $A$ consisting of two bosonic modes. Using $\cmat{Z} $ and $\cmat{X}  $ to denote Pauli matrices acting on $\mathbb{C} ^{ 2 }$, we focus on the dissipator generated by the following generalized unitary operator 
\begin{align}
\hat u _{\ell m} 
	& =  
	\left[ e ^{ - i 
\theta  \left(
\cmat{Z}  
\hat \sigma _{z } 
\cos \varphi  
+  \cmat{X} 
\hat {\mathbb{I}} _{B}
\sin \varphi \right)
} \right ]
	_{\ell m}  
	.  
	\label{seq:qnr.2comm.ulm}
\end{align} 
The corresponding system Lindbladian is given by  
\begin{align}
\mathcal{L}  \hat \rho  
	=  &  \Gamma   
	\mathcal{D}  
[   \hat{a} _{ 1 } 
    \left(  
    \cos \theta
    - i \hat \sigma _{z } 
    \cos \varphi  
    \sin \theta 
    \right) 
    - i 
    \hat{a} _{ 2 }  
    \sin \theta 
    \sin \varphi   ]
    \hat{\rho} 
    \nonumber \\
    &   + \Gamma \mathcal{D}  
[   \hat{a} _{ 2 } 
    \left(  
    \cos \theta
    + i \hat \sigma _{z } 
    \cos \varphi  
    \sin \theta 
    \right) 
    - i 
    \hat{a} _{ 1 } 
    \sin \theta 
    \sin \varphi  ]
    \hat{\rho} 
    . 
\end{align} 
In this case 
$\cmat{U} _{\uparrow (\downarrow)} 
= e ^{ - i 
\theta  \left(
\pm \cmat{Z}   
\cos \varphi  
+  \cmat{X}  
\sin \varphi \right)
} $, and we can generally rewrite the Lindbladian into the form in Eq.~\eqref{seq:qnr.22.decomp}. For the specific case with $\theta = \pi /2$, the relevant nonlocal basis simplifies into 
$\hat{a} _{ y,\pm } = (\hat{a} _{ 1 } 
\pm i \hat{a} _{ 2 } ) 
/ \sqrt{2} $, so that we have    
\begin{align}
\mathcal{L}  \hat \rho  
	= & \frac{\Gamma }{2} 
	\mathcal{D}  
    \left [ e ^{ i 
    (\pi /2 - \varphi )
	\hat \sigma _{z }  }  
    \hat{a} _{ y, + } 
    \right]
    \hat{\rho} 
    + \frac{\Gamma }{2} 
	\mathcal{D}  
    \left [ e ^{ i 
    (-\pi /2+ \varphi )
	\hat \sigma _{z }  }  
    \hat{a} _{ y, - } 
    \right]
    \hat{\rho}   
    . 
\end{align}

More generally, if we cannot diagonalize the generator (and hence the generalized unitary) via a local basis transformation on $A$, and if $ \hat u _{\ell m} $ operators do not commute or $B$ has more complicated level structure than a qubit, then there are cases where it is impossible to write the Lindbladian $\mathcal{L} 
_{\mathrm{multi}  }$ as Eq.~\eqref{seq:qnr.multi.decomp}, i.e.~incoherent sum of $N$ nonreciprocal dissipators. For example, We can again consider $2$-mode system $A$ with the generalized unitary 
\begin{align}
\hat u _{\ell m} 
	& =  
\left[ e ^{ - i  
\left(
\varphi  \cmat{Z}  
\hat \sigma _{z }  
+  \theta  \cmat{X} 
\hat \sigma _{x } 
\right)
} \right ]
	_{\ell m}  
	.  
\label{seq:qnr.2nonc.ulm}
\end{align} 
The corresponding system Lindbladian is given by  
\begin{align}
\mathcal{L}  \hat \rho  
	=  &  \Gamma   
	\mathcal{D}  
[   e ^{ -i \varphi  
\hat \sigma _{z }  
} \hat{a} _{ 1 } 
    \cos \theta
    - i  \hat \sigma _{x }  
e ^{ i \varphi 
\hat \sigma _{z }  
} \hat{a} _{ 2 } 
    \sin \theta
    ]
    \hat{\rho} 
    \nonumber \\
    &   + \Gamma \mathcal{D}  
[   e ^{ i \varphi 
\hat \sigma _{z }  
} \hat{a} _{ 2 } 
    \cos \theta
    - i  \hat \sigma _{x }  
e ^{ -i \varphi 
\hat \sigma _{z }  
} \hat{a} _{ 1 } 
    \sin \theta  ]
    \hat{\rho} 
    , 
    \label{seq:qnr.multi.2diss}
\end{align} 
which reproduces Eqs.~\eqref{eq:qnr.multi.2} and \eqref{eq:qnr.multi.z} in the main text. 
It is interesting to note that if we ignore all 
$\hat \sigma _{x } $ operators in  Eq.~\eqref{seq:qnr.multi.2diss}, the resulting Lindbladian is still fully nonreciprocal, and can be rewritten as $ \Gamma  
\left( \mathcal{D}  
[   e ^{ -i \varphi  
\hat \sigma _{z }  
} \hat{a} _{ 1 } ]
+ \mathcal{D}  
[  
e ^{ i \varphi 
\hat \sigma _{z }  
} \hat{a} _{ 2 }  ] 
\right ) 
\hat{\rho}  $, i.e.~equivalent to sum of single nonreciprocal dissipators.  The inclusion of $\hat \sigma _{x }  $ in the jump operators would not affect local dynamics of system $A$; however, those $\hat \sigma _{x }  $ operators matter for evolution of correlations within the bipartite system. As discussed in Sec.~\ref{sec:qnr.multi}, those correlations can be highly nonclassical and even generating entanglement between $A$ and $B$.

\subsection{Steady state of multi-dissipator unidirectional dynamics with bosonic lowering $A$ operators $\hat{A} _{ m } $  }

It is interesting to note that if system $A$ operators are bosonic lowering operators of cavity modes 
$a_{ m } $, i.e.~$\hat{A} _{ m } = \hat{a} _{ m } $, and if $A$ is initialized in a single-mode Fock state, we can further explicitly derive the long-time limit of the quantum map generated by 
$\mathcal{L} _{\mathrm{multi}  }$. To see this, we first expand the master equation in Eq.~\eqref{eq:qnr.multi.gen} and rewrite it as sum of quantum-jump, and no-jump contributions. Making use of the unitarity conditions in Eq.~\eqref{eq:qnr.multi.ortho}, the equation can be simplified as 
\begin{align}
\mathcal{L} 
_{\mathrm{multi}  } \hat \rho  
	& = \Gamma 
	\sum _{ m,m'=1}^{N}   
    \hat{A} _{ m } 
    \mathcal{E} _{ m m'} 
    ( \hat{\rho} )
    \hat{A} _{ m' } ^{\dag}
    - \frac{\Gamma }{2} 
    \left\{\sum _{n=1}^{N}  
    \hat{A} _{ n } ^{\dag}
    \hat{A} _{ n }  
    , \hat{\rho}  
    \right \}
    , \\
\mathcal{E} _{ m m'} 
    ( \hat{\rho} )
    & = \sum _{j =1}^{N} 
	\hat u _{j m} 
    \hat{\rho} 
    \hat u _{j m' } ^{\dag}
    .
\end{align} 
If $\hat{A} _{ m } = \hat{a} _{ m } $, it is straightforward to show that 
\begin{align} 
& \lim _{ t \to \infty} 
	e ^{  \mathcal{L}  _{\mathrm{multi}} t  }
	\left[  
	(\hat{a} _{ m  } ^{\dag}) ^{\ell}
	| 0 \rangle \langle 0 | 
	\hat{a} _{ m  } ^{\ell} 
	\otimes  \hat \rho _{B,\mathrm{i}} 
	\right] 
	=  \mathcal{E} _{ m m}  ^{\ell} 
    (\hat  \rho _{B,\mathrm{i}} 
    )
	.
\end{align} 
Thus, for general $\mathcal{L} 
_{\mathrm{multi}  }$ that cannot be rewritten as sum of single nonreciprocal dissipators, the long-time limit of dynamics generated by nonreciprocal interaction will be dissipative, in contrast to the single-dissipator case in Eq.~\eqref{eq:nl.nr.Lindbladian}.


\begin{thebibliography}{61}%
\makeatletter
\providecommand \@ifxundefined [1]{%
 \@ifx{#1\undefined}
}%
\providecommand \@ifnum [1]{%
 \ifnum #1\expandafter \@firstoftwo
 \else \expandafter \@secondoftwo
 \fi
}%
\providecommand \@ifx [1]{%
 \ifx #1\expandafter \@firstoftwo
 \else \expandafter \@secondoftwo
 \fi
}%
\providecommand \natexlab [1]{#1}%
\providecommand \enquote  [1]{``#1''}%
\providecommand \bibnamefont  [1]{#1}%
\providecommand \bibfnamefont [1]{#1}%
\providecommand \citenamefont [1]{#1}%
\providecommand \href@noop [0]{\@secondoftwo}%
\providecommand \href [0]{\begingroup \@sanitize@url \@href}%
\providecommand \@href[1]{\@@startlink{#1}\@@href}%
\providecommand \@@href[1]{\endgroup#1\@@endlink}%
\providecommand \@sanitize@url [0]{\catcode `\\12\catcode `\$12\catcode
  `\&12\catcode `\#12\catcode `\^12\catcode `\_12\catcode `\%12\relax}%
\providecommand \@@startlink[1]{}%
\providecommand \@@endlink[0]{}%
\providecommand \url  [0]{\begingroup\@sanitize@url \@url }%
\providecommand \@url [1]{\endgroup\@href {#1}{\urlprefix }}%
\providecommand \urlprefix  [0]{URL }%
\providecommand \Eprint [0]{\href }%
\providecommand \doibase [0]{https://doi.org/}%
\providecommand \selectlanguage [0]{\@gobble}%
\providecommand \bibinfo  [0]{\@secondoftwo}%
\providecommand \bibfield  [0]{\@secondoftwo}%
\providecommand \translation [1]{[#1]}%
\providecommand \BibitemOpen [0]{}%
\providecommand \bibitemStop [0]{}%
\providecommand \bibitemNoStop [0]{.\EOS\space}%
\providecommand \EOS [0]{\spacefactor3000\relax}%
\providecommand \BibitemShut  [1]{\csname bibitem#1\endcsname}%
\let\auto@bib@innerbib\@empty
\bibitem [{\citenamefont {You}\ \emph {et~al.}(2020)\citenamefont {You},
  \citenamefont {Baskaran},\ and\ \citenamefont {Marchetti}}]{Marchetti2020}%
  \BibitemOpen
  \bibfield  {author} {\bibinfo {author} {\bibfnamefont {Z.}~\bibnamefont
  {You}}, \bibinfo {author} {\bibfnamefont {A.}~\bibnamefont {Baskaran}},\ and\
  \bibinfo {author} {\bibfnamefont {M.~C.}\ \bibnamefont {Marchetti}},\
  }\bibfield  {title} {\bibinfo {title} {nonreciprocity as a generic route to
  traveling states},\ }\href {https://doi.org/10.1073/pnas.2010318117}
  {\bibfield  {journal} {\bibinfo  {journal} {Proc. Natl. Acad. Sci. U.S.A.}\
  }\textbf {\bibinfo {volume} {117}},\ \bibinfo {pages} {19767} (\bibinfo
  {year} {2020})}\BibitemShut {NoStop}%
\bibitem [{\citenamefont {Fruchart}\ \emph {et~al.}(2021)\citenamefont
  {Fruchart}, \citenamefont {Hanai}, \citenamefont {Littlewood},\ and\
  \citenamefont {Vitelli}}]{Vitelli2021}%
  \BibitemOpen
  \bibfield  {author} {\bibinfo {author} {\bibfnamefont {M.}~\bibnamefont
  {Fruchart}}, \bibinfo {author} {\bibfnamefont {R.}~\bibnamefont {Hanai}},
  \bibinfo {author} {\bibfnamefont {P.~B.}\ \bibnamefont {Littlewood}},\ and\
  \bibinfo {author} {\bibfnamefont {V.}~\bibnamefont {Vitelli}},\ }\bibfield
  {title} {\bibinfo {title} {nonreciprocal phase transitions},\ }\href
  {https://doi.org/10.1038/s41586-021-03375-9} {\bibfield  {journal} {\bibinfo
  {journal} {Nature}\ }\textbf {\bibinfo {volume} {592}},\ \bibinfo {pages}
  {363} (\bibinfo {year} {2021})}\BibitemShut {NoStop}%
\bibitem [{\citenamefont {Shankar}\ \emph {et~al.}(2021)\citenamefont
  {Shankar}, \citenamefont {Souslov}, \citenamefont {Bowick}, \citenamefont
  {Marchetti},\ and\ \citenamefont {Vitelli}}]{Vitelli2021review}%
  \BibitemOpen
  \bibfield  {author} {\bibinfo {author} {\bibfnamefont {S.}~\bibnamefont
  {Shankar}}, \bibinfo {author} {\bibfnamefont {A.}~\bibnamefont {Souslov}},
  \bibinfo {author} {\bibfnamefont {M.~J.}\ \bibnamefont {Bowick}}, \bibinfo
  {author} {\bibfnamefont {M.~C.}\ \bibnamefont {Marchetti}},\ and\ \bibinfo
  {author} {\bibfnamefont {V.}~\bibnamefont {Vitelli}},\ }\bibfield  {title}
  {\bibinfo {title} {Topological active matter},\ }\href
  {https://doi.org/10.48550/arXiv.2010.00364} {\bibfield  {journal} {\bibinfo
  {journal} {arXiv preprint arXiv:2010.00364}\ } (\bibinfo {year}
  {2021})}\BibitemShut {NoStop}%
\bibitem [{\citenamefont {Stannigel}\ \emph {et~al.}(2012)\citenamefont
  {Stannigel}, \citenamefont {Rabl},\ and\ \citenamefont
  {Zoller}}]{Zoller2012}%
  \BibitemOpen
  \bibfield  {author} {\bibinfo {author} {\bibfnamefont {K.}~\bibnamefont
  {Stannigel}}, \bibinfo {author} {\bibfnamefont {P.}~\bibnamefont {Rabl}},\
  and\ \bibinfo {author} {\bibfnamefont {P.}~\bibnamefont {Zoller}},\
  }\bibfield  {title} {\bibinfo {title} {Driven-dissipative preparation of
  entangled states in cascaded quantum-optical networks},\ }\href
  {https://doi.org/10.1088/1367-2630/14/6/063014} {\bibfield  {journal}
  {\bibinfo  {journal} {New J. Phys.}\ }\textbf {\bibinfo {volume} {14}},\
  \bibinfo {pages} {063014} (\bibinfo {year} {2012})}\BibitemShut {NoStop}%
\bibitem [{\citenamefont {Pichler}\ \emph {et~al.}(2015)\citenamefont
  {Pichler}, \citenamefont {Ramos}, \citenamefont {Daley},\ and\ \citenamefont
  {Zoller}}]{Zoller2015}%
  \BibitemOpen
  \bibfield  {author} {\bibinfo {author} {\bibfnamefont {H.}~\bibnamefont
  {Pichler}}, \bibinfo {author} {\bibfnamefont {T.}~\bibnamefont {Ramos}},
  \bibinfo {author} {\bibfnamefont {A.~J.}\ \bibnamefont {Daley}},\ and\
  \bibinfo {author} {\bibfnamefont {P.}~\bibnamefont {Zoller}},\ }\bibfield
  {title} {\bibinfo {title} {Quantum optics of chiral spin networks},\ }\href
  {https://doi.org/10.1103/PhysRevA.91.042116} {\bibfield  {journal} {\bibinfo
  {journal} {Phys. Rev. A}\ }\textbf {\bibinfo {volume} {91}},\ \bibinfo
  {pages} {042116} (\bibinfo {year} {2015})}\BibitemShut {NoStop}%
\bibitem [{\citenamefont {Lodahl}\ \emph {et~al.}(2017)\citenamefont {Lodahl},
  \citenamefont {Mahmoodian}, \citenamefont {Stobbe}, \citenamefont
  {Rauschenbeutel}, \citenamefont {Schneeweiss}, \citenamefont {Volz},
  \citenamefont {Pichler},\ and\ \citenamefont {Zoller}}]{Zoller2017}%
  \BibitemOpen
  \bibfield  {author} {\bibinfo {author} {\bibfnamefont {P.}~\bibnamefont
  {Lodahl}}, \bibinfo {author} {\bibfnamefont {S.}~\bibnamefont {Mahmoodian}},
  \bibinfo {author} {\bibfnamefont {S.}~\bibnamefont {Stobbe}}, \bibinfo
  {author} {\bibfnamefont {A.}~\bibnamefont {Rauschenbeutel}}, \bibinfo
  {author} {\bibfnamefont {P.}~\bibnamefont {Schneeweiss}}, \bibinfo {author}
  {\bibfnamefont {J.}~\bibnamefont {Volz}}, \bibinfo {author} {\bibfnamefont
  {H.}~\bibnamefont {Pichler}},\ and\ \bibinfo {author} {\bibfnamefont
  {P.}~\bibnamefont {Zoller}},\ }\bibfield  {title} {\bibinfo {title} {Chiral
  quantum optics},\ }\href {https://doi.org/10.1038/nature21037} {\bibfield
  {journal} {\bibinfo  {journal} {Nature}\ }\textbf {\bibinfo {volume} {541}},\
  \bibinfo {pages} {473} (\bibinfo {year} {2017})}\BibitemShut {NoStop}%
\bibitem [{\citenamefont {Fang}\ \emph {et~al.}(2012)\citenamefont {Fang},
  \citenamefont {Yu},\ and\ \citenamefont {Fan}}]{Fan2012}%
  \BibitemOpen
  \bibfield  {author} {\bibinfo {author} {\bibfnamefont {K.}~\bibnamefont
  {Fang}}, \bibinfo {author} {\bibfnamefont {Z.}~\bibnamefont {Yu}},\ and\
  \bibinfo {author} {\bibfnamefont {S.}~\bibnamefont {Fan}},\ }\bibfield
  {title} {\bibinfo {title} {Realizing effective magnetic field for photons by
  controlling the phase of dynamic modulation},\ }\href
  {https://doi.org/10.1038/nphoton.2012.236} {\bibfield  {journal} {\bibinfo
  {journal} {Nat. Photonics}\ }\textbf {\bibinfo {volume} {6}},\ \bibinfo
  {pages} {782} (\bibinfo {year} {2012})}\BibitemShut {NoStop}%
\bibitem [{\citenamefont {Estep}\ \emph {et~al.}(2014)\citenamefont {Estep},
  \citenamefont {Sounas}, \citenamefont {Soric},\ and\ \citenamefont
  {Al{\`{u}}}}]{Alu2014}%
  \BibitemOpen
  \bibfield  {author} {\bibinfo {author} {\bibfnamefont {N.~A.}\ \bibnamefont
  {Estep}}, \bibinfo {author} {\bibfnamefont {D.~L.}\ \bibnamefont {Sounas}},
  \bibinfo {author} {\bibfnamefont {J.}~\bibnamefont {Soric}},\ and\ \bibinfo
  {author} {\bibfnamefont {A.}~\bibnamefont {Al{\`{u}}}},\ }\bibfield  {title}
  {\bibinfo {title} {Magnetic-free nonreciprocity and isolation based on
  parametrically modulated coupled-resonator loops},\ }\href
  {https://doi.org/10.1038/nphys3134} {\bibfield  {journal} {\bibinfo
  {journal} {Nat. Phys.}\ }\textbf {\bibinfo {volume} {10}},\ \bibinfo {pages}
  {923} (\bibinfo {year} {2014})}\BibitemShut {NoStop}%
\bibitem [{\citenamefont {Nassar}\ \emph {et~al.}(2020)\citenamefont {Nassar},
  \citenamefont {Yousefzadeh}, \citenamefont {Fleury}, \citenamefont {Ruzzene},
  \citenamefont {Al{\`{u}}}, \citenamefont {Daraio}, \citenamefont {Norris},
  \citenamefont {Huang},\ and\ \citenamefont {Haberman}}]{Haberman2020}%
  \BibitemOpen
  \bibfield  {author} {\bibinfo {author} {\bibfnamefont {H.}~\bibnamefont
  {Nassar}}, \bibinfo {author} {\bibfnamefont {B.}~\bibnamefont {Yousefzadeh}},
  \bibinfo {author} {\bibfnamefont {R.}~\bibnamefont {Fleury}}, \bibinfo
  {author} {\bibfnamefont {M.}~\bibnamefont {Ruzzene}}, \bibinfo {author}
  {\bibfnamefont {A.}~\bibnamefont {Al{\`{u}}}}, \bibinfo {author}
  {\bibfnamefont {C.}~\bibnamefont {Daraio}}, \bibinfo {author} {\bibfnamefont
  {A.~N.}\ \bibnamefont {Norris}}, \bibinfo {author} {\bibfnamefont
  {G.}~\bibnamefont {Huang}},\ and\ \bibinfo {author} {\bibfnamefont {M.~R.}\
  \bibnamefont {Haberman}},\ }\bibfield  {title} {\bibinfo {title}
  {nonreciprocity in acoustic and elastic materials},\ }\href
  {https://doi.org/10.1038/s41578-020-0206-0} {\bibfield  {journal} {\bibinfo
  {journal} {Nat. Rev. Mater.}\ }\textbf {\bibinfo {volume} {5}},\ \bibinfo
  {pages} {667} (\bibinfo {year} {2020})}\BibitemShut {NoStop}%
\bibitem [{\citenamefont {Kamal}\ \emph {et~al.}(2011)\citenamefont {Kamal},
  \citenamefont {Clarke},\ and\ \citenamefont {Devoret}}]{Devoret2011}%
  \BibitemOpen
  \bibfield  {author} {\bibinfo {author} {\bibfnamefont {A.}~\bibnamefont
  {Kamal}}, \bibinfo {author} {\bibfnamefont {J.}~\bibnamefont {Clarke}},\ and\
  \bibinfo {author} {\bibfnamefont {M.~H.}\ \bibnamefont {Devoret}},\
  }\bibfield  {title} {\bibinfo {title} {Noiseless nonreciprocity in a
  parametric active~device},\ }\href {https://doi.org/10.1038/nphys1893}
  {\bibfield  {journal} {\bibinfo  {journal} {Nat. Phys.}\ }\textbf {\bibinfo
  {volume} {7}},\ \bibinfo {pages} {311} (\bibinfo {year} {2011})}\BibitemShut
  {NoStop}%
\bibitem [{\citenamefont {Abdo}\ \emph {et~al.}(2014)\citenamefont {Abdo},
  \citenamefont {Sliwa}, \citenamefont {Shankar}, \citenamefont {Hatridge},
  \citenamefont {Frunzio}, \citenamefont {Schoelkopf},\ and\ \citenamefont
  {Devoret}}]{Devoret2014diramp}%
  \BibitemOpen
  \bibfield  {author} {\bibinfo {author} {\bibfnamefont {B.}~\bibnamefont
  {Abdo}}, \bibinfo {author} {\bibfnamefont {K.}~\bibnamefont {Sliwa}},
  \bibinfo {author} {\bibfnamefont {S.}~\bibnamefont {Shankar}}, \bibinfo
  {author} {\bibfnamefont {M.}~\bibnamefont {Hatridge}}, \bibinfo {author}
  {\bibfnamefont {L.}~\bibnamefont {Frunzio}}, \bibinfo {author} {\bibfnamefont
  {R.}~\bibnamefont {Schoelkopf}},\ and\ \bibinfo {author} {\bibfnamefont
  {M.}~\bibnamefont {Devoret}},\ }\bibfield  {title} {\bibinfo {title}
  {Josephson directional amplifier for quantum measurement of superconducting
  circuits},\ }\href {https://doi.org/10.1103/PhysRevLett.112.167701}
  {\bibfield  {journal} {\bibinfo  {journal} {Phys. Rev. Lett.}\ }\textbf
  {\bibinfo {volume} {112}},\ \bibinfo {pages} {167701} (\bibinfo {year}
  {2014})}\BibitemShut {NoStop}%
\bibitem [{\citenamefont {Sliwa}\ \emph {et~al.}(2015)\citenamefont {Sliwa},
  \citenamefont {Hatridge}, \citenamefont {Narla}, \citenamefont {Shankar},
  \citenamefont {Frunzio}, \citenamefont {Schoelkopf},\ and\ \citenamefont
  {Devoret}}]{Devoret2015}%
  \BibitemOpen
  \bibfield  {author} {\bibinfo {author} {\bibfnamefont {K.~M.}\ \bibnamefont
  {Sliwa}}, \bibinfo {author} {\bibfnamefont {M.}~\bibnamefont {Hatridge}},
  \bibinfo {author} {\bibfnamefont {A.}~\bibnamefont {Narla}}, \bibinfo
  {author} {\bibfnamefont {S.}~\bibnamefont {Shankar}}, \bibinfo {author}
  {\bibfnamefont {L.}~\bibnamefont {Frunzio}}, \bibinfo {author} {\bibfnamefont
  {R.~J.}\ \bibnamefont {Schoelkopf}},\ and\ \bibinfo {author} {\bibfnamefont
  {M.~H.}\ \bibnamefont {Devoret}},\ }\bibfield  {title} {\bibinfo {title}
  {Reconfigurable josephson circulator/directional amplifier},\ }\href
  {https://doi.org/10.1103/PhysRevX.5.041020} {\bibfield  {journal} {\bibinfo
  {journal} {Phys. Rev. X}\ }\textbf {\bibinfo {volume} {5}},\ \bibinfo {pages}
  {041020} (\bibinfo {year} {2015})}\BibitemShut {NoStop}%
\bibitem [{\citenamefont {Ruesink}\ \emph {et~al.}(2016)\citenamefont
  {Ruesink}, \citenamefont {Miri}, \citenamefont {Al{\`{u}}},\ and\
  \citenamefont {Verhagen}}]{Verhagen2016}%
  \BibitemOpen
  \bibfield  {author} {\bibinfo {author} {\bibfnamefont {F.}~\bibnamefont
  {Ruesink}}, \bibinfo {author} {\bibfnamefont {M.-A.}\ \bibnamefont {Miri}},
  \bibinfo {author} {\bibfnamefont {A.}~\bibnamefont {Al{\`{u}}}},\ and\
  \bibinfo {author} {\bibfnamefont {E.}~\bibnamefont {Verhagen}},\ }\bibfield
  {title} {\bibinfo {title} {nonreciprocity and magnetic-free isolation based
  on optomechanical interactions},\ }\href
  {https://doi.org/10.1038/ncomms13662} {\bibfield  {journal} {\bibinfo
  {journal} {Nat. Commun.}\ }\textbf {\bibinfo {volume} {7}},\ \bibinfo {pages}
  {13662} (\bibinfo {year} {2016})}\BibitemShut {NoStop}%
\bibitem [{\citenamefont {Lecocq}\ \emph {et~al.}(2017)\citenamefont {Lecocq},
  \citenamefont {Ranzani}, \citenamefont {Peterson}, \citenamefont {Cicak},
  \citenamefont {Simmonds}, \citenamefont {Teufel},\ and\ \citenamefont
  {Aumentado}}]{Aumentado2017}%
  \BibitemOpen
  \bibfield  {author} {\bibinfo {author} {\bibfnamefont {F.}~\bibnamefont
  {Lecocq}}, \bibinfo {author} {\bibfnamefont {L.}~\bibnamefont {Ranzani}},
  \bibinfo {author} {\bibfnamefont {G.~A.}\ \bibnamefont {Peterson}}, \bibinfo
  {author} {\bibfnamefont {K.}~\bibnamefont {Cicak}}, \bibinfo {author}
  {\bibfnamefont {R.~W.}\ \bibnamefont {Simmonds}}, \bibinfo {author}
  {\bibfnamefont {J.~D.}\ \bibnamefont {Teufel}},\ and\ \bibinfo {author}
  {\bibfnamefont {J.}~\bibnamefont {Aumentado}},\ }\bibfield  {title} {\bibinfo
  {title} {nonreciprocal microwave signal processing with a field-programmable
  josephson amplifier},\ }\href
  {https://doi.org/10.1103/PhysRevApplied.7.024028} {\bibfield  {journal}
  {\bibinfo  {journal} {Phys. Rev. Applied}\ }\textbf {\bibinfo {volume} {7}},\
  \bibinfo {pages} {024028} (\bibinfo {year} {2017})}\BibitemShut {NoStop}%
\bibitem [{\citenamefont {Fang}\ \emph {et~al.}(2017)\citenamefont {Fang},
  \citenamefont {Luo}, \citenamefont {Metelmann}, \citenamefont {Matheny},
  \citenamefont {Marquardt}, \citenamefont {Clerk},\ and\ \citenamefont
  {Painter}}]{Painter2017}%
  \BibitemOpen
  \bibfield  {author} {\bibinfo {author} {\bibfnamefont {K.}~\bibnamefont
  {Fang}}, \bibinfo {author} {\bibfnamefont {J.}~\bibnamefont {Luo}}, \bibinfo
  {author} {\bibfnamefont {A.}~\bibnamefont {Metelmann}}, \bibinfo {author}
  {\bibfnamefont {M.~H.}\ \bibnamefont {Matheny}}, \bibinfo {author}
  {\bibfnamefont {F.}~\bibnamefont {Marquardt}}, \bibinfo {author}
  {\bibfnamefont {A.~A.}\ \bibnamefont {Clerk}},\ and\ \bibinfo {author}
  {\bibfnamefont {O.}~\bibnamefont {Painter}},\ }\bibfield  {title} {\bibinfo
  {title} {Generalized nonreciprocity in an optomechanical circuit via
  synthetic magnetism and reservoir engineering},\ }\href
  {https://doi.org/10.1038/nphys4009} {\bibfield  {journal} {\bibinfo
  {journal} {Nat. Phys.}\ }\textbf {\bibinfo {volume} {13}},\ \bibinfo {pages}
  {465} (\bibinfo {year} {2017})}\BibitemShut {NoStop}%
\bibitem [{\citenamefont {Peterson}\ \emph {et~al.}(2017)\citenamefont
  {Peterson}, \citenamefont {Lecocq}, \citenamefont {Cicak}, \citenamefont
  {Simmonds}, \citenamefont {Aumentado},\ and\ \citenamefont
  {Teufel}}]{Teufel2017}%
  \BibitemOpen
  \bibfield  {author} {\bibinfo {author} {\bibfnamefont {G.~A.}\ \bibnamefont
  {Peterson}}, \bibinfo {author} {\bibfnamefont {F.}~\bibnamefont {Lecocq}},
  \bibinfo {author} {\bibfnamefont {K.}~\bibnamefont {Cicak}}, \bibinfo
  {author} {\bibfnamefont {R.~W.}\ \bibnamefont {Simmonds}}, \bibinfo {author}
  {\bibfnamefont {J.}~\bibnamefont {Aumentado}},\ and\ \bibinfo {author}
  {\bibfnamefont {J.~D.}\ \bibnamefont {Teufel}},\ }\bibfield  {title}
  {\bibinfo {title} {Demonstration of efficient nonreciprocity in a microwave
  optomechanical circuit},\ }\href {https://doi.org/10.1103/PhysRevX.7.031001}
  {\bibfield  {journal} {\bibinfo  {journal} {Phys. Rev. X}\ }\textbf {\bibinfo
  {volume} {7}},\ \bibinfo {pages} {031001} (\bibinfo {year}
  {2017})}\BibitemShut {NoStop}%
\bibitem [{\citenamefont {Bernier}\ \emph {et~al.}(2017)\citenamefont
  {Bernier}, \citenamefont {T{\'{o}}th}, \citenamefont {Koottandavida},
  \citenamefont {Ioannou}, \citenamefont {Malz}, \citenamefont {Nunnenkamp},
  \citenamefont {Feofanov},\ and\ \citenamefont {Kippenberg}}]{Kippenberg2017}%
  \BibitemOpen
  \bibfield  {author} {\bibinfo {author} {\bibfnamefont {N.~R.}\ \bibnamefont
  {Bernier}}, \bibinfo {author} {\bibfnamefont {L.~D.}\ \bibnamefont
  {T{\'{o}}th}}, \bibinfo {author} {\bibfnamefont {A.}~\bibnamefont
  {Koottandavida}}, \bibinfo {author} {\bibfnamefont {M.~A.}\ \bibnamefont
  {Ioannou}}, \bibinfo {author} {\bibfnamefont {D.}~\bibnamefont {Malz}},
  \bibinfo {author} {\bibfnamefont {A.}~\bibnamefont {Nunnenkamp}}, \bibinfo
  {author} {\bibfnamefont {A.~K.}\ \bibnamefont {Feofanov}},\ and\ \bibinfo
  {author} {\bibfnamefont {T.~J.}\ \bibnamefont {Kippenberg}},\ }\bibfield
  {title} {\bibinfo {title} {nonreciprocal reconfigurable microwave
  optomechanical circuit},\ }\href {https://doi.org/10.1038/s41467-017-00447-1}
  {\bibfield  {journal} {\bibinfo  {journal} {Nat. Commun.}\ }\textbf {\bibinfo
  {volume} {8}},\ \bibinfo {pages} {604} (\bibinfo {year} {2017})}\BibitemShut
  {NoStop}%
\bibitem [{\citenamefont {Xu}\ \emph {et~al.}(2019)\citenamefont {Xu},
  \citenamefont {Jiang}, \citenamefont {Clerk},\ and\ \citenamefont
  {Harris}}]{Harris2019}%
  \BibitemOpen
  \bibfield  {author} {\bibinfo {author} {\bibfnamefont {H.}~\bibnamefont
  {Xu}}, \bibinfo {author} {\bibfnamefont {L.}~\bibnamefont {Jiang}}, \bibinfo
  {author} {\bibfnamefont {A.~A.}\ \bibnamefont {Clerk}},\ and\ \bibinfo
  {author} {\bibfnamefont {J.~G.~E.}\ \bibnamefont {Harris}},\ }\bibfield
  {title} {\bibinfo {title} {nonreciprocal control and cooling of phonon modes
  in an optomechanical system},\ }\href
  {https://doi.org/10.1038/s41586-019-1061-2} {\bibfield  {journal} {\bibinfo
  {journal} {Nature}\ }\textbf {\bibinfo {volume} {568}},\ \bibinfo {pages}
  {65} (\bibinfo {year} {2019})}\BibitemShut {NoStop}%
\bibitem [{\citenamefont {Malz}\ \emph {et~al.}(2018)\citenamefont {Malz},
  \citenamefont {T\'oth}, \citenamefont {Bernier}, \citenamefont {Feofanov},
  \citenamefont {Kippenberg},\ and\ \citenamefont
  {Nunnenkamp}}]{Nunnenkamp2018}%
  \BibitemOpen
  \bibfield  {author} {\bibinfo {author} {\bibfnamefont {D.}~\bibnamefont
  {Malz}}, \bibinfo {author} {\bibfnamefont {L.~D.}\ \bibnamefont {T\'oth}},
  \bibinfo {author} {\bibfnamefont {N.~R.}\ \bibnamefont {Bernier}}, \bibinfo
  {author} {\bibfnamefont {A.~K.}\ \bibnamefont {Feofanov}}, \bibinfo {author}
  {\bibfnamefont {T.~J.}\ \bibnamefont {Kippenberg}},\ and\ \bibinfo {author}
  {\bibfnamefont {A.}~\bibnamefont {Nunnenkamp}},\ }\bibfield  {title}
  {\bibinfo {title} {Quantum-limited directional amplifiers with
  optomechanics},\ }\href {https://doi.org/10.1103/PhysRevLett.120.023601}
  {\bibfield  {journal} {\bibinfo  {journal} {Phys. Rev. Lett.}\ }\textbf
  {\bibinfo {volume} {120}},\ \bibinfo {pages} {023601} (\bibinfo {year}
  {2018})}\BibitemShut {NoStop}%
\bibitem [{\citenamefont {Gardiner}(1993)}]{Gardiner1993}%
  \BibitemOpen
  \bibfield  {author} {\bibinfo {author} {\bibfnamefont {C.~W.}\ \bibnamefont
  {Gardiner}},\ }\bibfield  {title} {\bibinfo {title} {Driving a quantum system
  with the output field from another driven quantum system},\ }\href
  {https://doi.org/10.1103/PhysRevLett.70.2269} {\bibfield  {journal} {\bibinfo
   {journal} {Phys. Rev. Lett.}\ }\textbf {\bibinfo {volume} {70}},\ \bibinfo
  {pages} {2269} (\bibinfo {year} {1993})}\BibitemShut {NoStop}%
\bibitem [{\citenamefont {Carmichael}(1993)}]{Carmichael1993}%
  \BibitemOpen
  \bibfield  {author} {\bibinfo {author} {\bibfnamefont {H.~J.}\ \bibnamefont
  {Carmichael}},\ }\bibfield  {title} {\bibinfo {title} {Quantum trajectory
  theory for cascaded open systems},\ }\href
  {https://doi.org/10.1103/PhysRevLett.70.2273} {\bibfield  {journal} {\bibinfo
   {journal} {Phys. Rev. Lett.}\ }\textbf {\bibinfo {volume} {70}},\ \bibinfo
  {pages} {2273} (\bibinfo {year} {1993})}\BibitemShut {NoStop}%
\bibitem [{\citenamefont {Metelmann}\ and\ \citenamefont
  {Clerk}(2015)}]{Clerk2015}%
  \BibitemOpen
  \bibfield  {author} {\bibinfo {author} {\bibfnamefont {A.}~\bibnamefont
  {Metelmann}}\ and\ \bibinfo {author} {\bibfnamefont {A.~A.}\ \bibnamefont
  {Clerk}},\ }\bibfield  {title} {\bibinfo {title} {nonreciprocal photon
  transmission and amplification via reservoir engineering},\ }\href
  {https://doi.org/10.1103/PhysRevX.5.021025} {\bibfield  {journal} {\bibinfo
  {journal} {Phys. Rev. X}\ }\textbf {\bibinfo {volume} {5}},\ \bibinfo {pages}
  {021025} (\bibinfo {year} {2015})}\BibitemShut {NoStop}%
\bibitem [{\citenamefont {Metelmann}\ and\ \citenamefont
  {Clerk}(2017)}]{Clerk2017}%
  \BibitemOpen
  \bibfield  {author} {\bibinfo {author} {\bibfnamefont {A.}~\bibnamefont
  {Metelmann}}\ and\ \bibinfo {author} {\bibfnamefont {A.~A.}\ \bibnamefont
  {Clerk}},\ }\bibfield  {title} {\bibinfo {title} {nonreciprocal quantum
  interactions and devices via autonomous feedforward},\ }\href
  {https://doi.org/10.1103/PhysRevA.95.013837} {\bibfield  {journal} {\bibinfo
  {journal} {Phys. Rev. A}\ }\textbf {\bibinfo {volume} {95}},\ \bibinfo
  {pages} {013837} (\bibinfo {year} {2017})}\BibitemShut {NoStop}%
\bibitem [{\citenamefont {Verstraete}\ \emph {et~al.}(2009)\citenamefont
  {Verstraete}, \citenamefont {Wolf},\ and\ \citenamefont
  {Cirac}}]{Verstraete2009}%
  \BibitemOpen
  \bibfield  {author} {\bibinfo {author} {\bibfnamefont {F.}~\bibnamefont
  {Verstraete}}, \bibinfo {author} {\bibfnamefont {M.~M.}\ \bibnamefont
  {Wolf}},\ and\ \bibinfo {author} {\bibfnamefont {J.~I.}\ \bibnamefont
  {Cirac}},\ }\bibfield  {title} {\bibinfo {title} {Quantum computation and
  quantum-state engineering driven by dissipation},\ }\href
  {https://doi.org/10.1038/nphys1342} {\bibfield  {journal} {\bibinfo
  {journal} {Nat. Phys.}\ }\textbf {\bibinfo {volume} {5}},\ \bibinfo {pages}
  {633} (\bibinfo {year} {2009})}\BibitemShut {NoStop}%
\bibitem [{\citenamefont {Zanardi}\ and\ \citenamefont
  {Campos~Venuti}(2014)}]{CamposVenuti2014}%
  \BibitemOpen
  \bibfield  {author} {\bibinfo {author} {\bibfnamefont {P.}~\bibnamefont
  {Zanardi}}\ and\ \bibinfo {author} {\bibfnamefont {L.}~\bibnamefont
  {Campos~Venuti}},\ }\bibfield  {title} {\bibinfo {title} {Coherent quantum
  dynamics in steady-state manifolds of strongly dissipative systems},\ }\href
  {https://doi.org/10.1103/PhysRevLett.113.240406} {\bibfield  {journal}
  {\bibinfo  {journal} {Phys. Rev. Lett.}\ }\textbf {\bibinfo {volume} {113}},\
  \bibinfo {pages} {240406} (\bibinfo {year} {2014})}\BibitemShut {NoStop}%
\bibitem [{\citenamefont {Albert}\ \emph {et~al.}(2016)\citenamefont {Albert},
  \citenamefont {Bradlyn}, \citenamefont {Fraas},\ and\ \citenamefont
  {Jiang}}]{Jiang2016}%
  \BibitemOpen
  \bibfield  {author} {\bibinfo {author} {\bibfnamefont {V.~V.}\ \bibnamefont
  {Albert}}, \bibinfo {author} {\bibfnamefont {B.}~\bibnamefont {Bradlyn}},
  \bibinfo {author} {\bibfnamefont {M.}~\bibnamefont {Fraas}},\ and\ \bibinfo
  {author} {\bibfnamefont {L.}~\bibnamefont {Jiang}},\ }\bibfield  {title}
  {\bibinfo {title} {Geometry and response of lindbladians},\ }\href
  {https://doi.org/10.1103/PhysRevX.6.041031} {\bibfield  {journal} {\bibinfo
  {journal} {Phys. Rev. X}\ }\textbf {\bibinfo {volume} {6}},\ \bibinfo {pages}
  {041031} (\bibinfo {year} {2016})}\BibitemShut {NoStop}%
\bibitem [{\citenamefont {Arenz}\ and\ \citenamefont
  {Metelmann}(2020)}]{Metelmann2020}%
  \BibitemOpen
  \bibfield  {author} {\bibinfo {author} {\bibfnamefont {C.}~\bibnamefont
  {Arenz}}\ and\ \bibinfo {author} {\bibfnamefont {A.}~\bibnamefont
  {Metelmann}},\ }\bibfield  {title} {\bibinfo {title} {Emerging unitary
  evolutions in dissipatively coupled systems},\ }\href
  {https://doi.org/10.1103/PhysRevA.101.022101} {\bibfield  {journal} {\bibinfo
   {journal} {Phys. Rev. A}\ }\textbf {\bibinfo {volume} {101}},\ \bibinfo
  {pages} {022101} (\bibinfo {year} {2020})}\BibitemShut {NoStop}%
\bibitem [{\citenamefont {Misra}\ and\ \citenamefont
  {Sudarshan}(1977)}]{Sudarshan1977}%
  \BibitemOpen
  \bibfield  {author} {\bibinfo {author} {\bibfnamefont {B.}~\bibnamefont
  {Misra}}\ and\ \bibinfo {author} {\bibfnamefont {E.~C.~G.}\ \bibnamefont
  {Sudarshan}},\ }\bibfield  {title} {\bibinfo {title} {The zeno's paradox in
  quantum theory},\ }\href {https://doi.org/10.1063/1.523304} {\bibfield
  {journal} {\bibinfo  {journal} {J. Math. Phys.}\ }\textbf {\bibinfo {volume}
  {18}},\ \bibinfo {pages} {756} (\bibinfo {year} {1977})}\BibitemShut
  {NoStop}%
\bibitem [{\citenamefont {Facchi}\ and\ \citenamefont
  {Pascazio}(2002)}]{Pascazio2002}%
  \BibitemOpen
  \bibfield  {author} {\bibinfo {author} {\bibfnamefont {P.}~\bibnamefont
  {Facchi}}\ and\ \bibinfo {author} {\bibfnamefont {S.}~\bibnamefont
  {Pascazio}},\ }\bibfield  {title} {\bibinfo {title} {Quantum zeno
  subspaces},\ }\href {https://doi.org/10.1103/PhysRevLett.89.080401}
  {\bibfield  {journal} {\bibinfo  {journal} {Phys. Rev. Lett.}\ }\textbf
  {\bibinfo {volume} {89}},\ \bibinfo {pages} {080401} (\bibinfo {year}
  {2002})}\BibitemShut {NoStop}%
\bibitem [{\citenamefont {Facchi}\ and\ \citenamefont
  {Pascazio}(2008)}]{Pascazio2008}%
  \BibitemOpen
  \bibfield  {author} {\bibinfo {author} {\bibfnamefont {P.}~\bibnamefont
  {Facchi}}\ and\ \bibinfo {author} {\bibfnamefont {S.}~\bibnamefont
  {Pascazio}},\ }\bibfield  {title} {\bibinfo {title} {Quantum zeno dynamics:
  mathematical and physical aspects},\ }\href
  {https://doi.org/10.1088/1751-8113/41/49/493001} {\bibfield  {journal}
  {\bibinfo  {journal} {J. Phys. A}\ }\textbf {\bibinfo {volume} {41}},\
  \bibinfo {pages} {493001} (\bibinfo {year} {2008})}\BibitemShut {NoStop}%
\bibitem [{\citenamefont {Layden}\ \emph {et~al.}(2016)\citenamefont {Layden},
  \citenamefont {Mart\'{\i}n-Mart\'{\i}nez},\ and\ \citenamefont
  {Kempf}}]{Kempf2016}%
  \BibitemOpen
  \bibfield  {author} {\bibinfo {author} {\bibfnamefont {D.}~\bibnamefont
  {Layden}}, \bibinfo {author} {\bibfnamefont {E.}~\bibnamefont
  {Mart\'{\i}n-Mart\'{\i}nez}},\ and\ \bibinfo {author} {\bibfnamefont
  {A.}~\bibnamefont {Kempf}},\ }\bibfield  {title} {\bibinfo {title} {Universal
  scheme for indirect quantum control},\ }\href
  {https://doi.org/10.1103/PhysRevA.93.040301} {\bibfield  {journal} {\bibinfo
  {journal} {Phys. Rev. A}\ }\textbf {\bibinfo {volume} {93}},\ \bibinfo
  {pages} {040301} (\bibinfo {year} {2016})}\BibitemShut {NoStop}%
\bibitem [{\citenamefont {Nielsen}\ and\ \citenamefont
  {Chuang}(2010)}]{NielsenChuang2010}%
  \BibitemOpen
  \bibfield  {author} {\bibinfo {author} {\bibfnamefont {M.~A.}\ \bibnamefont
  {Nielsen}}\ and\ \bibinfo {author} {\bibfnamefont {I.~L.}\ \bibnamefont
  {Chuang}},\ }\href {https://doi.org/10.1017/CBO9780511976667} {\emph
  {\bibinfo {title} {Quantum Computation and Quantum Information: 10th
  Anniversary Edition}}}\ (\bibinfo  {publisher} {Cambridge University Press},\
  \bibinfo {year} {2010})\BibitemShut {NoStop}%
\bibitem [{\citenamefont {Watrous}(2018)}]{Watrous2018}%
  \BibitemOpen
  \bibfield  {author} {\bibinfo {author} {\bibfnamefont {J.}~\bibnamefont
  {Watrous}},\ }\href {https://doi.org/10.1017/9781316848142} {\emph {\bibinfo
  {title} {The Theory of Quantum Information}}}\ (\bibinfo  {publisher}
  {Cambridge University Press},\ \bibinfo {year} {2018})\BibitemShut {NoStop}%
\bibitem [{\citenamefont {Kitaev}(1997)}]{Kitaev1997}%
  \BibitemOpen
  \bibfield  {author} {\bibinfo {author} {\bibfnamefont {A.~Y.}\ \bibnamefont
  {Kitaev}},\ }\bibfield  {title} {\bibinfo {title} {Quantum computations:
  algorithms and error correction},\ }\href
  {https://doi.org/10.1070/rm1997v052n06abeh002155} {\bibfield  {journal}
  {\bibinfo  {journal} {Russ. Math. Surv.}\ }\textbf {\bibinfo {volume} {52}},\
  \bibinfo {pages} {1191} (\bibinfo {year} {1997})}\BibitemShut {NoStop}%
\bibitem [{dno()}]{dnorm_isom}%
  \BibitemOpen
  \href@noop {} {}\bibinfo {note} {This calculation amounts to computing the
  diamond norm between two unitary maps. An analytical derivation of diamond
  norm with such forms can be found in e.g.~Sec.~3.3 of
  Ref.~\cite{Watrous2018}}\BibitemShut {NoStop}%
\bibitem [{\citenamefont {K{\"{u}}mmerer}\ and\ \citenamefont
  {Maassen}(1987)}]{maassen1987}%
  \BibitemOpen
  \bibfield  {author} {\bibinfo {author} {\bibfnamefont {B.}~\bibnamefont
  {K{\"{u}}mmerer}}\ and\ \bibinfo {author} {\bibfnamefont {H.}~\bibnamefont
  {Maassen}},\ }\bibfield  {title} {\bibinfo {title} {The essentially
  commutative dilations of dynamical semigroups {onM} n},\ }\href
  {https://doi.org/10.1007/bf01205670} {\bibfield  {journal} {\bibinfo
  {journal} {Commun. Math. Phys.}\ }\textbf {\bibinfo {volume} {109}},\
  \bibinfo {pages} {1} (\bibinfo {year} {1987})}\BibitemShut {NoStop}%
\bibitem [{\citenamefont {Fagnola}\ \emph {et~al.}(2019)\citenamefont
  {Fagnola}, \citenamefont {Gough}, \citenamefont {Nurdin},\ and\ \citenamefont
  {Viola}}]{viola2019}%
  \BibitemOpen
  \bibfield  {author} {\bibinfo {author} {\bibfnamefont {F.}~\bibnamefont
  {Fagnola}}, \bibinfo {author} {\bibfnamefont {J.~E.}\ \bibnamefont {Gough}},
  \bibinfo {author} {\bibfnamefont {H.~I.}\ \bibnamefont {Nurdin}},\ and\
  \bibinfo {author} {\bibfnamefont {L.}~\bibnamefont {Viola}},\ }\bibfield
  {title} {\bibinfo {title} {Mathematical models of markovian dephasing},\
  }\href {https://doi.org/10.1088/1751-8121/ab38ec} {\bibfield  {journal}
  {\bibinfo  {journal} {J. Phys. A}\ }\textbf {\bibinfo {volume} {52}},\
  \bibinfo {pages} {385301} (\bibinfo {year} {2019})}\BibitemShut {NoStop}%
\bibitem [{\citenamefont {Wiseman}(1994)}]{Wiseman1994}%
  \BibitemOpen
  \bibfield  {author} {\bibinfo {author} {\bibfnamefont {H.~M.}\ \bibnamefont
  {Wiseman}},\ }\bibfield  {title} {\bibinfo {title} {Quantum theory of
  continuous feedback},\ }\href {https://doi.org/10.1103/PhysRevA.49.2133}
  {\bibfield  {journal} {\bibinfo  {journal} {Phys. Rev. A}\ }\textbf {\bibinfo
  {volume} {49}},\ \bibinfo {pages} {2133} (\bibinfo {year}
  {1994})}\BibitemShut {NoStop}%
\bibitem [{\citenamefont {Kono}\ \emph {et~al.}(2018)\citenamefont {Kono},
  \citenamefont {Koshino}, \citenamefont {Tabuchi}, \citenamefont {Noguchi},\
  and\ \citenamefont {Nakamura}}]{Nakamura2018}%
  \BibitemOpen
  \bibfield  {author} {\bibinfo {author} {\bibfnamefont {S.}~\bibnamefont
  {Kono}}, \bibinfo {author} {\bibfnamefont {K.}~\bibnamefont {Koshino}},
  \bibinfo {author} {\bibfnamefont {Y.}~\bibnamefont {Tabuchi}}, \bibinfo
  {author} {\bibfnamefont {A.}~\bibnamefont {Noguchi}},\ and\ \bibinfo {author}
  {\bibfnamefont {Y.}~\bibnamefont {Nakamura}},\ }\bibfield  {title} {\bibinfo
  {title} {Quantum non-demolition detection of an itinerant microwave photon},\
  }\href {https://doi.org/10.1038/s41567-018-0066-3} {\bibfield  {journal}
  {\bibinfo  {journal} {Nat. Phys.}\ }\textbf {\bibinfo {volume} {14}},\
  \bibinfo {pages} {546} (\bibinfo {year} {2018})}\BibitemShut {NoStop}%
\bibitem [{\citenamefont {Besse}\ \emph {et~al.}(2018)\citenamefont {Besse},
  \citenamefont {Gasparinetti}, \citenamefont {Collodo}, \citenamefont
  {Walter}, \citenamefont {Kurpiers}, \citenamefont {Pechal}, \citenamefont
  {Eichler},\ and\ \citenamefont {Wallraff}}]{Wallraff2018}%
  \BibitemOpen
  \bibfield  {author} {\bibinfo {author} {\bibfnamefont {J.-C.}\ \bibnamefont
  {Besse}}, \bibinfo {author} {\bibfnamefont {S.}~\bibnamefont {Gasparinetti}},
  \bibinfo {author} {\bibfnamefont {M.~C.}\ \bibnamefont {Collodo}}, \bibinfo
  {author} {\bibfnamefont {T.}~\bibnamefont {Walter}}, \bibinfo {author}
  {\bibfnamefont {P.}~\bibnamefont {Kurpiers}}, \bibinfo {author}
  {\bibfnamefont {M.}~\bibnamefont {Pechal}}, \bibinfo {author} {\bibfnamefont
  {C.}~\bibnamefont {Eichler}},\ and\ \bibinfo {author} {\bibfnamefont
  {A.}~\bibnamefont {Wallraff}},\ }\bibfield  {title} {\bibinfo {title}
  {Single-shot quantum nondemolition detection of individual itinerant
  microwave photons},\ }\href {https://doi.org/10.1103/PhysRevX.8.021003}
  {\bibfield  {journal} {\bibinfo  {journal} {Phys. Rev. X}\ }\textbf {\bibinfo
  {volume} {8}},\ \bibinfo {pages} {021003} (\bibinfo {year}
  {2018})}\BibitemShut {NoStop}%
\bibitem [{\citenamefont {Besse}\ \emph {et~al.}(2020)\citenamefont {Besse},
  \citenamefont {Gasparinetti}, \citenamefont {Collodo}, \citenamefont
  {Walter}, \citenamefont {Remm}, \citenamefont {Krause}, \citenamefont
  {Eichler},\ and\ \citenamefont {Wallraff}}]{Wallraff2020}%
  \BibitemOpen
  \bibfield  {author} {\bibinfo {author} {\bibfnamefont {J.-C.}\ \bibnamefont
  {Besse}}, \bibinfo {author} {\bibfnamefont {S.}~\bibnamefont {Gasparinetti}},
  \bibinfo {author} {\bibfnamefont {M.~C.}\ \bibnamefont {Collodo}}, \bibinfo
  {author} {\bibfnamefont {T.}~\bibnamefont {Walter}}, \bibinfo {author}
  {\bibfnamefont {A.}~\bibnamefont {Remm}}, \bibinfo {author} {\bibfnamefont
  {J.}~\bibnamefont {Krause}}, \bibinfo {author} {\bibfnamefont
  {C.}~\bibnamefont {Eichler}},\ and\ \bibinfo {author} {\bibfnamefont
  {A.}~\bibnamefont {Wallraff}},\ }\bibfield  {title} {\bibinfo {title} {Parity
  detection of propagating microwave fields},\ }\href
  {https://doi.org/10.1103/PhysRevX.10.011046} {\bibfield  {journal} {\bibinfo
  {journal} {Phys. Rev. X}\ }\textbf {\bibinfo {volume} {10}},\ \bibinfo
  {pages} {011046} (\bibinfo {year} {2020})}\BibitemShut {NoStop}%
\bibitem [{\citenamefont {Shi}\ \emph {et~al.}(2015)\citenamefont {Shi},
  \citenamefont {Yu},\ and\ \citenamefont {Fan}}]{Fan2015}%
  \BibitemOpen
  \bibfield  {author} {\bibinfo {author} {\bibfnamefont {Y.}~\bibnamefont
  {Shi}}, \bibinfo {author} {\bibfnamefont {Z.}~\bibnamefont {Yu}},\ and\
  \bibinfo {author} {\bibfnamefont {S.}~\bibnamefont {Fan}},\ }\bibfield
  {title} {\bibinfo {title} {Limitations of nonlinear optical isolators due to
  dynamic reciprocity},\ }\href {https://doi.org/10.1038/nphoton.2015.79}
  {\bibfield  {journal} {\bibinfo  {journal} {Nat. Photonics}\ }\textbf
  {\bibinfo {volume} {9}},\ \bibinfo {pages} {388} (\bibinfo {year}
  {2015})}\BibitemShut {NoStop}%
\bibitem [{\citenamefont {Sounas}\ and\ \citenamefont
  {Al{\'{u}}}(2018)}]{Alu2018}%
  \BibitemOpen
  \bibfield  {author} {\bibinfo {author} {\bibfnamefont {D.~L.}\ \bibnamefont
  {Sounas}}\ and\ \bibinfo {author} {\bibfnamefont {A.}~\bibnamefont
  {Al{\'{u}}}},\ }\bibfield  {title} {\bibinfo {title} {nonreciprocity based on
  nonlinear resonances},\ }\href {https://doi.org/10.1109/LAWP.2018.2866913}
  {\bibfield  {journal} {\bibinfo  {journal} {IEEE Antennas Wirel. Propag.
  Lett.}\ }\textbf {\bibinfo {volume} {17}},\ \bibinfo {pages} {1958} (\bibinfo
  {year} {2018})}\BibitemShut {NoStop}%
\bibitem [{\citenamefont {Fratini}\ \emph {et~al.}(2014)\citenamefont
  {Fratini}, \citenamefont {Mascarenhas}, \citenamefont {Safari}, \citenamefont
  {Poizat}, \citenamefont {Valente}, \citenamefont {Auff\`eves}, \citenamefont
  {Gerace},\ and\ \citenamefont {Santos}}]{Santos2014}%
  \BibitemOpen
  \bibfield  {author} {\bibinfo {author} {\bibfnamefont {F.}~\bibnamefont
  {Fratini}}, \bibinfo {author} {\bibfnamefont {E.}~\bibnamefont
  {Mascarenhas}}, \bibinfo {author} {\bibfnamefont {L.}~\bibnamefont {Safari}},
  \bibinfo {author} {\bibfnamefont {J.-P.}\ \bibnamefont {Poizat}}, \bibinfo
  {author} {\bibfnamefont {D.}~\bibnamefont {Valente}}, \bibinfo {author}
  {\bibfnamefont {A.}~\bibnamefont {Auff\`eves}}, \bibinfo {author}
  {\bibfnamefont {D.}~\bibnamefont {Gerace}},\ and\ \bibinfo {author}
  {\bibfnamefont {M.~F.}\ \bibnamefont {Santos}},\ }\bibfield  {title}
  {\bibinfo {title} {Fabry-perot interferometer with quantum mirrors: Nonlinear
  light transport and rectification},\ }\href
  {https://doi.org/10.1103/PhysRevLett.113.243601} {\bibfield  {journal}
  {\bibinfo  {journal} {Phys. Rev. Lett.}\ }\textbf {\bibinfo {volume} {113}},\
  \bibinfo {pages} {243601} (\bibinfo {year} {2014})}\BibitemShut {NoStop}%
\bibitem [{\citenamefont {Rosario~Hamann}\ \emph {et~al.}(2018)\citenamefont
  {Rosario~Hamann}, \citenamefont {M\"uller}, \citenamefont {Jerger},
  \citenamefont {Zanner}, \citenamefont {Combes}, \citenamefont {Pletyukhov},
  \citenamefont {Weides}, \citenamefont {Stace},\ and\ \citenamefont
  {Fedorov}}]{Fedorov2018}%
  \BibitemOpen
  \bibfield  {author} {\bibinfo {author} {\bibfnamefont {A.}~\bibnamefont
  {Rosario~Hamann}}, \bibinfo {author} {\bibfnamefont {C.}~\bibnamefont
  {M\"uller}}, \bibinfo {author} {\bibfnamefont {M.}~\bibnamefont {Jerger}},
  \bibinfo {author} {\bibfnamefont {M.}~\bibnamefont {Zanner}}, \bibinfo
  {author} {\bibfnamefont {J.}~\bibnamefont {Combes}}, \bibinfo {author}
  {\bibfnamefont {M.}~\bibnamefont {Pletyukhov}}, \bibinfo {author}
  {\bibfnamefont {M.}~\bibnamefont {Weides}}, \bibinfo {author} {\bibfnamefont
  {T.~M.}\ \bibnamefont {Stace}},\ and\ \bibinfo {author} {\bibfnamefont
  {A.}~\bibnamefont {Fedorov}},\ }\bibfield  {title} {\bibinfo {title}
  {nonreciprocity realized with quantum nonlinearity},\ }\href
  {https://doi.org/10.1103/PhysRevLett.121.123601} {\bibfield  {journal}
  {\bibinfo  {journal} {Phys. Rev. Lett.}\ }\textbf {\bibinfo {volume} {121}},\
  \bibinfo {pages} {123601} (\bibinfo {year} {2018})}\BibitemShut {NoStop}%
\bibitem [{\citenamefont {Nefedkin}\ \emph {et~al.}(2022)\citenamefont
  {Nefedkin}, \citenamefont {Cotrufo}, \citenamefont {Krasnok},\ and\
  \citenamefont {Al{\'{u}}}}]{Alu2022}%
  \BibitemOpen
  \bibfield  {author} {\bibinfo {author} {\bibfnamefont {N.}~\bibnamefont
  {Nefedkin}}, \bibinfo {author} {\bibfnamefont {M.}~\bibnamefont {Cotrufo}},
  \bibinfo {author} {\bibfnamefont {A.}~\bibnamefont {Krasnok}},\ and\ \bibinfo
  {author} {\bibfnamefont {A.}~\bibnamefont {Al{\'{u}}}},\ }\bibfield  {title}
  {\bibinfo {title} {Dark-state induced quantum nonreciprocity},\ }\href
  {https://doi.org/https://doi.org/10.1002/qute.202100112} {\bibfield
  {journal} {\bibinfo  {journal} {Adv. Quantum Technol.}\ ,\ \bibinfo {pages}
  {2100112}} (\bibinfo {year} {2022})}\BibitemShut {NoStop}%
\bibitem [{\citenamefont {Nielsen}(2002)}]{Nielsen2002}%
  \BibitemOpen
  \bibfield  {author} {\bibinfo {author} {\bibfnamefont {M.~A.}\ \bibnamefont
  {Nielsen}},\ }\bibfield  {title} {\bibinfo {title} {A simple formula for the
  average gate fidelity of a quantum dynamical operation},\ }\href
  {https://doi.org/https://doi.org/10.1016/S0375-9601(02)01272-0} {\bibfield
  {journal} {\bibinfo  {journal} {Phys. Lett. A}\ }\textbf {\bibinfo {volume}
  {303}},\ \bibinfo {pages} {249} (\bibinfo {year} {2002})}\BibitemShut
  {NoStop}%
\bibitem [{com()}]{comp_mat}%
  \BibitemOpen
  \href@noop {} {}\bibinfo {note} {We use script letters with overhead check
  mark to denote matrices acting on the complex linear space $\mathbb{C}
  ^{N}$.}\BibitemShut {Stop}%
\bibitem [{\citenamefont {Breuer}\ and\ \citenamefont
  {Petruccione}(2002)}]{oqs2002book}%
  \BibitemOpen
  \bibfield  {author} {\bibinfo {author} {\bibfnamefont {H.-P.}\ \bibnamefont
  {Breuer}}\ and\ \bibinfo {author} {\bibfnamefont {F.}~\bibnamefont
  {Petruccione}},\ }\href@noop {} {\emph {\bibinfo {title} {The theory of open
  quantum systems}}}\ (\bibinfo  {publisher} {Oxford University Press on
  Demand},\ \bibinfo {year} {2002})\BibitemShut {NoStop}%
\bibitem [{\citenamefont {Mirrahimi}\ \emph {et~al.}(2014)\citenamefont
  {Mirrahimi}, \citenamefont {Leghtas}, \citenamefont {Albert}, \citenamefont
  {Touzard}, \citenamefont {Schoelkopf}, \citenamefont {Jiang},\ and\
  \citenamefont {Devoret}}]{Devoret2014}%
  \BibitemOpen
  \bibfield  {author} {\bibinfo {author} {\bibfnamefont {M.}~\bibnamefont
  {Mirrahimi}}, \bibinfo {author} {\bibfnamefont {Z.}~\bibnamefont {Leghtas}},
  \bibinfo {author} {\bibfnamefont {V.~V.}\ \bibnamefont {Albert}}, \bibinfo
  {author} {\bibfnamefont {S.}~\bibnamefont {Touzard}}, \bibinfo {author}
  {\bibfnamefont {R.~J.}\ \bibnamefont {Schoelkopf}}, \bibinfo {author}
  {\bibfnamefont {L.}~\bibnamefont {Jiang}},\ and\ \bibinfo {author}
  {\bibfnamefont {M.~H.}\ \bibnamefont {Devoret}},\ }\bibfield  {title}
  {\bibinfo {title} {Dynamically protected cat-qubits: a new paradigm for
  universal quantum computation},\ }\href
  {https://doi.org/10.1088/1367-2630/16/4/045014} {\bibfield  {journal}
  {\bibinfo  {journal} {New J. Phys.}\ }\textbf {\bibinfo {volume} {16}},\
  \bibinfo {pages} {045014} (\bibinfo {year} {2014})}\BibitemShut {NoStop}%
\bibitem [{\citenamefont {Lebreuilly}\ \emph {et~al.}(2021)\citenamefont
  {Lebreuilly}, \citenamefont {Noh}, \citenamefont {Wang}, \citenamefont
  {Girvin},\ and\ \citenamefont {Jiang}}]{Jiang2021}%
  \BibitemOpen
  \bibfield  {author} {\bibinfo {author} {\bibfnamefont {J.}~\bibnamefont
  {Lebreuilly}}, \bibinfo {author} {\bibfnamefont {K.}~\bibnamefont {Noh}},
  \bibinfo {author} {\bibfnamefont {C.-H.}\ \bibnamefont {Wang}}, \bibinfo
  {author} {\bibfnamefont {S.~M.}\ \bibnamefont {Girvin}},\ and\ \bibinfo
  {author} {\bibfnamefont {L.}~\bibnamefont {Jiang}},\ }\bibfield  {title}
  {\bibinfo {title} {Autonomous quantum error correction and quantum
  computation},\ }\href {https://arxiv.org/abs/2103.05007} {\bibfield
  {journal} {\bibinfo  {journal} {arXiv preprint arXiv:2103.05007}\ } (\bibinfo
  {year} {2021})}\BibitemShut {NoStop}%
\bibitem [{\citenamefont {Lescanne}\ \emph {et~al.}(2020)\citenamefont
  {Lescanne}, \citenamefont {Villiers}, \citenamefont {Peronnin}, \citenamefont
  {Sarlette}, \citenamefont {Delbecq}, \citenamefont {Huard}, \citenamefont
  {Kontos}, \citenamefont {Mirrahimi},\ and\ \citenamefont
  {Leghtas}}]{Leghtas2020}%
  \BibitemOpen
  \bibfield  {author} {\bibinfo {author} {\bibfnamefont {R.}~\bibnamefont
  {Lescanne}}, \bibinfo {author} {\bibfnamefont {M.}~\bibnamefont {Villiers}},
  \bibinfo {author} {\bibfnamefont {T.}~\bibnamefont {Peronnin}}, \bibinfo
  {author} {\bibfnamefont {A.}~\bibnamefont {Sarlette}}, \bibinfo {author}
  {\bibfnamefont {M.}~\bibnamefont {Delbecq}}, \bibinfo {author} {\bibfnamefont
  {B.}~\bibnamefont {Huard}}, \bibinfo {author} {\bibfnamefont
  {T.}~\bibnamefont {Kontos}}, \bibinfo {author} {\bibfnamefont
  {M.}~\bibnamefont {Mirrahimi}},\ and\ \bibinfo {author} {\bibfnamefont
  {Z.}~\bibnamefont {Leghtas}},\ }\bibfield  {title} {\bibinfo {title}
  {Exponential suppression of bit-flips in a qubit encoded in an oscillator},\
  }\href {https://doi.org/10.1038/s41567-020-0824-x} {\bibfield  {journal}
  {\bibinfo  {journal} {Nat. Phys.}\ }\textbf {\bibinfo {volume} {16}},\
  \bibinfo {pages} {509} (\bibinfo {year} {2020})}\BibitemShut {NoStop}%
\bibitem [{\citenamefont {Grimm}\ \emph {et~al.}(2020)\citenamefont {Grimm},
  \citenamefont {Frattini}, \citenamefont {Puri}, \citenamefont {Mundhada},
  \citenamefont {Touzard}, \citenamefont {Mirrahimi}, \citenamefont {Girvin},
  \citenamefont {Shankar},\ and\ \citenamefont {Devoret}}]{Devoret2020}%
  \BibitemOpen
  \bibfield  {author} {\bibinfo {author} {\bibfnamefont {A.}~\bibnamefont
  {Grimm}}, \bibinfo {author} {\bibfnamefont {N.~E.}\ \bibnamefont {Frattini}},
  \bibinfo {author} {\bibfnamefont {S.}~\bibnamefont {Puri}}, \bibinfo {author}
  {\bibfnamefont {S.~O.}\ \bibnamefont {Mundhada}}, \bibinfo {author}
  {\bibfnamefont {S.}~\bibnamefont {Touzard}}, \bibinfo {author} {\bibfnamefont
  {M.}~\bibnamefont {Mirrahimi}}, \bibinfo {author} {\bibfnamefont {S.~M.}\
  \bibnamefont {Girvin}}, \bibinfo {author} {\bibfnamefont {S.}~\bibnamefont
  {Shankar}},\ and\ \bibinfo {author} {\bibfnamefont {M.~H.}\ \bibnamefont
  {Devoret}},\ }\bibfield  {title} {\bibinfo {title} {Stabilization and
  operation of a kerr-cat qubit},\ }\href
  {https://doi.org/10.1038/s41586-020-2587-z} {\bibfield  {journal} {\bibinfo
  {journal} {Nature}\ }\textbf {\bibinfo {volume} {584}},\ \bibinfo {pages}
  {205} (\bibinfo {year} {2020})}\BibitemShut {NoStop}%
\bibitem [{\citenamefont {Gertler}\ \emph {et~al.}(2021)\citenamefont
  {Gertler}, \citenamefont {Baker}, \citenamefont {Li}, \citenamefont {Shirol},
  \citenamefont {Koch},\ and\ \citenamefont {Wang}}]{Wang2021}%
  \BibitemOpen
  \bibfield  {author} {\bibinfo {author} {\bibfnamefont {J.~M.}\ \bibnamefont
  {Gertler}}, \bibinfo {author} {\bibfnamefont {B.}~\bibnamefont {Baker}},
  \bibinfo {author} {\bibfnamefont {J.}~\bibnamefont {Li}}, \bibinfo {author}
  {\bibfnamefont {S.}~\bibnamefont {Shirol}}, \bibinfo {author} {\bibfnamefont
  {J.}~\bibnamefont {Koch}},\ and\ \bibinfo {author} {\bibfnamefont
  {C.}~\bibnamefont {Wang}},\ }\bibfield  {title} {\bibinfo {title} {Protecting
  a bosonic qubit with autonomous quantum error correction},\ }\href
  {https://doi.org/10.1038/s41586-021-03257-0} {\bibfield  {journal} {\bibinfo
  {journal} {Nature}\ }\textbf {\bibinfo {volume} {590}},\ \bibinfo {pages}
  {243} (\bibinfo {year} {2021})}\BibitemShut {NoStop}%
\bibitem [{\citenamefont {Fredrickson}\ and\ \citenamefont
  {Andersen}(1984)}]{Andersen1984}%
  \BibitemOpen
  \bibfield  {author} {\bibinfo {author} {\bibfnamefont {G.~H.}\ \bibnamefont
  {Fredrickson}}\ and\ \bibinfo {author} {\bibfnamefont {H.~C.}\ \bibnamefont
  {Andersen}},\ }\bibfield  {title} {\bibinfo {title} {Kinetic ising model of
  the glass transition},\ }\href {https://doi.org/10.1103/PhysRevLett.53.1244}
  {\bibfield  {journal} {\bibinfo  {journal} {Phys. Rev. Lett.}\ }\textbf
  {\bibinfo {volume} {53}},\ \bibinfo {pages} {1244} (\bibinfo {year}
  {1984})}\BibitemShut {NoStop}%
\bibitem [{\citenamefont {Zohar}\ \emph {et~al.}(2015)\citenamefont {Zohar},
  \citenamefont {Cirac},\ and\ \citenamefont {Reznik}}]{Reznik2015}%
  \BibitemOpen
  \bibfield  {author} {\bibinfo {author} {\bibfnamefont {E.}~\bibnamefont
  {Zohar}}, \bibinfo {author} {\bibfnamefont {J.~I.}\ \bibnamefont {Cirac}},\
  and\ \bibinfo {author} {\bibfnamefont {B.}~\bibnamefont {Reznik}},\
  }\bibfield  {title} {\bibinfo {title} {Quantum simulations of lattice gauge
  theories using ultracold atoms in optical lattices},\ }\href
  {https://doi.org/10.1088/0034-4885/79/1/014401} {\bibfield  {journal}
  {\bibinfo  {journal} {Rep. Prog. Phys.}\ }\textbf {\bibinfo {volume} {79}},\
  \bibinfo {pages} {014401} (\bibinfo {year} {2015})}\BibitemShut {NoStop}%
\bibitem [{\citenamefont {Barbiero}\ \emph {et~al.}(2019)\citenamefont
  {Barbiero}, \citenamefont {Schweizer}, \citenamefont {Aidelsburger},
  \citenamefont {Demler}, \citenamefont {Goldman},\ and\ \citenamefont
  {Grusdt}}]{Grusdt2019}%
  \BibitemOpen
  \bibfield  {author} {\bibinfo {author} {\bibfnamefont {L.}~\bibnamefont
  {Barbiero}}, \bibinfo {author} {\bibfnamefont {C.}~\bibnamefont {Schweizer}},
  \bibinfo {author} {\bibfnamefont {M.}~\bibnamefont {Aidelsburger}}, \bibinfo
  {author} {\bibfnamefont {E.}~\bibnamefont {Demler}}, \bibinfo {author}
  {\bibfnamefont {N.}~\bibnamefont {Goldman}},\ and\ \bibinfo {author}
  {\bibfnamefont {F.}~\bibnamefont {Grusdt}},\ }\bibfield  {title} {\bibinfo
  {title} {Coupling ultracold matter to dynamical gauge fields in optical
  lattices: From flux attachment to \&\#x2124;<sub>2</sub> lattice gauge
  theories},\ }\href {https://doi.org/10.1126/sciadv.aav7444} {\bibfield
  {journal} {\bibinfo  {journal} {Sci. Adv.}\ }\textbf {\bibinfo {volume}
  {5}},\ \bibinfo {pages} {eaav7444} (\bibinfo {year} {2019})}\BibitemShut
  {NoStop}%
\bibitem [{\citenamefont {Lienhard}\ \emph {et~al.}(2020)\citenamefont
  {Lienhard}, \citenamefont {Scholl}, \citenamefont {Weber}, \citenamefont
  {Barredo}, \citenamefont {de~L\'es\'eleuc}, \citenamefont {Bai},
  \citenamefont {Lang}, \citenamefont {Fleischhauer}, \citenamefont
  {B\"uchler}, \citenamefont {Lahaye},\ and\ \citenamefont
  {Browaeys}}]{Browaeys2020}%
  \BibitemOpen
  \bibfield  {author} {\bibinfo {author} {\bibfnamefont {V.}~\bibnamefont
  {Lienhard}}, \bibinfo {author} {\bibfnamefont {P.}~\bibnamefont {Scholl}},
  \bibinfo {author} {\bibfnamefont {S.}~\bibnamefont {Weber}}, \bibinfo
  {author} {\bibfnamefont {D.}~\bibnamefont {Barredo}}, \bibinfo {author}
  {\bibfnamefont {S.}~\bibnamefont {de~L\'es\'eleuc}}, \bibinfo {author}
  {\bibfnamefont {R.}~\bibnamefont {Bai}}, \bibinfo {author} {\bibfnamefont
  {N.}~\bibnamefont {Lang}}, \bibinfo {author} {\bibfnamefont {M.}~\bibnamefont
  {Fleischhauer}}, \bibinfo {author} {\bibfnamefont {H.~P.}\ \bibnamefont
  {B\"uchler}}, \bibinfo {author} {\bibfnamefont {T.}~\bibnamefont {Lahaye}},\
  and\ \bibinfo {author} {\bibfnamefont {A.}~\bibnamefont {Browaeys}},\
  }\bibfield  {title} {\bibinfo {title} {Realization of a density-dependent
  peierls phase in a synthetic, spin-orbit coupled rydberg system},\ }\href
  {https://doi.org/10.1103/PhysRevX.10.021031} {\bibfield  {journal} {\bibinfo
  {journal} {Phys. Rev. X}\ }\textbf {\bibinfo {volume} {10}},\ \bibinfo
  {pages} {021031} (\bibinfo {year} {2020})}\BibitemShut {NoStop}%
\bibitem [{\citenamefont {Celi}\ \emph {et~al.}(2020)\citenamefont {Celi},
  \citenamefont {Vermersch}, \citenamefont {Viyuela}, \citenamefont {Pichler},
  \citenamefont {Lukin},\ and\ \citenamefont {Zoller}}]{Zoller2020}%
  \BibitemOpen
  \bibfield  {author} {\bibinfo {author} {\bibfnamefont {A.}~\bibnamefont
  {Celi}}, \bibinfo {author} {\bibfnamefont {B.}~\bibnamefont {Vermersch}},
  \bibinfo {author} {\bibfnamefont {O.}~\bibnamefont {Viyuela}}, \bibinfo
  {author} {\bibfnamefont {H.}~\bibnamefont {Pichler}}, \bibinfo {author}
  {\bibfnamefont {M.~D.}\ \bibnamefont {Lukin}},\ and\ \bibinfo {author}
  {\bibfnamefont {P.}~\bibnamefont {Zoller}},\ }\bibfield  {title} {\bibinfo
  {title} {Emerging two-dimensional gauge theories in rydberg configurable
  arrays},\ }\href {https://doi.org/10.1103/PhysRevX.10.021057} {\bibfield
  {journal} {\bibinfo  {journal} {Phys. Rev. X}\ }\textbf {\bibinfo {volume}
  {10}},\ \bibinfo {pages} {021057} (\bibinfo {year} {2020})}\BibitemShut
  {NoStop}%
\bibitem [{\citenamefont {Khaneja}\ and\ \citenamefont
  {Glaser}(2001)}]{Glaser2001}%
  \BibitemOpen
  \bibfield  {author} {\bibinfo {author} {\bibfnamefont {N.}~\bibnamefont
  {Khaneja}}\ and\ \bibinfo {author} {\bibfnamefont {S.~J.}\ \bibnamefont
  {Glaser}},\ }\bibfield  {title} {\bibinfo {title} {Cartan decomposition of
  su(2n) and control of spin systems},\ }\href
  {https://doi.org/https://doi.org/10.1016/S0301-0104(01)00318-4} {\bibfield
  {journal} {\bibinfo  {journal} {Chem. Phys.}\ }\textbf {\bibinfo {volume}
  {267}},\ \bibinfo {pages} {11} (\bibinfo {year} {2001})}\BibitemShut
  {NoStop}%
\bibitem [{\citenamefont {Gardiner}\ and\ \citenamefont
  {Zoller}(2004)}]{Gardiner2004book}%
  \BibitemOpen
  \bibfield  {author} {\bibinfo {author} {\bibfnamefont {C.}~\bibnamefont
  {Gardiner}}\ and\ \bibinfo {author} {\bibfnamefont {P.}~\bibnamefont
  {Zoller}},\ }\href@noop {} {\emph {\bibinfo {title} {Quantum Noise: A
  Handbook of Markovian and Non-Markovian Quantum Stochastic Methods with
  Applications to Quantum Optics}}},\ Springer Series in Synergetics\ (\bibinfo
   {publisher} {Springer},\ \bibinfo {address} {Berlin},\ \bibinfo {year}
  {2004})\BibitemShut {NoStop}%
\end{thebibliography}
\end{document}